\documentclass{aa}

\usepackage{verbatim}
\usepackage{amsmath}
\usepackage{hyperref}
\usepackage{breakurl}
\hypersetup{draft} 
\usepackage{float}
\usepackage{color}
\usepackage[dvipsnames]{xcolor}
\usepackage{longtable}
\usepackage{caption}

\usepackage{graphicx}
\usepackage[normalem]{ulem}

\usepackage{txfonts}
\begin{document}

	\title{3D chemical structure of diffuse turbulent ISM}
	\subtitle{I - Statistics of the HI-to-H$_2$ transition}
	\author{
	E.~Bellomi \inst{\ref{lerma}, \ref{ENS}}
	\and  B.~Godard \inst{\ref{lerma},\ref{ENS}}
	\and  P.~Hennebelle \inst{\ref{CEA}}
	\and  V.~Valdivia \inst{\ref{CEA}}
	\and  G.~Pineau~des~For\^ets \inst{\ref{IAS},\ref{lerma}}
	\and  P.~Lesaffre \inst{\ref{ENS}}
	\and  M.~P\'erault \inst{\ref{ENS}}
}


\institute{
Observatoire de Paris, PSL University, Sorbonne Université, LERMA, 75014 Paris, France
	\label{lerma}\\
	\email{elena.bellomi@obspm.fr}
	\and
Laboratoire de Physique de l’Ecole normale supérieure, ENS, Université PSL, CNRS, Sorbonne Université, Université de Paris, F-75005 Paris, France
	\label{ENS}	
	\and
Laboratoire AIM, CEA/IRFU, CNRS/INSU, Universit\'e Paris Diderot, CEA-Saclay, 91191 Gif-Sur-Yvette, France 
  \label{CEA}
    \and
Université Paris-Saclay, CNRS, Institut d’Astrophysique Spatiale, 91405, Orsay, France
  \label{IAS}
}

	\date{}
	
	
	\abstract
	    {The amount of data collected by spectrometers from radio to ultraviolet (UV) wavelengths opens a new era where the statistical and chemical information contained in the observations can be used concomitantly to investigate the thermodynamical state and the evolution of the interstellar medium (ISM).
	    }
		{In this paper, we study the statistical properties of the HI-to-H$_2$ transition observed in absorption in the local diffuse and multiphase ISM. Our goal is to identify the physical processes that control the probability of occurrence of any line of sight and the origins of the variations of the integrated molecular fraction from one line of sight to another.}
		{The turbulent diffuse ISM is modeled using the RAMSES code, which includes detailed treatments of the magnetohydrodynamics (MHD), the thermal evolution of the gas, and the chemistry of H$_2$. The impacts of the UV radiation field, the mean density, the turbulent forcing, the integral scale, the magnetic field, and the gravity on the molecular content of the gas are explored through a parametric study that covers a wide range of physical conditions. The statistics of the HI-to-H$_2$ transition are interpreted through analytical prescriptions and compared with the observations using a modified and robust version of the Kolmogorov-Smirnov test.
		}
	    {The analysis of the observed background sources shows that the lengths of the lines of sight follow a flat distribution in logarithmic scale from $\sim 100$ pc to $\sim 3$ kpc. Without taking into account any variation of the parameters along a line of sight or from one line of sight to another, the results of one simulation, convolved with the distribution of distances of the observational sample, are able to simultaneously explain the position, the width, the dispersion, and most of the statistical properties of the HI-to-H$_2$ transition observed in the local ISM. The tightest agreement is obtained for a neutral diffuse gas modeled over $\sim 200$ pc, with a mean density $\overline{n_{\rm H}} =1-2$ cm$^{-3}$, illuminated by the standard interstellar UV radiation field, and stirred up by a large-scale compressive turbulent forcing. Within this configuration, the 2D  probability histogram (PH) of the column densities of H and H$_2$, poetically called the kingfisher diagram, is remarkably stable and is almost unaltered by gravity, the strength of the turbulent forcing, the resolution of the simulation, or the strength of the magnetic field $B_x$, as long as $B_x < 4$ $\mu$G. The weak effect of the resolution and our analytical prescription suggest that the column densities of HI are likely built up in large-scale warm neutral medium (WNM) and cold neutral medium (CNM) structures correlated in density over $\sim 20$ pc and $\sim 10$ pc, respectively, while those of H$_2$ are built up in CNM structures between $\sim 3$ pc and $\sim 10$ pc.
    	}
	    {Combining the chemical and statistical information contained in the observations of HI and H$_2$ sheds new light on the study of the diffuse matter. Applying this new tool to several atomic and molecular species is a promising perspective to understanding the effects of turbulence, magnetic field, thermal instability, and gravity on the formation and evolution of molecular clouds.
	    }

	
	\maketitle
	%

\section{Introduction}
The multiphase nature of the interstellar medium (ISM) is at the root of the regulation of star formation in galaxies (e.g., \citealt{hill_effect_2018}). As shown by the emission 
profiles of the HI 21 cm line \citep{heiles_millennium_2003,heiles_vizier_2003,murray_21-sponge_2015,murray_21-sponge_2018}, the diffuse neutral ISM is composed of two stable thermal states at thermal pressure equilibrium \citep{jenkins_distribution_2011}, the warm neutral medium (WNM, $T \sim 7000$ K) and the cold neutral medium (CNM, $T \sim 70$ K), coexisting with a third unstable state, the lukewarm neutral medium (LNM), whose temperature is comprised between those of the CNM and the WNM (e.g., \citealt{marchal_rohsa_2019}). Through condensation and evaporation processes, turbulent transport, and turbulent mixing, the diffuse matter flows from one stable state to the other eventually leading to the formation of dense and cold clouds massive enough to trigger gravitational collapse (e.g., \citealt{ostriker_regulation_2010}). While this picture is widely accepted, the intricated effects of turbulence, gravity, radiation field, and magnetic field on the exchange of mass and energy between the different phases and on the formation of structures at all scales has yet to be unveiled.

Following the illustrious analytical descriptions of the thermal instability process \citep{field_thermal_1965,wolfire_neutral_1995,wolfire_neutral_2003,bialy_thermal_2019}, several analytical and numerical studies have been dedicated to understand the dynamical evolution of the gas, focusing 
on the formation of CNM structures, molecular clouds, and collapsing cores (e.g., \citealt{hennebelle_dynamical_1999,koyama_origin_2002,koyama_production_2002,audit_thermal_2005,Vazquez_2007,hennebelle_warm_2008}), as well as on the stability of clouds of various geometries under evaporation and condensation 
conditions (e.g., \citealt{inoue_structure_2006,stone_mhd_2009,kim_instability_2013,nagashima_evaporation_2005,iwasaki_self-sustained_2014}). These show that large-scale turbulence combined with thermal instability is sufficient to explain several features of the neutral ISM, including the fractions of mass observed in the different thermal states \citep{seifried_forced_2011,Hennebelle_2014,hill_effect_2018}, the distribution of thermal pressure \citep{saury_structure_2014}, and the mass spectrum, the mass-size relation, and the velocity dispersion-size relation of molecular clouds (e.g., \citealt{audit_structure_2010,Padoan_2016,Iffrig_2017}). 

To extend the predictions of simulations to a larger set of observational diagnostics, recent numerical studies have undertaken the challenging task of solving the chemical evolution of turbulent and/or multiphase environments. Originally dedicated to the formation of CO in molecular clouds and to the analysis of the CO-to-H$_2$ conversion factor in galaxies and the CO dark-gas (e.g., \citealt{glover_modelling_2010,Smith_2014,Richings_2016a,Seifried_2017,Gong_2018}), numerical simulations are now used to study a variety of 
atomic and molecular tracers, including CII, CI, CH$^+$, OH$^+$, H$_2$O$^+$, and ArH$^+$ (e.g., \citealt{Richings_2016b,Valdivia_2017,Clark_2019,Bialy_2019a}).
All these works demonstrate the predictive power of astrochemistry. The column density distribution of each atom and molecule has a unique signature that provides detailed information on the thermodynamical state of the diffuse matter \citep{Clark_2019}. In turn, the confrontation with the predictions of numerical simulations can be used to estimate the scale and strength of the injection of mechanical energy by stellar feedback \citep{Bialy_2019a}, the large-scale turbulent transport and the interfaces between CNM and WNM \citep{Valdivia_2017}, and the nature of the turbulent dissipation processes \citep{Lesaffre_2020}.

	In this context, understanding the formation and survival of molecular hydrogen has long been recognized as a major topic of investigation. As the most abundant molecule in space, H$_2$ is at the root of interstellar chemistry and the growth of molecular complexity. In addition, and because its formation preferentially occurs in dense environments, H$_2$ naturally correlates with the star formation rate of galaxies (e.g., \citealt{lupi_simplified_2017}) and therefore offers a valuable proxy to understand the limit in the Kennicutt-Schmidt relation above which star formation occurs (e.g., \citealt{bigiel_star_2008,bigiel_scaling_2011,schruba_molecular_2011,leroy_molecular_2013}).
	
	Over the last decades, great efforts have thus been devoted to propose analytical descriptions of the HI-to-H$_2$ transition in homogeneous clouds with plane-parallel or spherical geometries (e.g., \citealt{sternberg_infrared_1988,krumholz_atomicmolecular_2008,mckee_atomic--molecular_2010,sternberg_h_2014} and references therein), compute this transition in detailed 1D chemical models assuming chemical equilibrium (e.g., \citealt{van_dishoeck_comprehensive_1986,abgrall_photodissociation_1992,le_bourlot_surface_2012,bron_surface_2014}) or not (e.g., \citealt{lee_photodissociation_1996,Goldsmith_2007,Lesaffre_2007}), treat the chemistry of H and H$_2$ in subgrid models applied to simulations of galaxy formations (e.g., \citealt{gnedin_modeling_2009,christensen_implementing_2012,thompson_molecular_2014,diemer_modeling_2018}), or solve it in 3D isothermal or multiphase simulations of the diffuse ISM using various treatments of the radiative transfer (e.g., \citealt{glover_modelling_2010,valdivia_h2_2016,hu_star_2016,bialy_h_2017,nickerson_simple_2018}).
	
	Thanks to all these works, a global picture of the formation of H$_2$ in galaxies is now emerging. At the scale of a homogeneous cloud, the molecular content, the sharpness of the HI-to-H$_2$ transition, and the asymptotic column density of HI are controlled by the ratio of the intensity of the ultraviolet (UV) field to the gas density and the dust-to-gas ratio, or equivalently, the metallicity \citep{sternberg_h_2014}. At larger scales, the integrated column densities of HI and H$_2$ also depend on the distribution of clouds of various densities along the line of sight and on the porosity to the UV radiation field. Because of these effects, the statistical properties of the total column density are found to depend on the strength, the scale, and the compressibility of the turbulent forcing in simulations of CNM gas \citep{micic_modelling_2012,bialy_h_2017}. The amount of molecular gas depends on the "clumpiness factor" used for the subgrid models in simulations of galaxy formation \citep{gnedin_modeling_2009,christensen_implementing_2012}.
	
	Despite these achievements, very few works have been dedicated so far to the analysis of the HI-to-H$_2$ transition in a turbulent multiphase medium at a scale sufficient to resolve the formation of CNM structures. In addition and while the predictions of analytical models (e.g., \citealt{krumholz_atomicmolecular_2008}) and simulations \citep{gnedin_modeling_2009,valdivia_h2_2016} were able to reproduce the trend of the HI-to-H$_2$ transition observed by Copernicus and FUSE in the local ISM (e.g., \citealt{savage_survey_1977,gillmon_fuse_2006,rachford_molecular_2009}), the LMC and the SMC (e.g., \citealt{browning_inferring_2003,gillmon_fuse_2006,leroy_spitzer_2007}), no detailed comparison with the statistical properties of these observations have been proposed. As a result, the occurrence of lines of sight with large molecular fractions predicted by numerical simulations often exceed what is deduced from the observations \citep{valdivia_h2_2016}. Finally, and while statistical studies of 1D probability distribution functions (PDF) have become a common tool to understand the formation and the dynamics of molecular clouds \citep{kortgen_shape_2019}, few statistical studies have been performed to date on 2D probability distribution functions using combined observations of different molecular tracers. In that perspective, the recent work of \citet{Bialy_2019a} opens new horizons for the analysis of chemistry in the diffuse matter.

	In the first paper of this series, we extend these pioneer statistical studies to the measurements of the atomic-to-molecular transition observed in the diffuse and translucent ISM located in a radius of $\sim 3$ kpc around the sun. We perform a parametric exploration of numerical simulations of the multiphase ISM and compare the results with the observed 2D probability histogram (PH) of total and molecular hydrogen column densities in order to identify the physical processes that control the molecular content of CNM clouds and the probability of occurrence of lines of sight. The observational dataset and the distribution of sizes of the sampled medium are presented in Sect. \ref{sec:h2}. The different setups of the simulations and the method used to reconstruct the 2D PH are described in Sect. \ref{sec:physics}. The comparisons	with the observations are shown in Sect. \ref{sec:results} which also highlights the influences of the different parameters. The paper finally ends with Sects. \ref{sec:discussion} and \ref{sec:conclusion} where we discuss the validity of our approach and summarize our main conclusions.

\section{Observations of the HI-to-H$_2$ transition}\label{sec:h2}

	\begin{figure}
		\includegraphics[width=1\linewidth]{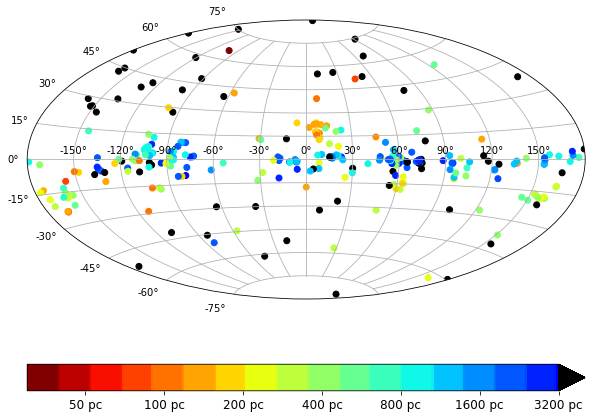}
		\caption{Aitoff projection, in Galactic longitude and latitude coordinates, of the background sources of the observational sample of HI and H$_2$ deduced from absorption studies and used in this work (see Appendix \ref{app:sources} and Table \ref{table:obs}). The color code indicates the distance of the source. All unknown distances correspond to extragalactic sources (see Table \ref{table:obs}): these are arbitrarily set to 1 Mpc and indicated with black points.}
		\label{fig:Ra_Dec_distribution}
	\end{figure}

	\subsection{Observational sample and distances}
	
	The observational sample studied in this work is built from the database of \citet{gudennavar_compilation_2012} who compiled existing data of atomic and molecular lines observed in absorption toward several thousand sources, including stars and AGNs.
	Limiting this catalog to observations or tentative detections of HI, H$_2$, and of the reddening  E(B-V), and removing the data associated to the Magellanic Cloud or high redshift extragalactic environments (e.g., \citealt{tumlinson_far_2002,cartledge_fuse_2005,welty_interstellar_2010,noterdaeme_physical_2007}), we obtain a sample of 360 sources which form, to date, the most complete set of observations of the HI-to-H$_2$ transition in the local diffuse ISM. A more detailed description of this set, the list of the background sources, and the values of the column densities of HI and H$_2$ toward each source, $N({\rm H})$ and $N({\rm H}_2)$ are given in Appendix \ref{app:sources} and Table \ref{table:obs}.

	The positions of the sources in the sky and their distance deduced from GAIA and Hipparcos measurements of parallaxes \citep{1997A&A...323L..49P,2018A&A...616A...1G} are shown in Fig. \ref{fig:Ra_Dec_distribution}, where unknown distances of extragalactic sources (see Table \ref{table:obs}) are arbitrarily set to 1 Mpc. With comparable numbers of observations in all Galactic quadrants, the sources appear to be well distributed in Galactic longitudes.
	Oppositely, and while the sources cover almost all Galactic latitudes, about two-thirds of them are located toward the Galactic disk with latitudes smaller than 15$^{\circ}$, and only one third is located above, crossing the Galactic halo. Since the sample contains extragalactic sources, and since the amount of molecular gas in the Milky Way decreases exponentially as a function  of the distance from the midplane, the length of the line of sight $l_{\rm los}$ occupied by the observed diffuse gas cannot always be identified to the distance of the background source. For the sake of simplicity, we assume here a molecular height above the midplane of 100 pc and compute the length of the observed diffuse Galactic material as 
	\begin{equation}
	\label{eq:l_los_obs}
	l_{\rm los} = {\rm min} \left( \frac{1"}{p}, \frac{100}{{\rm sin}(|b|)} \right) \,\, {\rm pc}, 
	\end{equation}
	where $b$ is the Galactic latitude of the background source and $p$ is its parallax. 
	
\begin{figure}
		\includegraphics[width=1\linewidth]{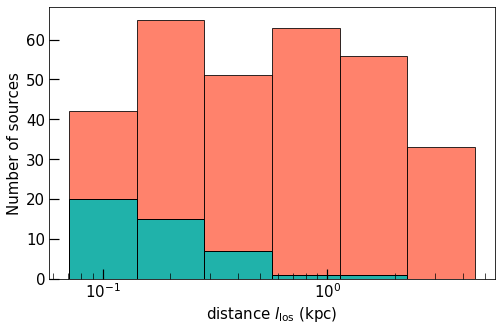}
		\caption{Distribution of lengths $l_{\rm los}$ of the intercepted diffuse material computed with Eq. \ref{eq:l_los_obs} along all lines of sight of the observational sample. The orange sample corresponds to lines of sight where H$_2$ is detected and the green sample to those for which an upper limit on $N({\rm H}_2)$ has been derived (see Table \ref{table:obs}).}
		\label{fig:distr_lengths}
	\end{figure}

	The resulting distribution of the lengths of the lines of sight is shown in Fig. \ref{fig:distr_lengths}. The shortest lines of sight are found to extend over $\sim 100$ pc and the largest over $\sim 3.5$ kpc. Remarkably, and because of the combined distributions of distances and Galactic latitudes of the sources, we find that ${\rm log}({l_{\rm los}})$ follows a flat distribution up to $l_{\rm los} \sim 2$ kpc with about 50 sources per bin and drops by about a factor of two for $l_{\rm los} \sim 3$ kpc. Oppositely, and	as expected, the distribution of lengths of non-detections of H$_2$ is not flat but decreases rapidly up to 1 kpc. Long lines of sight are finally not limited to the	first and fourth Galactic quadrants but are found to spread over all Galactic longitudes and mostly depend on the Galactic latitude of the background source.
	
	\subsection{Physical and statistical properties}
	\label{subsec:obshh2}

\begin{figure}[!t]
		\includegraphics[width=1\linewidth,trim = 1.5cm 2.7cm 1.5cm 2cm, clip,angle=0]{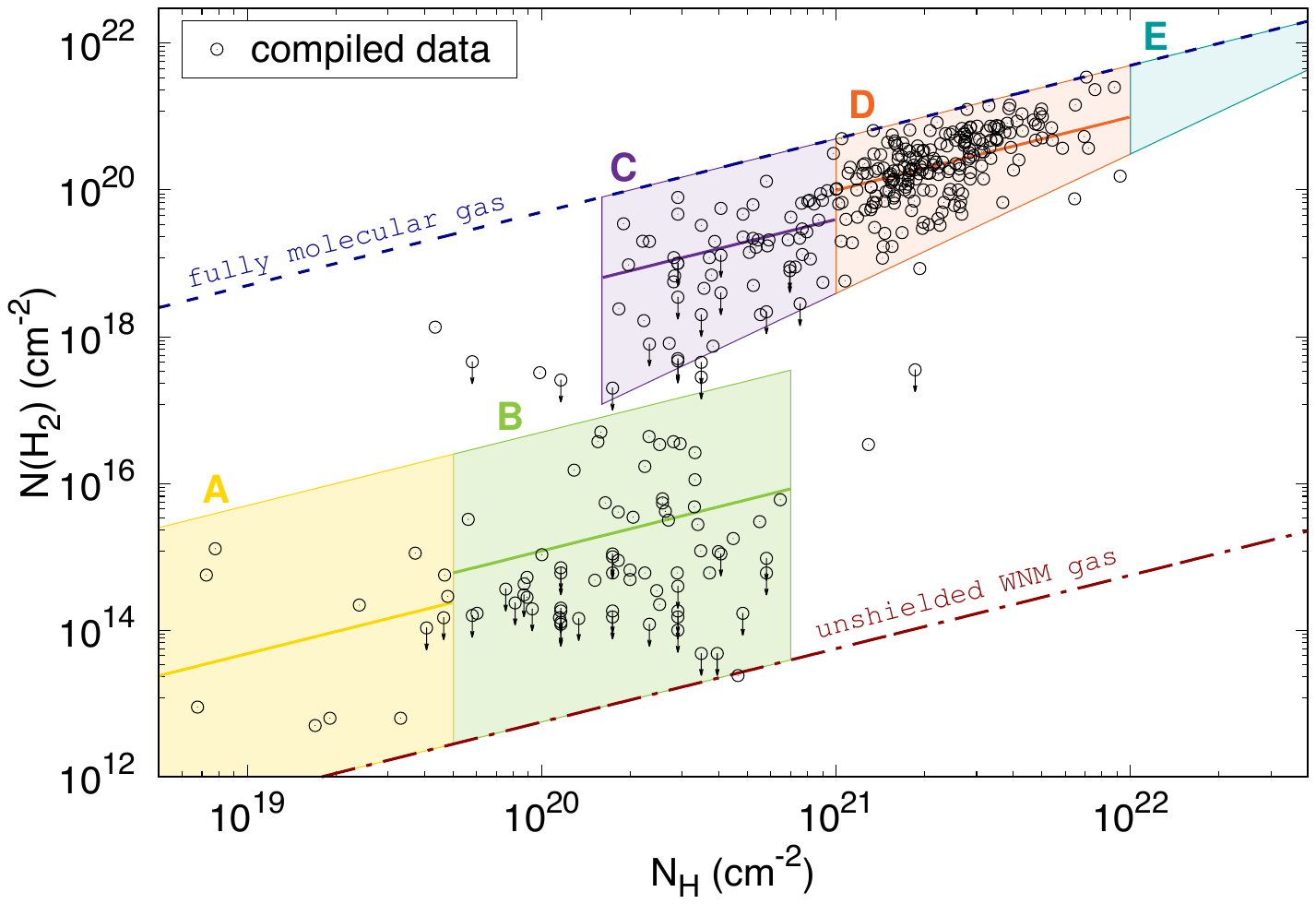}
		\caption{H$_2$ column density as a function  of the total column density of protons $N_{\rm H}$. Open circles correspond to detections of H$_2$ while arrows correspond to upper limits	(see Table \ref{table:obs}). The blue dashed line indicates the maximum value of N(H$_2$) derived from a purely molecular medium with an integrated molecular fraction $f_{{\rm H}_2} = 1$ (Eq. \ref{eq:fh2}). The red dashed-dotted line indicates the theoretical molecular fraction derived in an unshielded WNM-type environment with a density of 0.5 cm$^{-3}$ and a temperature of 8000 K, illuminated by a UV photon flux of $10^8$ cm$^{-2}$ s$^{-1}$ (see Eqs. \ref{eq:h2form} \& \ref{eq:photodiss}). The regions A, B, C, D, and E defined in Table \ref{Tab-regionHH2} correspond to an arbitrary separation of the observational sample used for quantitative comparisons with the results of simulations (see Sect. \ref{sec:results}).}
		\label{fig:regions}
	\end{figure}

	The compiled data are shown in Fig. \ref{fig:regions} which displays the observed column densities of H$_2$ as functions of the total proton column densities of the gas $N_{\rm H} = N({\rm H}) + 2 N({\rm H}_2)$.
	As shown in Fig. \ref{fig:regions} and as already noted by \citet{2009ASPC..417..177G}, almost no line of sight is either purely WNM or purely molecular. This implies that the observed gas is necessarily composed of a combination of phases and clouds of different extinctions with various contributions to the volume spanned by the different lines of sight. As a result, the integrated molecular fraction computed as
	\begin{equation}
	\label{eq:fh2}
	f_{{\rm H}_2} = \frac{2 N({\rm H}_2)}{N_{\rm H}}
	\end{equation}
	shows a large dispersion in the observational sample covering about seven orders of magnitude. While the molecular fraction averaged over all the lines of sight is found to be 0.20, the mass averaged molecular fraction is 0.27, a value similar to the results obtained by \citet{miville-deschenes_physical_2017} at 8.5 kpc based on the analysis of molecular clouds observed in CO in the entire Galactic disk. The position of the HI-to-H$_2$ transition, on the other hand, is found to extend over about one order of magnitude of total column density from $N_{\rm H} \sim 10^{20}$ to $10^{21}$ cm$^{-2}$ and occurs, on average, at $N_{\rm H} \sim 3 \times 10^{20}$ cm$^{-2}$ \citep{gillmon_fuse_2006}.
	
	In order to highlight the statistical features of this transition, we divide the observational sample into 5 subsamples A, B, C, D, and E, shown in Fig. \ref{fig:regions}, which encompass almost all the observational points and whose statistical properties are summarized in Table \ref{Tab-regionHH2}. 48 lines of sight out of 360 are found to be not detected in H$_2$. 
	Most of these upper limits are obtained for a total column density smaller than $10^{21}$ cm$^{-2}$ and about half of them provide strong constraints on the molecular fraction with $f_{{\rm H}_2} \leqslant 10^{-5}$. 3\% of the 312 detections belong to the subsample A, 13\% to subsample B, 16\% to subsample C, and 65\% to subsample D. Interestingly, the subsample E is empty and no line of sight is observed with $N_{\rm H} > 10^{22}$ cm$^{-2}$. While the mean value of the logarithm of the molecular fraction strongly increases from subsamples A to D, the dispersion simultaneously decreases by about a factor of three, probably revealing an effect of average over long distances.
	
	All these statistical properties, and more precisely the probability of occurrence of a given line of sight, are the subject of this paper. What physical processes control the HI-to-H$_2$ transition? How does the distribution of lengths of the lines of sight influence its observed statistical properties? What are the origins of the variations of the molecular fraction from one line of sight to another?

\begin{table}
\begin{center}
\caption{Statistical properties of H$_2$ observations in the subsamples A, B, C, D, and E defined in footnote and shown in Fig. \ref{fig:regions}. Only the lines of sight where H$_2$ has been detected (312 sources out of 360, see Table \ref{table:obs}) are considered. The mean and dispersion values, $\mu$ and $\sigma$ are computed on the logarithm of the molecular fraction $f_{{\rm H}_2}$ observed in the corresponding subsample.}
\label{Tab-regionHH2}
\begin{tabular}{l r r c c}
\hline
region & number & \% & $\mu \left[{\rm log}(f_{{\rm H}_2}) \right]$ & $\sigma \left[{\rm log}(f_{{\rm H}_2}) \right]$ \\
\hline
A      &  10    &  3 & -5.02 & 0.99 \\
B      &  41    & 13 & -4.62 & 0.70 \\
C      &  50    & 16 & -1.10 & 0.49 \\
D      & 204    & 65 & -0.70 & 0.37 \\
E      &   0    &  0 & $-$   & $-$  \\
\hline
\end{tabular}
\end{center}
{\tiny
Definitions of subsamples:
\begin{itemize}
\item[$\bullet$] region A: $5  \times 10^{18} \leqslant N_{\rm H} \leqslant 5 \times 10^{19}$, $5.6 \times 10^{-8} \leqslant f_{{\rm H}_2}  \leqslant 10^{-3}$\\[-13pt]
\item[$\bullet$] region B: $5  \times 10^{19} \leqslant N_{\rm H} \leqslant 7 \times 10^{20}$, $5.6 \times 10^{-8} \leqslant f_{{\rm H}_2}  \leqslant 10^{-3}$\\[-13pt]
\item[$\bullet$] region C: $1.6\times 10^{20} \leqslant N_{\rm H} \leqslant          10^{21}$, $10^{-3} \left(\frac{N_{\rm H}}{10^{20}}\right)^{0.9} \leqslant f_{{\rm H}_2}  \leqslant 1$\\[-13pt]
\item[$\bullet$] region D: $          10^{21} \leqslant N_{\rm H} \leqslant          10^{22}$, $10^{-3} \left(\frac{N_{\rm H}}{10^{20}}\right)^{0.9} \leqslant f_{{\rm H}_2}  \leqslant 1$\\[-13pt]
\item[$\bullet$] region E: $          10^{22} \leqslant N_{\rm H} \leqslant          10^{23}$, $10^{-3} \left(\frac{N_{\rm H}}{10^{20}}\right)^{0.9} \leqslant f_{{\rm H}_2}  \leqslant 1$\\[-13pt]
\end{itemize}
}
\end{table}

\section{Physics and numerical method}\label{sec:physics}

		\begin{table*}[!ht]
			\begin{center}
			\caption{Fiducial model and range of parameters explored in this work} \label{table:grids}
			\begin{tabular}{ l | c | c | c | c }
				\hline \hline	
				\textbf{Parameter}		& \textbf{Notation} &\textbf{Ref} & \textbf{Range} & \textbf{Units}\\ 		
				\hline
                				
				box size & $L$ & 200 & 20 - 200 & pc\\
				mean density & $\overline{n_H}$ & 2 & 0.5 - 4 & cm$^{-3}$\\
				UV radiation field & $G_0$ & 1 & 0.5- 4 & Habing field\\
				resolution & $R$   & 256$^3$ & 64$^3$ - 512$^3$ & --\\
				turbulent forcing 	& $F$   & 9 $\times$ 10$^{-4}$ & 10$^{-5}$ - $10^{-2}$& kpc Myr$^{-2}$\\
				compressive ratio & $\zeta$   & 0.1 & 0.1 - 0.9 & --\\
				Doppler broadening & $b_D$ & 8 & 1 - 8 & km s$^{-1}$\\
				initial magnetic field & $B_x$   & 3.8 & 0 - 40& $\mu$G\\
				self-gravity & -- & on & on - off & --\\
				Galactic well & --  &on & on - off & --\\
				\hline \hline 
			\end{tabular}
            \end{center}
		\end{table*}

		To study the physical processes at play in the HI-to-H$_2$ transition, we performed numerical simulations of the multiphase diffuse ISM, using the RAMSES code \citep{teyssier_cosmological_2002,Fromang_2006}, a grid-based solver with adaptative mesh refinement \citep{berger_adaptive_1984}. 	
		The methodology applied in this paper follows the works of \citet{seifried_forced_2011}, \citet{saury_structure_2014}, and \citet{ valdivia_h2_2015,valdivia_h2_2016}.
		
		The diffuse matter in the Solar Neighborhood of our galaxy is simulated over a box of size $L$ with periodic boundary conditions. The matter, defined by a mean proton density $\overline{n_\mathrm{H}}$, is assumed to be illuminated on all sides by an isotropic spectrum of UV photons set to the standard interstellar radiation field \citep{habing_interstellar_1968} and scaled with a factor $G_0$. 
		
		\subsection{Fluid equations}
		Within this framework, RAMSES computes the evolution of the gas solving the classic equations of ideal magnetohydrodynamics (MHD):
		
		\begin{align}
		\frac{\partial \rho}{\partial t} + \nabla \cdot (\rho \vec{v}) &= 0,\\
		\frac{\partial \rho \vec{v}}{\partial t} + \nabla \cdot (\rho \vec{vv} - \vec{BB}) + \nabla P &= -\rho \nabla \Phi + \rho \vec{f},\\	
		\frac{\partial E}{\partial t} + \nabla \cdot [(E+P)\vec{v} - \vec{B(Bv)}] &= -\rho \vec{v} \cdot\nabla \Phi + \rho \vec{f} \cdot \vec{v} -\rho \mathcal{L}, \,\, {\rm and}\\	
		\frac{\partial \vec{B}}{\partial t} + \nabla \cdot (\vec{vB - Bv}) &= 0,
		\end{align}
   where $\rho$, $\vec{v}$, $\vec{B}$, $P$ and $E$ are the mass density, the velocity field, the magnetic field, the total pressure, and the total energy density, respectively. 
   The net cooling function per unit mass, $\mathcal{L}$, and the acceleration due to the turbulent driving, \vec{f}, are described in Sections \ref{sec:cooling} and \ref{sec:turb_forcing}. The axis $x$, $y$, and $z$ are chosen so that $z$ corresponds to the direction perpendicular to the Galactic disk, and $x$ corresponds to the direction of the mean magnetic field initially parametrized by a constant value $B_x$.\\
		
		To take into account all gravitational forces, including self-gravity and the action of stars and dark matter, the gravitational potential $\Phi$ is divided into two terms:
		\begin{equation}
		\Phi = \phi_\text{gas} + \phi_\text{gal}.
		\end{equation}
		The self-gravity potential, $\phi_\text{gas}$, is deduced from the Poisson's equation:
		\begin{equation}			
		\nabla^2 \phi = 4 \pi G \rho.
		\end{equation}	
		Following \citet{kuijken_mass_1989} and \citet{joung_turbulent_2006}, we assume that the Galactic potential along the direction $z$ perpendicular to the Galactic disk can be written as
		\begin{equation}
		\phi_\text{gal}(z) = a_1 \left(\sqrt{z^2 + z_0^2} -z_0\right) + 2a_2 z^2,
		\end{equation}
		where the first term is the contribution of the stellar disk parametrized by $z_0 = 0.18$ kpc and $a_1 = 1.42 \times 10^{-3}$ kpc Myr$^{-2}$ and the second term is the contribution of the spherical dark halo parametrized by $a_2 = 5.49 \times 10^{-4}$ Myr$^{-2}$.

		\subsection{Thermal processes and radiative transfer}\label{sec:cooling}
		As shown by \citet{field_thermal_1965}, the multiphase nature of the ISM results from the thermal balance of the gas and thus from its net cooling function $\mathcal{L}$ defined by
		\begin{equation}
		\rho \mathcal{L} = n_{\rm H}^2\Lambda - n_{\rm H}\Gamma,
		\end{equation}
		where $n_{\rm H}^2\Lambda$ and $n_{\rm H}\Gamma$ are the cooling and heating rates of the medium (in erg cm$^{-3}$ s$^{-1}$) and $n_{\rm H}$ is the proton density.
		To correctly describe the thermal state of the diffuse ISM, we include in this work the heating induced by the photoelectric effect and the decay of cosmic ray particles and the cooling induced by Lyman $\alpha$ photons, the recombination of electrons onto grains, and the fine structure lines of OI and CII.
		All these processes, described in Appendix \ref{app:cooling}, are modeled with the analytical formulae given by \citet{wolfire_neutral_2003}.
		
		The absorption of UV photons by dust, and its subsequent impact on the photoelectric effect, is treated with the tree-based method proposed by \citet{valdivia_fast_2014}. At each point the effective radiation field $G_{\rm eff}$ (in Habing units) is computed as 
		\begin{equation}
		G_{\rm eff} = G_0 \,\,\langle \textrm{e}^{-2.5 A_V}\rangle,
		\end{equation}
		where $A_V$ is the visual extinction along a given ray, deduced from the integrated proton column density\footnote{In this work we use the relation between $N_\text{H}$ and $A_V$ deduced from the observations of the mean Galactic extinction curve \citep{Fitzpatrick_1986}.} computed from the border of the box to the current point
		\begin{equation}
		A_V = 5.34 \times 10^{-22} \left(\frac{N_\text{H}}{{\rm cm}^{-2}}\right)\,,
		\end{equation}
		and $\langle \textrm{e}^{-2.5 A_V}\rangle$ is an average performed over 12 directions, treated as solid angles evenly spread in polar coordinates. 

	\subsection{Turbulence forcing}\label{sec:turb_forcing}
		To mimic the injection of mechanical energy in the diffuse ISM, a large scale turbulent forcing is applied. 
		Following \citet{schmidt_numerical_2009} and \citet{federrath_comparing_2010}, this forcing, modeled by an acceleration $\vec{f}$ in the momentum conservation equation, is driven through an Ornstein-Uhlenbeck process
		using a pseudo-spectral method.
		At regular time intervals $\Delta \tau$, random fluctuations of the forcing term are generated and applied over an autocorrelation timescale $\tau$. To excite only large scale modes, the forcing is modeled as a paraboloid in Fourier space covering a small interval of wavenumbers $1 \leqslant k \leqslant 3$ and centered on $k = 2$. 
		Using the notations of \citet{seifried_forced_2011} and \citet{saury_structure_2014}, the total magnitude of these perturbations is set with either an acceleration parameter $F$ or, equivalently, a velocity parameter $V$ related by 
		$F = V^2/L_{\rm drive}$, where $L_{\rm drive}$ is the main driving scale, $L_{\rm drive}=L/2$.
		A Helmholtz decomposition is finally applied, in order to control the powers injected in compressive and solenoidal modes. Using the classical notation, these powers are set with a parameter\footnote{We note that for $\zeta = 0.5$, the power of the compressive forcing corresponds to 1/3 of the total power.} $\zeta$ ranging from a pure solenoidal field ($\zeta = 1$) to a pure compressive field ($\zeta = 0$).
		
		Throughout this work, we adopt $\Delta \tau \sim 0.4$ Myr which roughly corresponds to the time interval separating two supernova events occurring in a volume of (200 pc)$^3$. The characteristic damping time of the turbulence $\tau$ is approximately set to the turnover timescale of the diffuse ISM $\tau \sim 33 (L/200 \text{pc})^{0.6}$ Myr \citep{larson_turbulence_1981, hennebelle_turbulent_2012}.
		$F$ (or $V$) and $\zeta$ are left as free parameters.
		
		\subsection{H$_2$ chemistry}
		The timescale required for the abundance of molecular hydrogen to reach its equilibrium value is known to range over several orders of magnitude, depending on the physical conditions of the ISM (e.g., \citealt{Goldsmith_2007,Tabone_2018}).
		In the diffuse gas, this timescale varies typically between a few 10$^3$ yr and a few 10$^7$ yr \citep{valdivia_h2_2016}, hence over a range of values that often exceeds the dynamical timescales. To take into account this important aspect of the diffuse interstellar chemistry, the out-of-equilibrium abundance of H$_2$ is computed self-consistently in the simulation, using the formalism introduced in RAMSES by \citet{valdivia_h2_2015,valdivia_h2_2016}.
		
		The formation of H$_2$ onto grains in physisorption and chemisorption sites is modeled with the simplified rate of \citet{le_bourlot_surface_2012}	
		\begin{equation} \label{eq:h2form}
		k_{f} = 3\times 10^{-17} n_{\rm H} n({\rm H}) \sqrt{\frac{T}{100\text{ K}}} S(T) \,\, \text{ cm}^{-3}\text{s}^{-1},
		\end{equation}
		where $n_{\rm H}$ and $n({\rm H}) $ are the local proton and atomic hydrogen densities, and	
		\begin{equation}
		S(T) = \frac{1}{1+(\frac{T}{T_2})^\beta}
		\end{equation}
		is the sticking coefficient of H onto grain, parametrized by $T_2$ = 464 K and $\beta$ = 1.5.
		
		The destruction of H$_2$ by UV photons is computed using the formalism, described by \citet{draine_structure_1996} and \citet{sternberg_h_2014}, which is classically introduced in many astrochemical models (e.g., \citealt{lesaffre_low-velocity_2013,gong_simple_2017,bialy_h_2017}). In each cell, the photodestruction rate of H$_2$ is modeled as 
		\begin{equation}
		k_{d} = k_{d}^0\, n({\rm H}_2) G_0 \langle e^{-\sigma_{d} N_{\rm H}}\rangle \, \left\langle f_{\rm shield}(x)\right\rangle \,\, {\rm cm }^{-3} {\rm s}^{-1},
		\label{eq:photodiss}
		\end{equation}	
		where $x=N({\rm H}_2)/5 \times 10^{14} {\rm cm}^{-2}$, $k_{d,0} = 3.3\times 10^{-11} \text{ s}^{-1}$ is the inverse freespace dissociation timescale of H$_2$ in an isotropic Habing field, $n({\rm H}_2)$ is the local density of the molecular hydrogen, $ e^{-\sigma_{d} N_{\rm H}}$ is the shielding induced by dust and $f_{\rm shield}$ the self-shielding function.
		We adopt here an effective dust attenuation cross section at $\lambda$ = 1000 \AA, $\sigma_d = 2 \times 10^{-21}$ cm$^2$ \citep{sternberg_h_2014}.
		Following \citet{draine_structure_1996}, the self-shielding function is computed as
		\begin{equation}\label{eq:f_shield}
		f_{\rm shield}(x) = \frac{0.965}{(1+x/b_D)^2}+\frac{0.035 \text{ e}^{-8.5 \times 10^{-4}\sqrt{1+x}}}{\sqrt{1+x}},
		\end{equation}
		where $b_D$ is the Doppler broadening parameter expressed in km s$^{-1}$. 
		As done for the photoelectric heating rate (see section 3.2), both the shielding by dust and the self-shielding are calculated along 12 different directions and then averaged to obtain the photodissociation rate of Eq. \ref{eq:photodiss}.
		
	\subsection{Fiducial model and grids of parameters} \label{sec:grid}

		The framework described above lays on several independent parameters, which are all related to key physical ingredients of the ISM. The influence of each ingredient on the HI-to-H$_2$ transition is studied here through several grids of simulations $-$ including a total of 305 runs $-$ covering a broad range of physical conditions and centered around a fiducial setup\footnote{The grids have been run on the computing cluster Totoro funded by the ERC Advanced Grant MIST. The computational time of the standard simulation is $\sim 6$ days using 40 processors.}.
		The reference value adopted for each parameter, and the range of values explored in this work, are summarized in Table \ref{table:grids}. Among all parameters, $L$, $\overline{n_{\rm H}}$ and $G_0$ are of particular importance. 
		
		With our assumptions, $L$ simultaneously corresponds to the scale of illumination of the gas by UV photons and twice the integral scale of turbulence, that is twice the scale of injection of mechanical energy $L_{\rm drive}$. OB stars, which are the dominant sources of the interstellar UV field, are not uniformly distributed in the sky but are known to be clustered in associations \citep{ambartsumian_evolution_1947}. As shown by the recent 3D studies of the distributions of stars based on the Hipparcos and GAIA Catalogs (e.g., \citealt{bouy_cosmography_2015} \citealt{zari_3d_2018}), the typical distances separating two associations in the Solar Neighborhood range between 50 pc and a few hundreds of pc, that is several times the mean distance deduced from the integrated surface densities of OB stars ($\sim 1.6 \times 10^{-3}$ pc$^{-2}$, \citealt{maiz-apellaniz_spatial_2001}). Interestingly, such distances are not only comparable to the heights of the molecular gas ($\sim 75$ pc) and the cold HI gas ($\sim 150$ pc) above the Galactic plane deduced from CO and HI all-sky surveys (e.g., \citealt{dame_milky_2001}, \citealt{dickey_h_1990}, \citealt{kalberla_hi_2009}), but they also correspond to the typical size of HI superclouds \citep{elmegreen_h_1987}.
		For all these reasons, we, therefore, adopt a fiducial simulation with $L$ = 200 pc and explore values down to a few tens of pc. 
		
		The mean density of the gas, $\overline{n_H}$, represents the mass of the diffuse neutral ISM contained in a volume $L^3$, and also controls the porosity of the matter to the impinging radiation field.
		In this work, we adopt a fiducial value $\overline{n_{\rm H}}$ = 2 cm$^{-3}$, a value slightly larger than the standard Galactic midplane density of HI at a galactocentric distance of 8.5 kpc \citep{kalberla_hi_2009}. With our fiducial value of $L$, the mass of gas contained in the box accounts for all the mass surface density of HI in the Solar Neighborhood ($\Sigma_{\rm HI} \sim 10$ M$_\odot$ pc$^{-2}$, \citealt{nakanishi_three-dimensional_2016}, \citealt{miville-deschenes_physical_2017}).
		
		$G_0$ controls the intensity of the radiation field, hence the thermodynamical state of the gas. Since $G_0$ is normalized to the Habing field, we choose a fiducial value of $G_0 = 1$. The corresponding UV energy density of the standard setup is therefore slightly smaller than that contained in the standard UV radiation fields given by \citet{Draine_1978} and \citet{Mathis_1983}.

        The standard values of the two parameters $F$ and $B_x$ are set to $9 \times 10^{-4}$ kpc Myr$^{-2}$ and 3.8 $\mu$G, respectively.
   As we will show later, those values are chosen so that the velocity dispersion of the gas and the strength of the magnetic field are close to the values observed in the diffuse ISM.
		
\subsection{Steady-state}

\begin{figure*}[!h]
\begin{center}
\includegraphics[width=17cm,trim = 0cm 2cm 0cm 0cm, clip,angle=0]{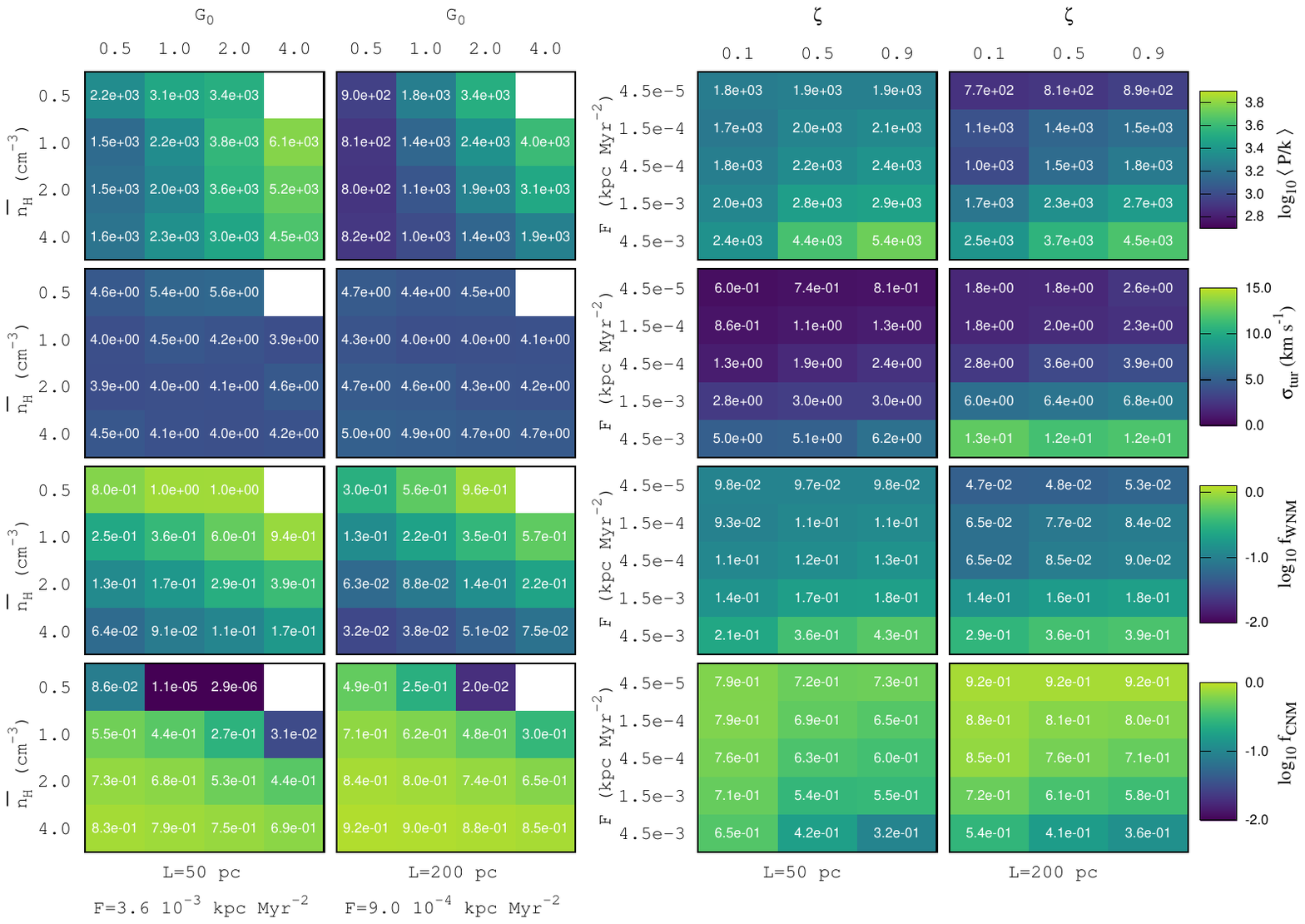}
\caption{Colored tables of the mean pressure expressed in K cm$^{-3}$ (first line), the turbulent velocity dispersion $\sigma_{\rm tur}$ (second line), and the fractions of mass $f_{\rm WNM}$ and $f_{\rm CNM}$ contained in the WNM phase (third line) and the CNM phase (fourth line). The first and second columns display these quantities as functions of $\overline{n_{\rm H}}$ and $G_0$, for $L=50$ pc and $F=3.6 \times 10^{-3}$ kpc Myr$^{-2}$ (first column) and for $L=200$ pc and $F=9 \times 10^{-4}$ kpc Myr$^{-2}$ (second column). The third and fourth columns display these quantities as functions of the acceleration parameter $F$ and the compressive ratio $\zeta$, for
$L=50$ pc (third column) and $L=200$ pc (fourth column). All other parameters are set to their standard values (see Table \ref{table:grids}).}
\label{Fig-prop-WNM_CNM}
\end{center}
\end{figure*}

The evolution of the multiphase environments simulated here is identical to the description already given in many papers (e.g., \citealt{seifried_forced_2011,saury_structure_2014,valdivia_h2_2015,valdivia_h2_2016}). Starting from an homogeneous density $n_{\rm H} = \overline{n_{\rm H}}$, the gas evolves under the joint actions of turbulence, gravity, and thermal instability, and splits up in three different phases at thermal pressure equilibrium, the WNM, the CNM, and the LNM. The formation of dense environments well shielded from the destructive UV radiation field triggers the formation of H$_2$ and the medium jointly evolves from a purely atomic state to a partly molecular state. If the mass of the gas is conserved, as imposed by the periodic boundary conditions, the medium progressively tends toward a steady-state where the mean pressure, the volume filling factors of the different phases, their velocity dispersion, and their mean molecular fractions are roughly constant. This steady-state is typically reached after a few turnover timescales, providing that the corresponding time is longer than the damping time (see Sect. \ref{sec:turb_forcing}).

Because of turbulence and thermal instability, the steady-state has a statistical nature. The turbulent forcing and the subsequent turbulent cascade induce pressure variations and shear motions at all scales which trigger mass exchanges between the different phases. Any pressure or density structure is therefore a transient system which is usually described by its contribution to probability distribution functions. At steady-state, only PDFs remain constant. This steady yet ever changing environment is the reason for the sustained presence of a substantial amount of gas in the LNM at densities and temperatures out of thermal equilibrium (e.g., \citealt{marchal_rohsa_2019}). All the results shown throughout this paper are taken at steady-state, at times ranging from a few tens of Myr up to 100 Myr depending on the strength of the turbulent forcing. 

\subsection{Properties of the multiphase medium} \label{sec:propbiphase}

The steady-state values of the mean pressure $\langle P / k \rangle$, the turbulent velocity dispersion $\sigma_{\rm tur}$, and the fractions of mass of the WNM and the CNM, $f_{\rm WNM}$ and $f_{\rm CNM}$, are shown in Fig. \ref{Fig-prop-WNM_CNM} for a set of 60 different simulations. For the sake of simplicity, we assume that the WNM is composed of all cells with a temperature $T \geqslant 3000$ K, the CNM of all cells with $T \leqslant 300$ K, and the LNM of all cells with $300\,{\rm K} < T < 3000$ K, hence
\begin{equation}
f_{\rm WNM} = \frac{\sum \rho_{|T \geqslant 3000\,{\rm K}}}{\sum \rho}
\end{equation}
and
\begin{equation}
f_{\rm CNM} = \frac{\sum \rho_{|T \leqslant 300\,{\rm K}}}{\sum \rho}.
\end{equation}
The mean pressure $\langle P / k \rangle$  (with $k$ the Boltzmann constant) is classically computed as an average over the entire volume. While the above conventions are well established, there are many ways to define the turbulent velocity dispersion $\sigma_{\rm tur}$. It could be defined as the volume-weighted or mass-weighted velocity dispersion computed over the entire volume \citep{audit_thermal_2005,federrath_comparing_2010}, the average of the mass-weighted velocity dispersion computed along individual lines of sight \citep{Miville_2007,saury_structure_2014}, or the dispersion of the mass-weighted velocity centroids computed along independent directions \citep{Henshaw_2019}. All these definitions give velocity dispersions that differ from one another and are not equally relevant for the comparison with observed quantities. 
To relate the velocity dispersion to a kinetic energy and provide values that could be compared to the observations of broad HI emission profiles at high Galactic latitude (see below), we chose to compute $\sigma_{\rm tur}$ in the WNM only as 
\begin{equation}
\sigma_{\rm tur}^{2} = \frac{1}{3} \frac{\sum \rho ||\vec{v} - \overline{\vec{v}}||^2}{\sum \rho}
\end{equation}
with
\begin{equation}
\overline{\vec{v}} = \frac{\sum \rho \vec{v}}{\sum \rho}
\end{equation}
and where the sums are performed over all cells with $T \geqslant 3000$ K.

The results displayed in Fig. \ref{Fig-prop-WNM_CNM} are very similar to those obtained in the parametric studies of \citet{seifried_forced_2011} and \citet{saury_structure_2014} and are in line with the expectations of models of turbulent multiphase environments \citep{wolfire_neutral_2003,ostriker_regulation_2010}. The mean pressure of the gas is primarily regulated by the thermal equilibrium curve (see Appendix \ref{app:cooling}) which depends on the local illumination of the gas by the UV radiation field $G_{\rm eff}$. Since $G_{\rm eff}$ results from the absorption of the external UV field by the surrounding environments, $\langle P / k \rangle$ is not only sensitive to $G_0$ but also to the total mass of the simulation set by $\overline{n_{\rm H}}$ and $L$. Larger values of $G_0$ or smaller values of $L$ or $\overline{n_{\rm H}}$ leads to larger $\langle P / k \rangle$. In turn, the fractions of mass contained in the WNM and the CNM are controlled by the mean pressure and the total mass of the gas. Larger pressure implies larger densities of the WNM (and the CNM). The fraction of mass of the CNM therefore decreases as $\langle P / k \rangle$ increases; eventually, if the density of the WNM becomes comparable to the mean density $\overline{n_{\rm H}}$, the CNM almost entirely disappears (see bottom left panels of Fig. \ref{Fig-prop-WNM_CNM}). At last, and because the WNM occupies most of the volume, $f_{\rm CNM}$ necessarily increases as a function  of the total mass of the gas, or equivalently $\overline{n_{\rm H}}$, even at constant pressure.

The turbulent forcing induces pressure fluctuations and shearing motions at all scales. As shown by \citet{seifried_forced_2011}, this not only frequently perturbs the gas out of the thermal equilibrium states but also strongly reduces the times spent by any fluid elements in the WNM, LNM, and CNM. Because of these two aspects, increasing the turbulent forcing reduces the mass of the CNM to the benefit of those of the LNM and WNM (right panels of Fig. \ref{Fig-prop-WNM_CNM} and Fig. 10 of \citealt{seifried_forced_2011}). The mean pressure therefore increases, the 1D PDF of the density broadens and its bimodal nature progressively disappears \citep{piontek_saturated-state_2005,walch_turbulent_2011}. These effects can be magnified depending on the nature of the turbulent forcing and the power injected in the compressive and solenoidal modes. Because solenoidal motions are more efficient to prevent the gas to condensate back to the CNM phase, a pure solenoidal forcing naturally leads to larger pressure and smaller CNM fractions than those obtained with an equivalent kinetic energy injected in pure compressive modes.

As expected, the velocity dispersion of the WNM is mostly given by the strength of the turbulent forcing and the driving scale ($L_{\rm drive} \sim L /2$, see Sect. \ref{sec:turb_forcing}), with a slight dependence on $\overline{n_{\rm H}}$ and $\zeta$. As proposed by \citet{saury_structure_2014}, a realistic value for the turbulent velocity dispersion of the WNM can be estimated by looking at the HI 21 cm emission spectra with the fewest components observed at high Galactic latitude \citep{kalberla2005}. Toward these directions, \citet{Haud2007} derive a total velocity dispersion $\sigma_{\rm tot} = (\sigma_{\rm tur}^2 + \sigma_{\rm thr}^2)^{1/2} \sim 10$ km s$^{-1}$, where the $\sigma_{\rm thr}$ is the 1D thermal velocity dispersion ($\sim 8.2$ km s$^{-1}$ for the WNM). In the present paper, the turbulent forcing applied to the standard simulation (see Table. \ref{table:grids} and Fig. \ref{Fig-prop-WNM_CNM}) is chosen so that $\sigma_{\rm tur} \sim 4-5$ km s$^{-1}$, in fair agreement with the observations at high Galactic latitude. While this value is chosen as a reference, the velocity dispersions obtained in all the simulations explored in this work range between 1 and 15 km s$^{-1}$ (see Fig. \ref{Fig-prop-WNM_CNM}). 

	\subsection{Reconstruction of lines of sight} \label{sec:reconstruction}

			 \begin{figure}[h!]
			 	\centering
			 	\includegraphics[width=1.\linewidth]{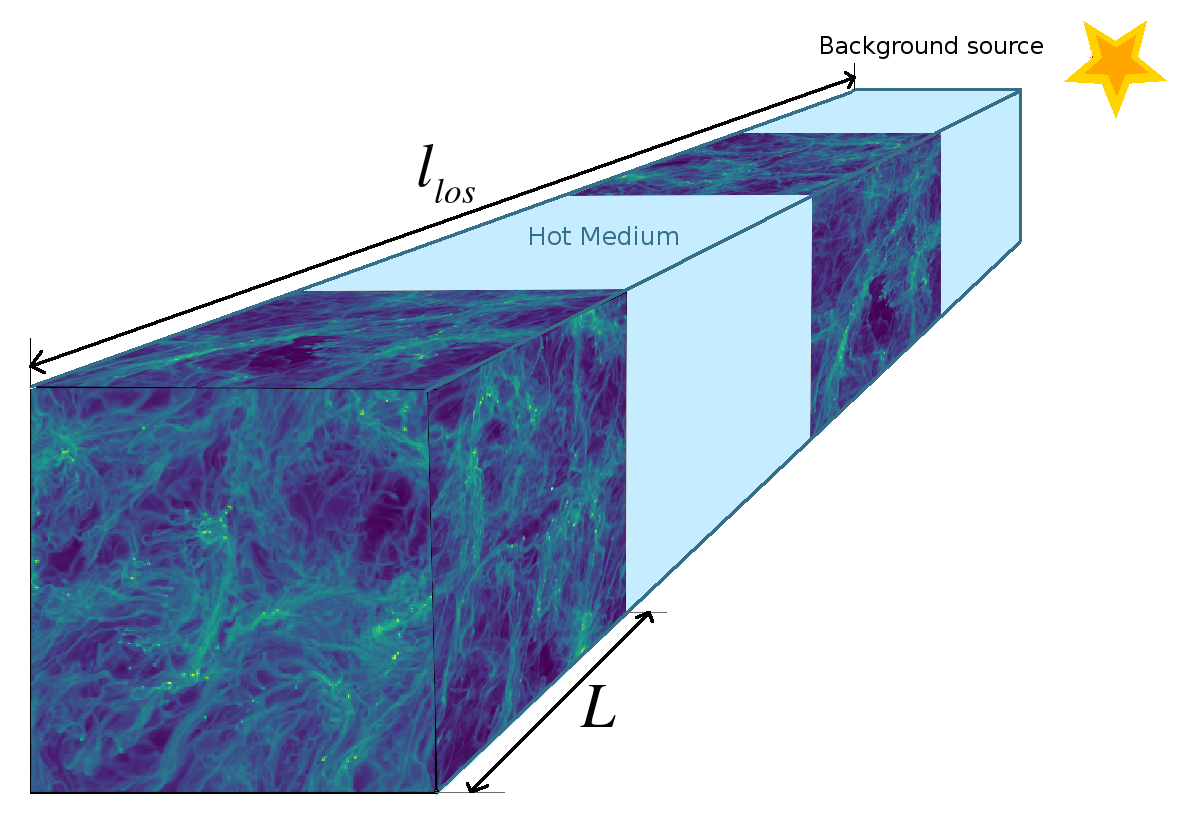}
			 	\caption{Schematic view of the reconstruction of individual lines of sight over a distance $l_{\rm los}$. The medium between the observer and the source is assumed to be composed of hot and warm ionized material (light blue cubes) with a volume filling factor $\varphi$ and of uncorrelated pieces of diffuse neutral gas of individual size $L$ (simulated boxes) with a volume filling factor (1-$\varphi$).}
			 	\label{fig:model}
			 \end{figure}

			As shown in Sect. \ref{sec:h2}, the medium observed in absorption at UV and visible wavelengths extends over a very broad range of distances, from $\sim$ 100 pc to several kpc (see Fig. \ref{fig:distr_lengths}). The targeted lines of sight may therefore contain several isolated diffuse neutral phases but also hot and warm ionized material \citep{mckee_evaporation_1977, de_avillez_volume_2004}. Indeed, such a superimposition of independent components is particularly well seen in submillimeter and infrared observations of the Galactic disk where the gas seen in absorption is found to cluster in several velocity components associated to known Galactic structures (e.g., \citealt{Gerin_2016}). Since our setup only follows a piece of diffuse neutral material of size $L$, an additional treatment regarding the 
			lengths of the lines of sight is therefore required in order to compare the results of the simulations to the distribution of observations. We apply here a methodology similar to that proposed by \citet{Bialy_2019a} and schematized in Fig. \ref{fig:model}.
			We assume that a given simulation corresponds to a building block of neutral diffuse ISM. 
			Depending on its length, any random line of sight necessarily intercepts parts or several of these building elements and an unknown mass of diffuse ionized gas parametrized by its volume filling factor $\varphi$. A total sample of N simulated lines of sight is then generated as follows.

			\begin{figure}[!t]
			\centering
			\includegraphics[width=1\linewidth]{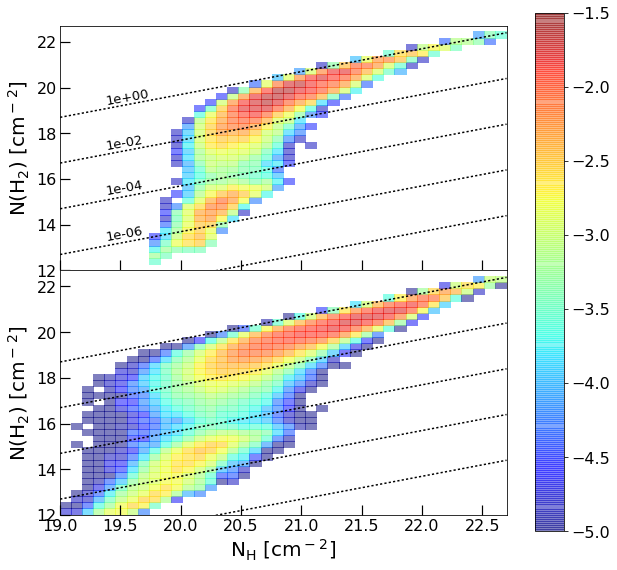}
			\caption{2D probability histogram of the total proton column density $N_{\rm H}$ and the column density of molecular hydrogen $N({\rm H}_2)$ obtained with the standard simulation (see Table \ref*{table:grids}). The upper panel shows the original data where all lines of sight have a size $L=200$ pc. The bottom panel shows the outcome of the reconstruction algorithm described in Sect. \ref{sec:reconstruction} that produces a sample of lines of sight ranging from 100 pc to 3200 pc. The color code indicates the fraction of lines of sight (in logarithmic scale) contained in each bin. Dotted lines are isocontours of the molecular fraction for $f_{{\rm H}_2}=10^{-8}$, $10^{-6}$, $10^{-4}$, $10^{-2}$, and $1$.}
			\label{fig:stat_h2}
			\end{figure}

			Based on the results of Sect. \ref{sec:h2} (Fig. \ref{fig:distr_lengths}), we consider six lengths of lines of sight homogeneously distributed in log space: $l_{los}$ = 100, 200, 400, 800, 1600, and 3200 pc.
			For each length, we generate a sample of
			$N_l = \frac{1}{6} w_l N $ lines of sight, where $w_l$ are normalized weights deduced from the distribution of distances in the observed sample: w$_1$ = 0.14, w$_2$ =  0.21, w$_3$ = 0.16, w$_4$ = 0.20, w$_5$ = 0.18, w$_6$ = 0.11 (see Fig. \ref{fig:distr_lengths}). 
			The column densities of H and H$_2$ along each lines of sight are finally reconstructed 
			by comparing the length occupied by the neutral medium $(1-\varphi) l_{los}$ (see Fig. \ref{fig:model}) and the 
			size of the box $L$. If $(1-\varphi) l_{los} = L$, we draw a random line of sight in the simulation
			and extract the corresponding column densities. If $(1-\varphi) l_{los} < L$, we draw a random line of sight and integrate the column density over a reduced distance of $(1-\varphi) l_{los}$. If $(1-\varphi) l_{los} > L$, we draw $L/\left[(1-\varphi) l_{los}\right]$ random lines of sight and add the respective individual column densities.

			For the sake of simplicity, we assume here that any line of sight intercepts a constant fraction of diffuse ionized gas with $\varphi = 0.5$ \citep{hill_effect_2018}. Similarly, we note that, while spatially uncorrelated, the pieces of diffuse neutral gas used in the reconstruction algorithm correspond to random realizations obtained with a single simulation. Potential
			variations of the mean density, of the external radiation field, or of the turbulent  forcing that naturally follow the Galactic structure depending on the position of the source (see Appendix \ref{app:sources}) are not taken into account. All these limitations are discussed in Sect. \ref{sec:discussion}.
			
			The outcome of the reconstruction algorithm is shown in Fig. \ref{fig:stat_h2} which displays the 2D PH of the total proton column density $N_{\rm H}$ and the column density of molecular hydrogen $N({\rm H}_2)$ obtained with the standard simulation. Because of the flat distribution of distances in log space (see Fig. \ref{fig:distr_lengths}), the peaks of the reconstructed PH are found to be shifted toward both the large and the low values of $N_{\rm H}$ compared to those of the initial distribution (top panel of Fig. \ref{fig:stat_h2}). This naturally enhances the initial bimodality and many lines of sight are found to be either at low ($\sim 10^{-5}$) or large ($\sim 10^{-1}$) integrated molecular fractions. In addition, it induces an inclination toward large column densities; more than half of the lines of sight are found to have $N_{\rm H} > 10^{21}$ cm$^{-2}$.
			By virtue of the central limit theorem, the molecular fraction obtained in those lines of sight tends toward the mean H$_2$ molecular fraction of the initial simulation with a dispersion that decreases as a function  of $N_{\rm H}$.
			
			Building simulated lines of sight from the observed distribution of sizes allows us to perform a statistical comparison of both samples and to limit the impact of observational biases to lines of sight at large column density (region E, see Sect. \ref{sec:discussion}). The combined 2D distributions of $N_{\rm H}$ and $N({\rm H}_2)$ deduced from the simulations are the main focus of this paper and will be shown several times in the following sections. For obvious aesthetical reasons, and to simplify the descriptions of the figures, we will often refer to this representation as a kingfisher diagram.

\subsection{Interpretative modeling} \label{sec:interpretmodel}

\begin{figure}[!h]
\begin{center}
\includegraphics[width=9cm,trim = 1.5cm 0cm 0.5cm 0cm, clip,angle=0]{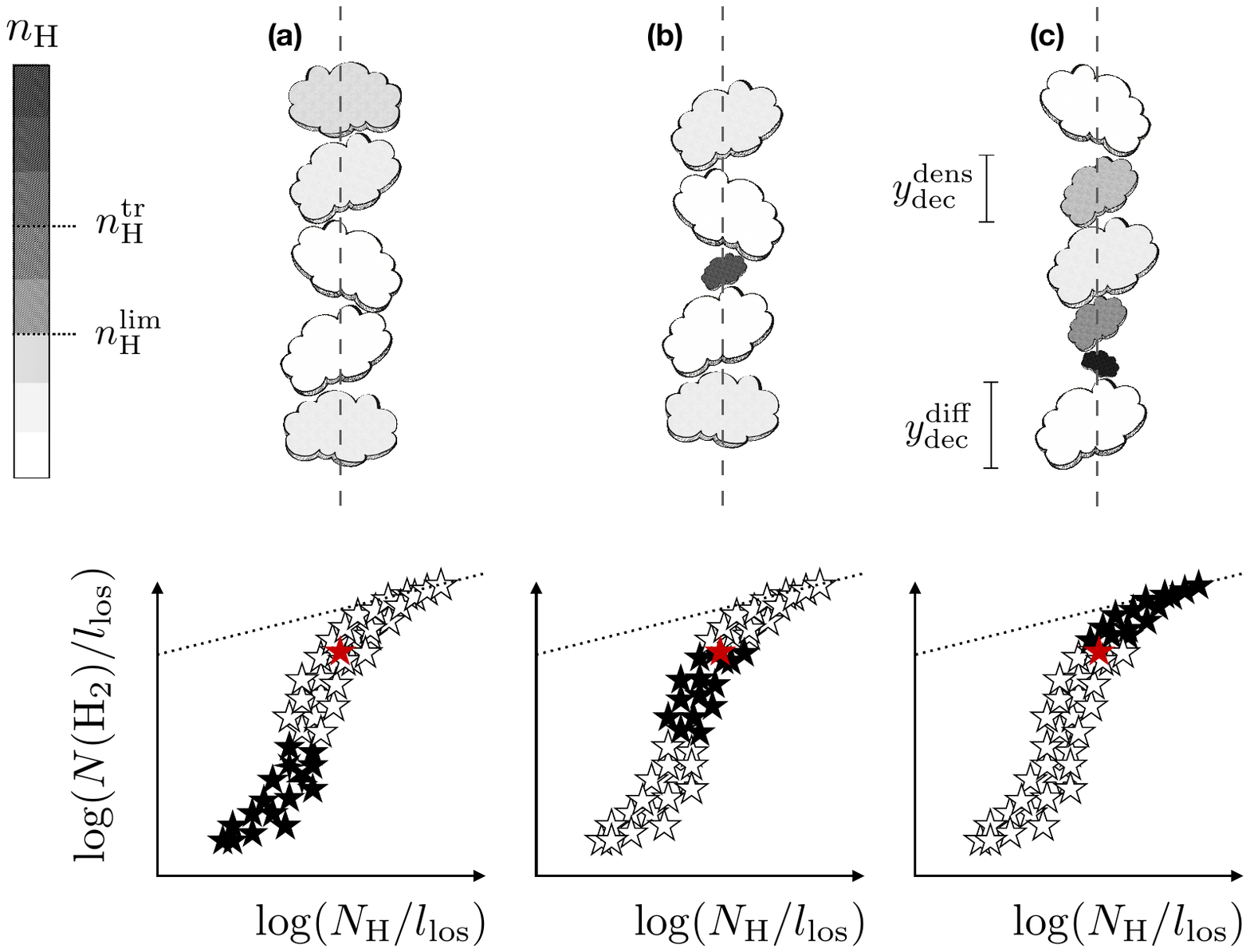}
\caption{Schematic view of lines of sight of fixed length $l_{\rm los}$ inferred from the analytical model described in Appendix \ref{app:analytical}, and corresponding contributions to the histogram of the normalized column densities $N_{\rm H}/l_{\rm los}$ and $N({\rm H}_2)/l_{\rm los}$. In the bottom panels, the white stars correspond to all lines of sight while the black stars correspond to the specific cases illustrated above. Any line of sight is intercepting several components of constant density $n_{\rm H}$. Diffuse components ($n_{\rm H} < n_{\rm H}^{\rm lim}$) have a size $y_{\rm dec}^{\rm diff}$; dense components ($n_{\rm H} \geqslant n_{\rm H}^{\rm lim}$) 
have a distribution of sizes $y_{\rm dec} \leqslant y_{\rm dec}^{\rm dens}$. Only components with densities larger than $n_{\rm H}^{\rm tr}$ are molecular (see main text). The red star in the bottom panels indicates the mean value of $N_{\rm H}/l_{\rm los}$ and $N({\rm H}_2) /l_{\rm los}$ computed over a large sample of lines of sight (white stars), hence the expected mean molecular fraction. The dotted line indicates a fully molecular medium.}
\label{Fig-scheme-los}
\end{center}
\end{figure}

The computation of column densities over various lines of sight and the subsequent kingfisher diagrams are the outcome of three main factors: the local conditions of the gas ($n_{\rm H}$, $G_{\rm eff}$, $T$, and the self-shielding) which control the local abundance of H$_2$; the probabilistic ordering of these local conditions along any random line of sight of size $l_{\rm los}$; and finally, the distribution of sizes $l_{\rm los}$ used for the reconstruction algorithm (see Sect. \ref{sec:reconstruction}).
In order to separate these effects, in particular the impact of local conditions from the probabilistic aspects, and propose a physical interpretation of the behaviors shown in this paper, we developed a semi-analytical approach. The resulting model and the confrontations of its predictions to the results of numerical simulations are described in details in Appendix \ref{app:analytical}. To keep the paper concise, we only summarize here its basic ingredients and our main deductions.

Following the works of \citet{vazquez-semadeni_probability_2001}, \citet{bialy_h_2017}, we assume that any line of sight can be understood as a succession of density fluctuations. These fluctuations are supposed to be fully correlated over a distance called the "decorrelation scale", and completely uncorrelated over larger distances. Because of the biphasic nature of the neutral gas, we adopt two different decorrelation scales: $y_{\rm dec}^{\rm diff}$ if $n_{\rm H} < n_{\rm H}^{\rm lim}$ and $y_{\rm dec}^{\rm dens}$ if $n_{\rm H} \geqslant n_{\rm H}^{\rm lim}$. The limit $n_{\rm H}^{\rm lim}$ separating the diffuse and dense components is chosen as the inflection point between the two log-normal distributions classically found in the PH of the gas density (see Fig. \ref{Fig-PDF-dens} of Appendix \ref{app:analytical}). Analyzing the production of H$_2$ and the integration of column densities in this framework leads to the following conclusions.

\begin{enumerate}
\item The comparisons with numerical simulations performed at four different scales and for fifteen different simulations show that the analytical model reproduces to an outstanding level the 1D PHs of the total column density $N_{\rm H}$ (see Fig. \ref{Fig-compar-sim-mod-Ht}), assuming that the diffuse gas is correlated over a scale $y_{\rm dec}^{\rm diff} = 0.2\,L_{\rm drive}$ (i.e., 20 pc for the fiducial simulation), and that the dense gas is correlated over a scale $y_{\rm dec}^{\rm dens} = 10\,\,{\rm pc}\,\,(\overline{n_{\rm H}}/ 2\,\,{\rm cm}^{-3})^{1/3}$. Interestingly, the value of $y_{\rm dec}^{\rm diff}$ is similar to that obtained by \citet{Bialy2020} in a set of isothermal MHD simulations with different driving scales and Mach numbers. This confirms that the WNM, which fills most of the volume, behaves like an isothermal gas perturbed by a sustained turbulent forcing.
\item Similar comparisons performed on the 2D PHs of $N_{\rm H}$ and $N({\rm H}_2)$ (Figs. \ref{Fig-compar-sim-mod-H2} and \ref{Fig-compar-sim-mod-H2_2} and Sect. \ref{Sect-distrib-sizes}) indicate that the dense gas is necessarily composed of a distribution of structures of different sizes. Indeed, while most of the mass and volume of the cold HI can be accurately modeled with a single scale $y_{\rm dec}^{\rm dens} = 10\,\,{\rm pc}\,\,(\overline{n_{\rm H}}/ 2\,\,{\rm cm}^{-3})^{1/3}$, H$_2$ is required to be built up in smaller components. $y_{\rm dec}^{\rm dens}$ should therefore be considered as the maximum decorrelation scale of the dense gas.
\item The local production of H$_2$ mostly depends on the density: low density components are atomic while high density components are molecular. The threshold $n_{\rm H}^{\rm tr}$ triggering the transition between the two regimes is set by the local self-shielding (induced by the component itself) and the large-scale self-shielding (induced by the surrounding environment). The local self-shielding alone in a component of size $y_{\rm dec} \leqslant y_{\rm dec}^{\rm dens}$ implies 
\begin{equation}
n_{\rm H}^{\rm tr} \sim 8\,\,{\rm cm}^{-3}\,\,G_0^{1/2} (y_{\rm dec}/10\,\,{\rm pc})^{-1/2}.
\end{equation}
The large-scale self-shielding can be seen as a stochastic process that lowers this limit: for $L=200$ pc, $n_{\rm H}^{\rm tr}$ is found to be reduced by a factor of two on average.
\item As schematized in Fig. \ref{Fig-scheme-los}, the distribution of normalized column densities $N_{\rm H}/l_{\rm los}$ and $N({\rm H}_2)/l_{\rm los}$ obtained for a given $l_{\rm los}$ can be divided in three categories. Lines of sight with low integrated molecular fraction ($f_{{\rm H}_2} < 10^{-4}$) exclusively contain components with $n_{\rm H} < n_{\rm H}^{\rm tr}$ (case a). In contrast, lines of sight with high integrated molecular fraction ($f_{{\rm H}_2} > 10^{-2}$) necessarily intercept at least one large or several small components at high density $n_{\rm H} > n_{\rm H}^{\rm tr}$ (case c). In spite of what intuition dictates, lines of sight with intermediate integrated molecular fraction 
($10^{-4} \leqslant f_{{\rm H}_2} \leqslant  10^{-2}$) do not result from components at moderate densities ($n_{\rm H} \sim n_{\rm H}^{\rm tr}$) but from the combination of low density material and a small number of small components at high density $n_{\rm H} > n_{\rm H}^{\rm tr}$ (case b).
\item The proportions of lines of sight of types (a), (b), and (c) (Fig. \ref{Fig-scheme-los}) are given by the volume filling factors of the diffuse and dense gas and the length of the lines of sight $l_{\rm los}$. If $l_{\rm los}$ is comparable to $y_{\rm dec}^{\rm diff}$ and $y_{\rm dec}^{\rm dens}$, the 2D PH of $N_{\rm H}/l_{\rm los}$ and $N({\rm H}_2) /l_{\rm los}$ is spread and contains a large number of lines of sight of type (a). Larger $l_{\rm los}$ naturally favor lines of sight of types (b) and (c). If $l_{\rm los}$ becomes large compared to to $y_{\rm dec}^{\rm diff}$ and $y_{\rm dec}^{\rm dens}$, the central limit theorem applies. The spread PH described above is squeezed along the x and y axis as both $N_{\rm H}/l_{\rm los}$ and $N({\rm H}_2)/l_{\rm los}$ progressively tend toward Gaussian distributions centered on the means (red star in Fig. \ref{Fig-scheme-los}).
\end{enumerate}

\section{Comparison with observations}\label{sec:results}

		\subsection{Fiducial simulation} \label{sec:res_fidsimu}

\begin{figure}
\centering
\includegraphics[width=9cm,trim = 1.0cm 2.0cm 1.0cm 1cm, clip,angle=0]{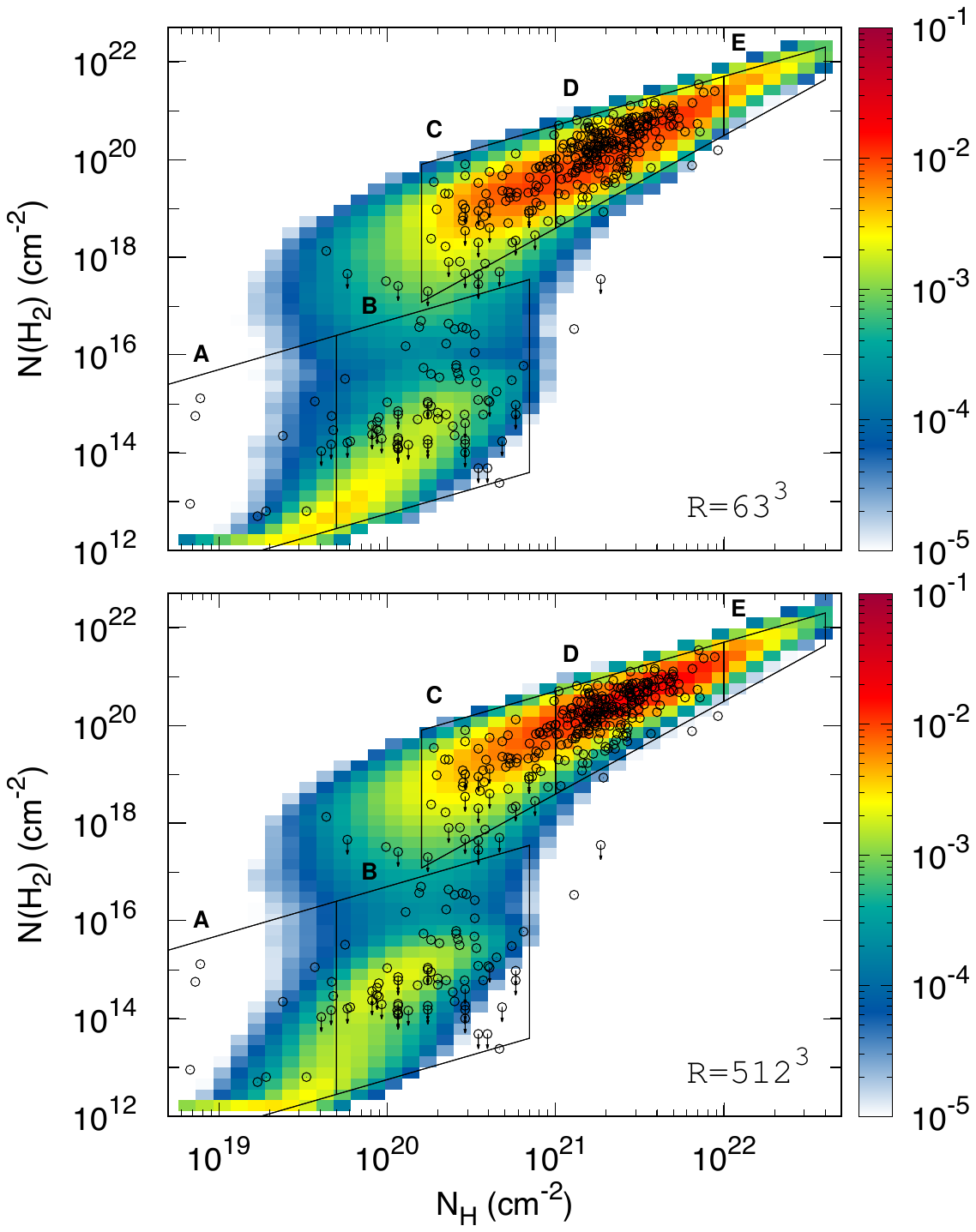}
\caption{Comparison of the observational dataset (black points) to the 2D probability histogram reconstruction algorithm (see Sect. \ref{sec:reconstruction}) applied to the fiducial simulation (colored histogram). Results are shown for two resolutions, $R=64^3$ (top panel) and $R=512^3$ (bottom panel). Observations include detections of H$_2$ (circles) and upper limits on N(H$_2$) (arrows). The color code indicates the fraction of lines of sight (in logarithmic scale) contained in each bin. As a reminder, contours of the regions A, B, C, D, and E defined in Sect. \ref{sec:h2} (see Table \ref{Tab-regionHH2}) are also displayed.
}
\label{fig:resolution}
\end{figure}

In Fig. \ref{fig:resolution}, we compare the observational dataset to the 2D PH of $N_{\rm H}$ and $N({\rm H}_2)$ (i.e., the kingfisher diagram) obtained with the reconstruction algorithm applied to the fiducial simulation for two resolutions, $R=64^3$ and $R=512^3$. In each panel, the color code indicates, in logarithmic scale, the fraction of lines of sight predicted for any couple ($N_{\rm H}$,$N({\rm H}_2)$). Quantitative comparisons of the observed and predicted fractions of lines of sight, mean molecular fractions, and dispersions in the regions A, B, C, D, and E defined in Sect. \ref{sec:h2} (see Table \ref{Tab-regionHH2}) are given in Fig. \ref{Fig-compar-sim-obs-L2p2}. Unexpectedly, the sample of lines of sight built from the fiducial simulation reproduces, to an outstanding level, the global trend of the  HI-to-H$_2$ transition and its statistical properties. Without taking into account any possible variation of the parameters along the lines of sight or from one line of sight to the next, the structures induced by the joint actions of turbulence and thermal instability alone are found to produce a wealth of lines of sight whose probabilities of occurrence match those derived from the observations. 

In particular, the integrated molecular fraction is predicted to have a bimodal distribution with a transition occurring at $N_{\rm H} \sim 3 \times 10^{20}$ cm$^{-2}$ and extending over one order of magnitude of total column density. More quantitatively, the fractions of lines of sight simulated and observed in regions A, B, C, and D are found to differ by 50\% at the most (see Fig. \ref{Fig-compar-sim-obs-L2p2}). Similarly, the observed and simulated mean molecular fractions and their corresponding dispersions are found to be comparable and to differ by less than a factor of three in region B, and less than a factor of two in regions C, and D. Because of the distribution of background sources, the reconstructed ensemble predicts a strong increase in probability from region C to D, which contains a large fraction of the entire sample, and a decrease in the dispersion of $f_{{\rm H}_2}$ as a function  of $N_{\rm H}$. At last, the probability of occurrence of lines of sight with $N_{\rm H} < 3 \times 10^{19}$ and $f_{{\rm H}_2} > 10^{-3}$ or $N_{\rm H} > 3 \times 10^{21}$ and $f_{{\rm H}_2} < 10^{-3}$ are rare to nonexistent. All these features are also found in the observational dataset.

Notwithstanding, Figs. \ref{fig:resolution} and \ref{Fig-compar-sim-obs-L2p2} also reveal a few discrepancies. Firstly, about 16 observed lines of sight (out of 360) lay at the border of the simulated distribution, in regions where the predicted probability is $\leqslant 10^{-4}$ (or even smaller than $10^{-5}$ for the white regions of Fig. \ref{fig:resolution}), a value far smaller than the inverse number of observations $\sim 3 \times 10^{-3}$. This implies that the simulation used here somehow fails to explain on its own part of the diversity observed in the Solar Neighborhood. Secondly, the mean molecular fractions predicted in regions A and B are found to be noticeably smaller than that derived from the observations, by about a factor of ten and three, respectively. It is important to note, however, that these molecular fractions deduced from the observations are probably overestimated as a third of the observed lines of sight contained in these regions correspond to upper limits on $N({\rm H}_2)$. Finally, the simulated sample clearly shows that a substantial fraction of the lines of sight are in the region E, with an integrated probability of 4\%. If the observational sample of 360 targets is unbiased, between 10 and 18 lines of sight should have been observed in this region, which is not the case. These discrepancy will be discussed in more details in Sect.~\ref{sec:discussion}.

\subsection{Impact of the resolution} \label{sec:impact_res}

The comparison of the two panels of Fig. \ref{fig:resolution} shows that the kingfisher diagram is independent of the resolution over about one order of magnitude of scales, from $64^3$ to $512^3$. Even if not systematically shown, this unusual result is not limited to the fiducial setup but is a general feature of all the simulations explored in this work (see Fig. \ref{fig:KSstat_Resolutions} of Appendix \ref{app:KStest} for instance). Our interpretation is based on the analytical model presented in Appendix \ref{app:analytical} and summarized in the previous section.

Evidently, high resolution simulations are important to accurately describe small scale structures. In particular, large resolution are required for modeling the formation of dense and gravitationally bound environments and follow their collapse. The fact that the kingfisher diagrams are independent of resolution for $R \geqslant 64^3$ therefore suggests that the cold and dense structures between 0.4 and 3 pc have no influence on the distributions of $N_{\rm H}$ and $N({\rm H}_2)$ for the fiducial simulations and are insignificant in the mass and the volume budgets of H and H$_2$. This is in line with the conclusions deduced from the analytical model. Indeed, as explained in the previous section, the turbulent forcing at large scale induces density fluctuations in the diffuse gas that extend over $\sim 20$ pc and a distribution of dense structures with sizes smaller than $\sim 10$ pc. While the total quantity of matter is inferred to be contained in the diffuse and the largest dense components, H$_2$ is exclusively built up in dense components smaller than $\sim 10$ pc. The results of Fig. \ref{fig:resolution} combined with conclusions deduced from the analytical model therefore imply that the structures contributing the most to the mass and volume of H$_2$ are above 3 pc and below $10$ pc for the fiducial simulation.

Beside the physical insights on the typical scales participating to the build-up of column densities, this result also provides a strong justification for using simulations with moderate numerical resolution for the study of the HI-to-H$_2$ transition. In all this work we therefore adopt a standard resolution of $R=256^3$ unless indicated otherwise.

\subsection{Impact of $G_0$ and $\overline{n_{\rm H}}$} 
\label{sec:impact_dens}

\begin{figure*}[!ht]
\begin{center}
\includegraphics[width=14.5cm,trim = 1.5cm 1.8cm 0.5cm 1cm, clip,angle=0]{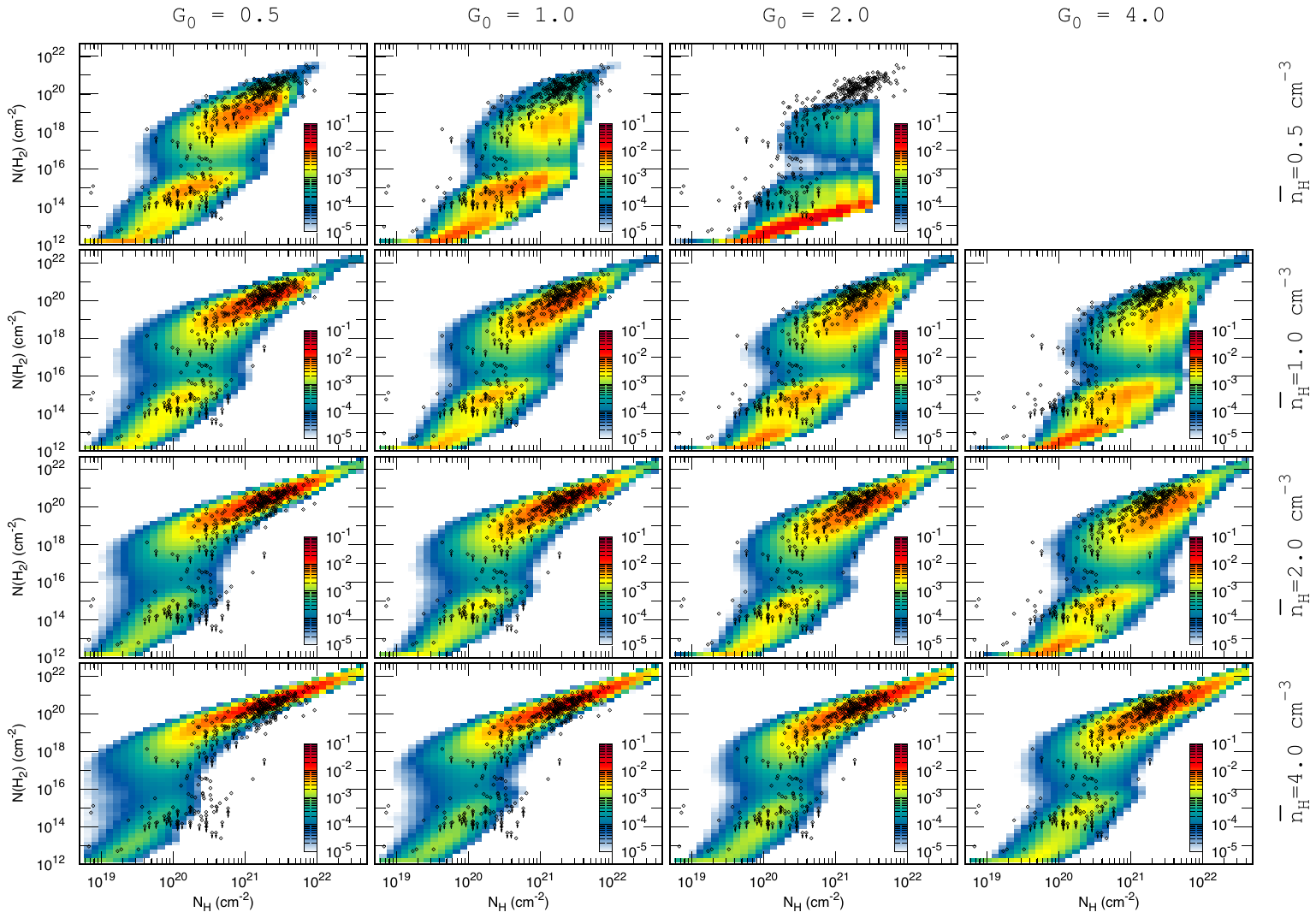}
\includegraphics[width=14.5cm,trim = 1.5cm 1.8cm 0.5cm 1cm, clip,angle=0]{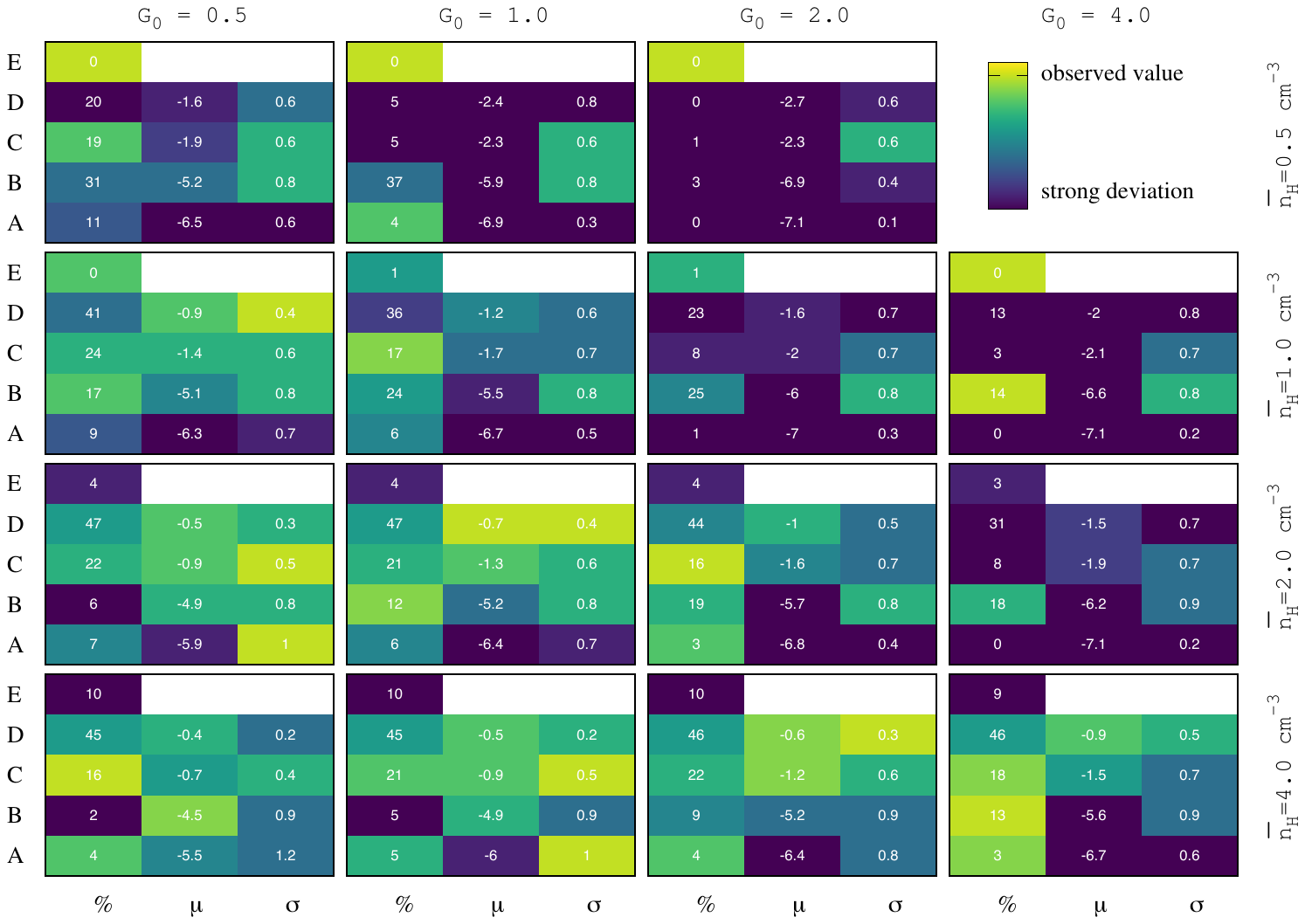}
\caption{Top frame: comparisons of the observational dataset (black points) with the 2D probability histograms of $N_{\rm H}$ and $N({\rm H}_2)$ computed from the reconstruction algorithm (see Sect. \ref{sec:reconstruction}) applied to the simulations. Bottom frame: fraction of lines of sight (\%), and mean value $\mu$ and dispersion $\sigma$ of the logarithm of the molecular fraction computed from the simulated histograms in the regions A, B, C, D, and E defined in Table \ref{Tab-regionHH2}.  Numbers correspond to the values of \%, $\mu$, and $\sigma$. The color code indicates a measure of distance (in arbitrary units) between the observed and simulated values in order to guide the eye. These comparisons are shown in each frame for 15 different simulations with $G_0$ varying from 0.5 (left panels) to 4 (right panels) and $\overline{n_{\rm H}}$ varying from 0.5 cm$^{-3}$ (top panels) to 4 cm$^{-3}$ (bottom panels). All other parameters are set to their fiducial values (see Table \ref{table:grids}).}
\label{Fig-compar-sim-obs-L2p2}
\end{center}
\end{figure*}
		
\begin{figure*}[!ht]
\begin{center}
\includegraphics[width=14.5cm,trim = 1.5cm 1.8cm 0.5cm 1cm, clip,angle=0]{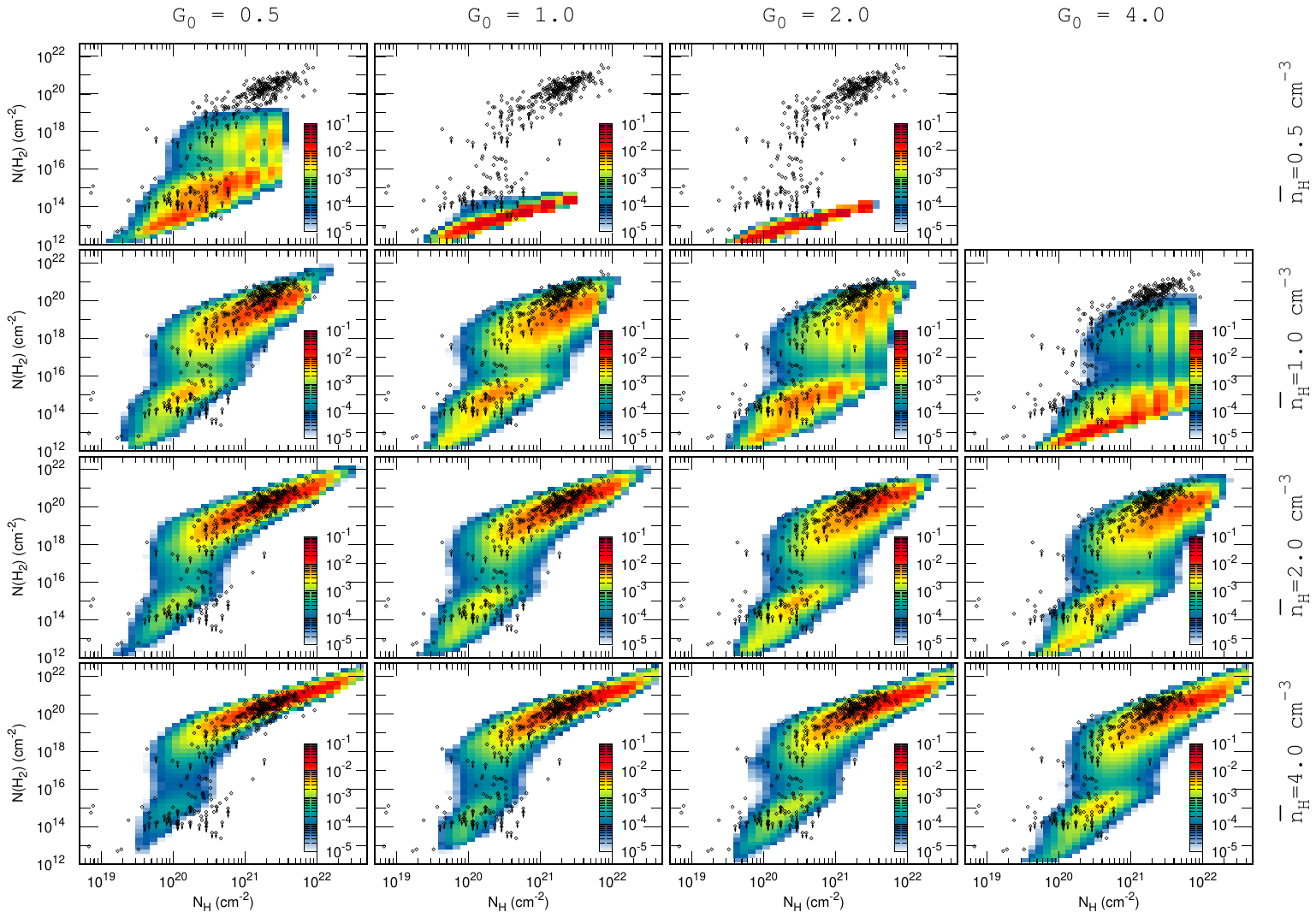}
\includegraphics[width=14.5cm,trim = 1.5cm 1.8cm 0.5cm 1cm, clip,angle=0]{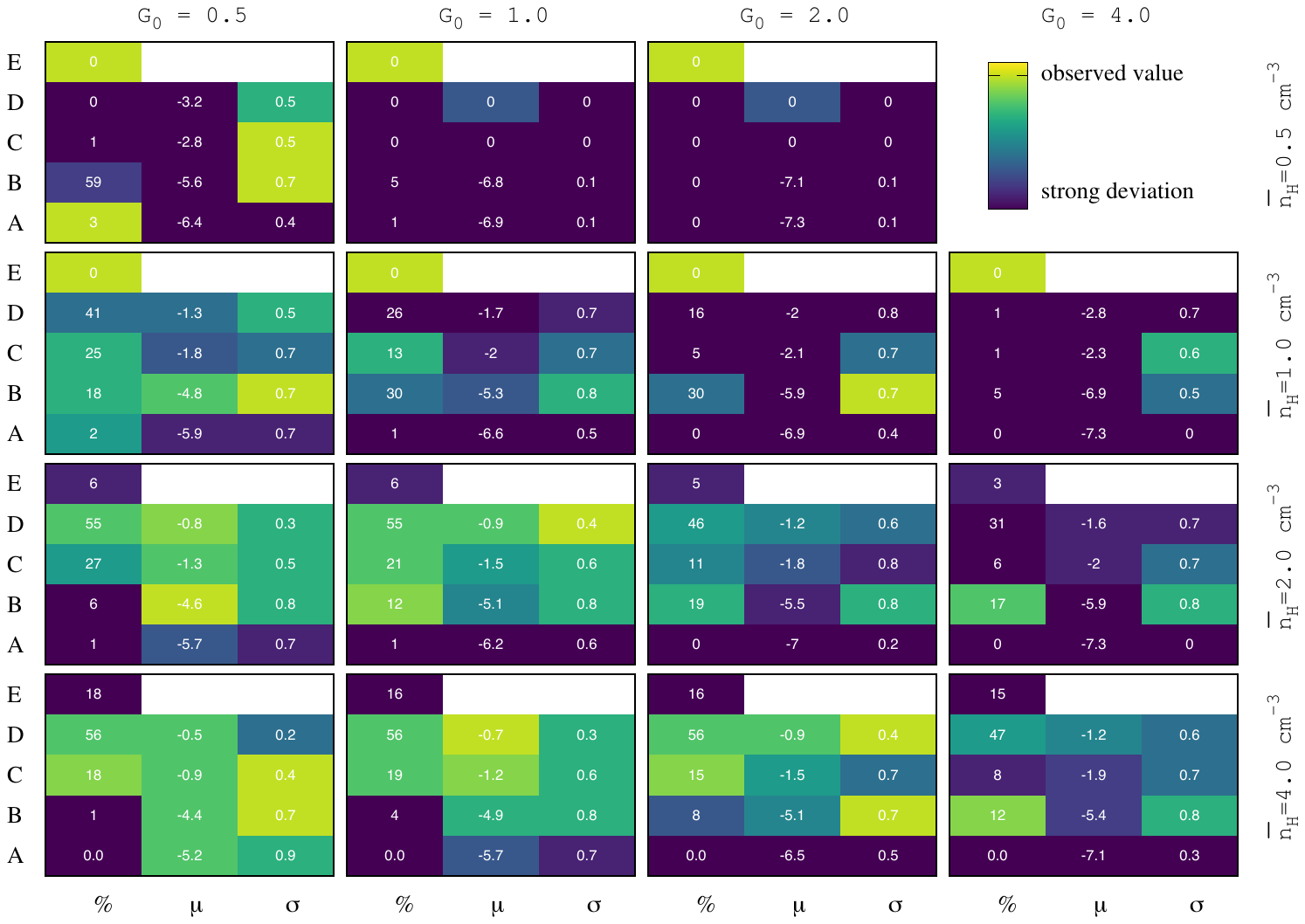}
\caption{Same as in Fig. \ref{Fig-compar-sim-obs-L2p2} for simulations with a box size of $L=50$ pc instead of $200$ pc. The turbulent forcing is adjusted as in Fig. \ref{Fig-prop-WNM_CNM} ($F=3.6 \times 10^{-3}$ kpc Myr$^{-2}$) to obtain similar velocity dispersions for $L=50$ pc and $L=200$ pc.}
\label{Fig-compar-sim-obs-L5p1}
\end{center}
\end{figure*}

Figs. \ref{Fig-compar-sim-obs-L2p2} and \ref{Fig-compar-sim-obs-L5p1} show comparisons between the observational dataset and the 2D PHs extracted from the simulations for 0.5 cm$^{-3}$ $\leqslant \overline{n_{\rm H}} \leqslant 4$ cm$^{-3}$, 0.5 $\leqslant G_0 \leqslant 4$, and two sets of values of the box size and the turbulent forcing, $L=200$ pc and $F=9 \times 10^{-4}$ kpc Myr$^{-2}$ (Fig. \ref{Fig-compar-sim-obs-L2p2}) and  $L=50$ pc and $F=3.6 \times 10^{-3}$ kpc Myr$^{-2}$ (Fig. \ref{Fig-compar-sim-obs-L5p1}). 
While the trend and the statistics of the HI-to-H$_2$ transition weakly depend on the resolution, they strongly depend on the total mass of the gas parametrized by $\overline{n_{\rm H}}$ and on the UV illumination factor. As $G_0$ increases or $\overline{n_{\rm H}}$ decreases, (1) the fraction of lines of sight with large $f({\rm H}_2)$ drops to the benefit of lines of sight with low $f({\rm H}_2)$, (2) the transition is shifted toward larger total column density and its width increases, and (3) the molecular fraction globally decreases over all lines of sight while its dispersion increases. Interestingly, and because $\overline{n_{\rm H}}$ and $G_0$ have opposite effects, the simulations that reproduce the most accurately the observed statistics of the HI-to-H$_2$ transition follow a trend with $G_0/\overline{n_{\rm H}} \sim 0.5-1$. While similar, the effect of these two parameters are, however, not symmetrical. In particular, $\overline{n_{\rm H}}$ has an obvious and strong impact on the fraction of lines of sight in region E, regardless of $G_0$. Likewise, the fraction of lines of sight at low column densities (regions A and B) and the mean molecular fraction at large column densities (regions C and D) are not constant for a given $G_0/\overline{n_{\rm H}}$ ratio but respectively decrease and increase with $\overline{n_{\rm H}}$. All these properties effectively break the degeneracies between the two parameters. All things considered, the tightest concordance between observed and simulated data is obtained for $\overline{n_{\rm H}} \sim 1 - 2$ cm$^{-3}$, in agreement with the Galactic midplane density deduced from HI surveys in the Solar Neighborhood \citep{kalberla_hi_2009}.
		
At first sight, the results described above seem obvious as they somehow mimic the dependencies of the HI-to-H$_2$ transition found with detailed models of photodissociation regions (e.g., \citealt{krumholz_atomic--molecular_2009,sternberg_h_2014}). Such an interpretation is, however, a dangerous misconception. Indeed, while $G_0$ is tightly linked to the effective radiation field $G_{\rm eff}$ that locally illuminates the gas, the mean density $\overline{n_{\rm H}}$ should not be mistaken for the local density. Moreover, the results displayed in Fig. \ref{Fig-compar-sim-obs-L2p2} and \ref{Fig-compar-sim-obs-L5p1} cannot be compared to PDR models because they are statistical in nature. For instance, simulations at high $G_0$ do not preclude the existence of clouds with high molecular fractions. Indeed, increasing $G_0$ may lead to denser local environments with larger molecular fractions whose probability of occurrence along a line of sight is simply reduced. It implies that the results shown here are very different from PDR model predictions. They rather reflect the complex link between the global properties of the simulation (mass, illumination, driving scale) on the one side, and the local conditions and their probability distribution functions on the other side. 

Because the local abundance of H$_2$ is inversely proportional to $G_0$, increasing $G_0$ naturally reduces the local self-shielding. Similarly, increasing $G_0$ or decreasing $\overline{n_{\rm H}}$ reduce the large-scale self-shielding. As a result, the density threshold $n_{\rm H}^{\rm tr}$ required to produce highly molecular environments (see item 3. of Sect. \ref{sec:interpretmodel}) rises by a factor of three when either $G_0$ is multiplied or $\overline{n_{\rm H}}$ is divided by a factor of eight. While significant, such an effect on the local conditions of the gas is, however, too shallow to fully explain the variations observed in Figs. \ref{Fig-compar-sim-obs-L2p2} and \ref{Fig-compar-sim-obs-L5p1}.

Indeed, regardless of local conditions, $G_0$ and $\overline{n_{\rm H}}$ have a major impact on the probabilistic reconstruction of lines of sight. As shown in Sect. \ref{sec:propbiphase}, increasing $G_0$ or decreasing $\overline{n_{\rm H}}$ strongly reduce the fractions of mass and volume occupied by the dense and cold gas. This not only reduces the molecular fraction averaged over the entire simulation (red star in Fig. \ref{Fig-scheme-los}) but also favors the occurrence of lines of sight with low or intermediate $f_{{\rm H}_2}$ (case (a) and (b) in Fig. \ref{Fig-scheme-los}). When combined with the distribution of sizes $l_{\rm los}$, the HI-to-H$_2$ transition is naturally shifted toward larger $N_{\rm H}$, and the asymptotic molecular fraction at high $N_{\rm H}$ drops. Moreover, because the central limit theorem requires larger lines of  sight to apply, the HI-to-H$_2$ transition is naturally wider and the dispersion of lines of sight at large molecular fraction increases. This final property is particularly well seen in the kingfisher diagram obtained for the simulation at $G_0=0.5$ and $\overline{n_{\rm H}} =4$ cm$^{-3}$ where most of the lines of sight follow an homothetic transformation of the mean normalized column densities $N_{\rm H}/l_{\rm los}$ and $N({\rm H}_2)/l_{\rm los}$ (red star in Fig. \ref{Fig-scheme-los}).

\subsection{Impact of the box size $L$} \label{sec:impact_size}

The impacts of the box size revealed by comparing Figs. \ref{Fig-compar-sim-obs-L2p2} and \ref{Fig-compar-sim-obs-L5p1} partly follow the results of the previous section. Reducing $L$ by a factor of four drastically reduces the total amount of matter in the simulation, hence the absorption of the impinging UV radiation field and the large-scale self-shielding. As shown in Sect. \ref{sec:propbiphase}, the mean pressure of the gas rises while the mass and volume occupied by the CNM decreases. All the local and statistical effects described in the previous section therefore apply and modify the kingfisher diagrams accordingly. Changing $L$ has, however, two additional and specific consequences.

Because $L_{\rm drive}$ is four times smaller in the simulations displayed in Fig. \ref{Fig-compar-sim-obs-L5p1} than in those displayed in Fig. \ref{Fig-compar-sim-obs-L2p2}, the decorrelation scales $y_{\rm dec}^{\rm diff}$ and $y_{\rm dec}^{\rm dens}$ are correspondingly smaller. According to the interpretation given in Sect. \ref{sec:interpretmodel}, all the reconstructed lines of sight are therefore considerably larger than individual density fluctuations. This not only favors the occurrence of lines of sight at intermediate and high molecular fractions (cases (b) and (c) of Fig. \ref{Fig-scheme-los}) but also magnify the impact of the central limit theorem (see item 5. of Sect. \ref{sec:interpretmodel}), as already illustrated by \citet{Bialy_2019a}. The kingfisher diagrams shown in Fig. \ref{Fig-compar-sim-obs-L5p1} are thus squeezed along the x and y axis compared to those of Fig. \ref{Fig-compar-sim-obs-L2p2}. Consequently, the simulations at $L=50$ pc predict almost no line of sight at low total column density ($N_{\rm H} \leqslant 3 \times 10^{19}$ cm$^{-2}$) and systematically underestimate the proportion of line of sight in region A compared to the observations.

Because it controls the total amount of matter, changing $L$ has finally a major effect on gravitational forces. For $L=50$ pc, the size of the largest dense clouds are almost always smaller than the Jean length computed in the CNM. Consequently, self-gravity plays almost no role in the physical state or the evolution of the gas for all simulations at $L=50$ pc. The specific impact of gravity on the probability histograms of column densities is described in more details in the following section.

\subsection{Impact of gravity} \label{sec:impact_grav}

\begin{figure}[ht!]
\centering
\includegraphics[width=1\linewidth]{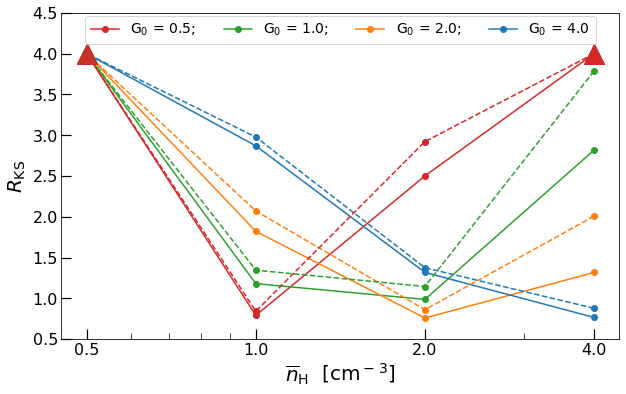}
\caption{KS distance between the simulations and the observational sample as a function of the mean density $\overline{n_{\rm H}}$ and the UV scaling factor $G_0 = 0.5$ (red), 1 (green), 2 (orange), and 4 (blue). Results obtained with and without gravity are shown with solid lines and dashed lines, respectively. All other parameters are set to their standard values (see Table \ref{table:grids}). Points correspond to reliable measurements of the KS distances. Triangles indicate lower limits corresponding to simulations where the upper error bar on $R_{\rm KS}$ tends toward infinity (see Appendix \ref{app:KStest}).}
\label{fig:KStest_WONandWof}
\end{figure}

To facilitate the comparison between simulations and observations and avoid a 
pedestrian repetition of the kingfisher diagrams, we developed a modified version of the Kolmogorov-Smirnov test. This test, fully explained in Appendix \ref{app:KStest} and validated over 30 different simulations, defines a value $R_{\rm KS}$, called the KS distance, that measures how two 2D PDFs (or 2D PHs) differ from one another. In a nutshell, any observational datapoint in a 2D diagram can be used to divide the diagram into four different regions which each contain different fractions of observed and simulated data. The modified KS test simply searches for the observational datapoint and the region that maximize the ratio between the fractions of simulated and observed lines of sight. The distance $R_{\rm KS}$ is the absolute value of the logarithm of this ratio: A value $R_{\rm KS}=1$ therefore implies that some region in the 2D diagram contains tens times more or ten times less simulated data than required to explain the observations, and that all the other regions have smaller 
ratios.

The results of the KS test applied to 15 simulations run with and without gravity are shown in Fig. \ref{fig:KStest_WONandWof}. As expected from Sect. \ref{sec:impact_dens}, the KS distance strongly depends on both $\overline{n_{\rm H}}$ and $G_0$. In comparison, gravity appears to have a limited impact on the kingfisher diagrams. While including gravity seems to slightly reduce the KS distance, the trends as functions of $\overline{n_{\rm H}}$ and $G_0$ and the set of simulations found to minimize $R_{\rm KS}$ remain unaltered. These results can be understood by simple statistical considerations.

The impact of gravity in multiphase simulations is to produce self-gravitating environments which appear as a high density tail in the PDF of the gas density. Because these self-gravitating clumps are dense and fully molecular, they often dominate the integrated column densities of both HI and H$_2$ along any line of sight that intercept them, and therefore favor lines of sight at high molecular fraction (case (c) in Fig. \ref{Fig-scheme-los}). This not only induces a tail at high $N_{\rm H}$ and $N({\rm H}_2)$ in the kingfisher diagram but may also contribute to shift the HI-to-H$_2$ transition to lower $N_{\rm H}$ as faint but highly molecular lines of sight starts to appear. The importance of these two effects depends on the area filling factor of self-gravitating clumps and on whether they carry or not a substantial fraction of the mass of the dense gas. For $\overline{n_{\rm H}}=0.5$ cm$^{-3}$, self-gravitating clumps occupy less than 0.001\% of the entire volume and carry less than a percent of the mass of the gas. These fractions increase, however, as functions of $\overline{n_{\rm H}}$: for $\overline{n_{\rm H}}=4$ cm$^{-3}$, self-gravitating environments occupy 0.01\% of the volume and contain as much as  30\% of the total mass. Therefore, while the impacts of gravity on the kingfisher diagram is negligible for most simulations, they become important at high $\overline{n_{\rm H}}$: this property is effectively captured by the KS distances displayed in Fig. \ref{fig:KStest_WONandWof}.

\subsection{Impact of turbulent forcing} \label{sec:impact_turb}

\begin{figure}[ht!]
\centering
\includegraphics[width=1\linewidth]{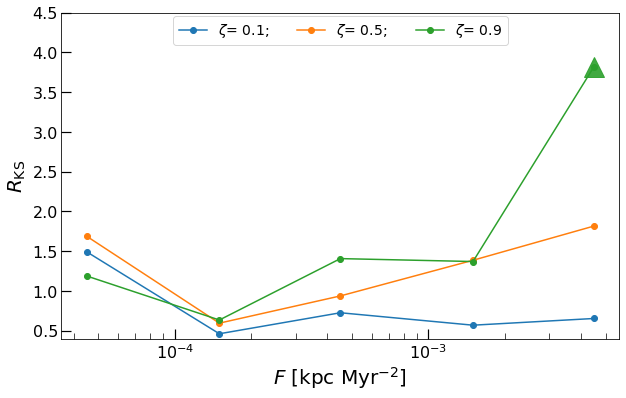}
\caption{KS distances between the simulations and the observational sample computed for five values of the acceleration parameter $F=4.5 \times 10^{-5}$, $1.5 \times 10^{-4}$, $4.5 \times 10^{-4}$, $1.5 \times 10^{-3}$, and $4.5 \times 10^{-3}$ kpc Myr$^{-2}$, and three values of the compressive ratio $\zeta=0.1$ (blue points), 0.5 (orange points), and 0.9 (green points), which set the balance between compressive and solenoidal forcing (see Sect. \ref{sec:turb_forcing}). All other parameters are set to their standard values (see Table \ref{table:grids}). Points correspond to reliable measurements of the KS distances. The triangle indicates a lower limit corresponding to a simulation where the upper error bar on $R_{\rm KS}$ tends toward infinity (see Appendix \ref{app:KStest}).}
\label{fig:KStest_turbulence}
\end{figure}

As done in the previous section, the impact of the turbulent forcing is discussed through the results of the Kolmogorov-Smirnov test. The KS distance obtained for various configurations of the turbulent forcing (Fig. \ref{fig:KStest_turbulence}) shows that the strength of turbulence affects differently the HI-to-H$_2$ transition depending on its nature: highly compressive turbulent forcing produces virtually identical column density distributions over almost two decades of the turbulent acceleration parameter $F$; oppositely the strength of the forcing significantly modifies these distributions if a substantial fraction of the kinetic energy is injected in pure solenoidal modes. The tightest agreement with the observational sample is obtained for $F \sim 1.5 \times 10^{-4}$ kpc Myr$^{-2}$ if $\zeta \geqslant 0.5$ and for all $F \geqslant 1.5 \times 10^{-4}$ kpc Myr$^{-2}$ if $\zeta = 0.1$. According to Sect. \ref{sec:propbiphase} (Fig. \ref{Fig-prop-WNM_CNM}), these values corresponds to a WNM velocity dispersion $\sigma_{\rm tur} \sim 2$ km s$^{-1}$ if $\zeta \geqslant 0.5$, and $\sigma_{\rm tur} \geqslant 2$ km s$^{-1}$ for $\zeta = 0.1$. In any case, a small amount of turbulence is always required.

All these characteristics are consequences of the fractions of volume and mass contained in the CNM structures and their size distribution. Without turbulence or with a weak turbulent forcing, CNM clouds are found to evolve toward large-scale entities which evaporate slowly. Such a configuration favor lines of sight at high molecular fraction (type (c) in Fig. \ref{Fig-scheme-los}) and leads to an overestimation of the global amount of H$_2$ in the local diffuse ISM. If the turbulent forcing increases, the CNM becomes progressively organized into a distribution of structures of different sizes down to the numerical resolution. Simultaneously, the fraction of mass located in the CNM phase diminishes to the benefit of the LNM (see Sect. \ref{sec:propbiphase}). Both effects reduce the mean molecular fraction of the gas and favor the occurrence of lines of sight of type (b) (see Fig. \ref{Fig-scheme-los}), in better agreement with the observational sample. Larger turbulent forcing ultimately lead to an underestimation of the global amount of H$_2$ in the local ISM. However, and as shown in Sect. \ref{sec:propbiphase}, this last effect is much more pronounced for a pure solenoidal forcing which efficiently prevents the unstable gas to condensate back to cold and dense environments compared to a pure compressive forcing.

Here again, the probabilistic information contained in the kingfisher diagram proves to provide a valuable tool to analyze the nature of turbulence in the diffuse local ISM. Indeed, the velocity dispersion deduced from the comparison of observations and simulations at high solenoidal forcing is substantially smaller than the velocity dispersion observed at high Galactic latitude (see Sect. \ref{sec:propbiphase}) and the velocity dispersion deduced from the observations of CO at a scale of 200 pc \citep{hennebelle_turbulent_2012}. The fact that the observed statistics of the HI-to-H$_2$ transition is reproduced over a broader range of velocity dispersion for a compressive forcing suggests that the large-scale turbulence of the diffuse local ISM is dominated by compressive modes. This picture is coherent with the results of \citet{saury_structure_2014} who found that a large-scale compressive forcing induces a distribution of thermal pressure in excellent agreement with that derived from the observations of the fine structure lines of CI \citep{jenkins_distribution_2011}. It is also coherent with the fact that 200 pc corresponds to the width of the Galactic disk seen in CO and therefore to the limit above which the equipartition between compressive and solenoidal modes switches from the values expected in a 3D fluid to those expected in a 2D medium, hence values of $\zeta$ smaller than 0.5.
At last, it concurs with the conclusions of \citet{Iffrig_2017} who found that compressible modes dominates at low altitudes, close to the equatorial plane, in simulations where the ISM turbulence is self-consistently driven by supernovae explosions.

\subsection{Impact of the magnetic field} \label{sec:impact_mag}

\begin{figure}[ht!]
\centering
\includegraphics[width=1\linewidth]{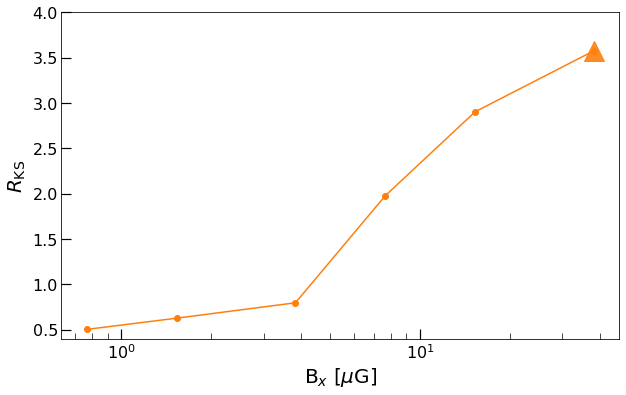}
\caption{KS distances between the simulations and the observational sample computed for six values of the initial magnetic field $B_x$. All other parameters are set to their standard values (see Table \ref{table:grids}). Points correspond to reliable measurements of the KS distances. The triangle indicates a lower limit corresponding to a simulation where the upper error bar on $R_{\rm KS}$ tends toward infinity (see Appendix \ref{app:KStest}).}
\label{fig:KStest_magnetic}
\end{figure}

The impact of the initial magnetic field on the kingfisher diagram simply reveals the competition between thermal instability which induces the production of dense environments, and the magnetic pressure which acts against this evolution. As shown in Fig. \ref{fig:KStest_magnetic}, the KS distance obtained for different magnetic field intensities is found to be constant until a critical value of $B_x \sim 4$ $\mu$G above which the predicted and observed distribution of $N_{\rm H}$ and  $N({\rm H}_2)$ significantly differ from one another. The initial homogeneous magnetic field adopted in the standard simulation, $B_x=3.8$ $\mu$G, is just under this critical value and corresponds to the case where the energy density of the large scale coherent magnetic field and the thermal energy density of the WNM are equivalent. It follows that reproducing the observed 2D PH of $N_{\rm H}$ and  $N({\rm H}_2)$ requires a magnetic field intensity below or at equipartition.

Interestingly, the value of $B_x$ adopted for the fiducial simulation leads, at steady-state, to a constant magnetic field intensity $B \sim 5$ $\mu$G for $n_{\rm H} < 10$ cm$^{-3}$, and a field intensity that scales as $B \propto (n_{\rm H})^{0.3}$ for $n_{\rm H} > 10$ cm$^{-3}$. Those values are comparable with the magnetic field intensities obtained in diffuse and molecular environments from Zeeman measurements \citep{Crutcher_2010} and those obtained in the most diffuse phases from Faraday rotation measurements toward extragalactic radio sources (e.g., \citealt{Frick_2001}). It appears that these information on the interstellar magnetic field are encoded, at least partly, in the statistical properties of the HI-to-H$_2$ transition.
	
\section{Discussion}\label{sec:discussion}
	
	\subsection{Observational biases}
		All the results presented in Sect. \ref{sec:results} are discussed under the assumption that the observational dataset is unbiased, meaning that the underlying lines of sight correspond to a random sample with no selection effect. This is not true. As \citet{krumholz_atomicmolecular_2008} already noticed, {\it"the FUSE and the Copernicus lines of sight have been specifically chosen to probe a certain range of column densities with a selection biased against high extinction which makes determining column densities very costly or altogether impossible."} Indeed, very few stars emit a UV radiation field strong enough to be detected through large visual extinction material. The problem is not due to the increase in H$_2$ absorption, which can be overcome by focusing on fainter bands, but to dust absorption itself and the complication of the structure of NaI often used as a proxy to derive the column density of HI \citep{rachford_far_2002}. Moreover, because such bias depends on the sensitivity of the instrument, it necessarily applies at different column densities for FUSE and \textit{Copernicus}.
		
		This selection effect obviously complicates the comparison between simulations and observations. As shown in Sect. \ref{sec:impact_dens} and Figs. \ref{Fig-compar-sim-obs-L2p2} and \ref{Fig-compar-sim-obs-L5p1}, several simulations,	including the fiducial setup, predict a significant fraction of lines of sight at high column densities ($N_{\rm H} \geqslant 10^{22}$ cm$^{-2}$, region E), in apparent	contradiction with the observations. However, the fact that the fraction of stars dismissed by selection is unknown makes it difficult to assess whether this result reveals an actual and important statistical discrepancy or a simple observational limit. It implies that the likelihood of a simulation to be representative of the local diffuse gas cannot be estimated from this criterion alone. It must involve other observational signatures such as the average and the dispersion of the molecular fraction in regions A, B, C, and D, or even the chemical and statistical signatures of other atomic and molecular lines. This latter aspect is currently under development and will be the subject of the next paper of this series.
		
		The fact that the limitations differ between FUSE and \textit{Copernicus} surveys finally raises the question of the validity of studying the two samples simultaneously. We find, 	however, that performing comparisons with simulations on the two samples separately gives very similar results and does not impact any of our conclusions. It is so because the observational bias discussed above occurs at an extinction which increases as the natural logarithm of the square root of the instrument sensitivity. The largest total column density probed by FUSE is thus only three times larger than that observed by \textit{Copernicus} \citep{gillmon_fuse_2006,rachford_molecular_2009}. Moreover, the number of lines of sight observed by FUSE that are above the maximum extinction seen by Copernicus corresponds to a small fraction of the entire sample \citep{gillmon_fuse_2006}.

	\subsection{Variations of physical conditions in the local ISM}
		The reconstruction of the simulated sample of lines of sight and the subsequent comparisons with the observations are done assuming that the local ISM can be built out of a single simulation (see Sect. \ref{sec:reconstruction}). This approach was chosen in order to highlight the natural variations induced by turbulence and thermal instability alone in a diffuse neutral gas with a known averaged density and UV illumination factor. However, because the medium probed by the observations extends in all directions around the sun over a maximum distance of 3 kpc (see Fig. \ref{fig:distr_lengths}), it stands to reason that potential variations of all parameters should be taken into account, not only from one line of sight to another but also along a single and outstretched line of sight. Indeed, such considerations would offer a natural explanation for observations whose existence is in contradiction with the corresponding simulated probability of occurrence (see Sect. \ref{sec:res_fidsimu}). 
		
		The total proton mass surface density deduced from HI and CO all sky surveys appears to be rather constant in the Galactic layer located between 5 and 12 kpc from the Galactic center (Fig. 9 of \citealt{miville-deschenes_physical_2017}). Taking into account variations of the ISM scale height above the Galaxy \citep{kalberla_hi_2009}, the midplane proton density is expected to vary by less than a factor of two over the corresponding volume.

        Using the Galactic star distribution of \citet{wainscoat_model_1992} and the grains composition and size distribution of \citet{weingartner_dust_2001}, \citet{porter_new_2005} and \citet{moskalenko_attenuation_2006} estimated the energy density of the radiation field across the Galaxy. Similarly to the midplane density, the mean UV radiation field is estimated to vary by a factor two over the volume considered in this paper (Fig. 2 of \citealt{porter_new_2005}). Interestingly, this estimation is far smaller than the variations derived by \citet{jenkins_distribution_2011} (Fig. 8 in their paper) from the observation of the fine structure line of CI in the local gas. Such discrepancies could be explained by the fact that \citet{jenkins_distribution_2011} perform local measurements: the observed gas could thus be located close to or far from 	an irradiating star. Alternatively, we note that the results obtained by \citet{jenkins_distribution_2011} are derived from models at equilibrium which do not take into account the uncorrelated perturbations of pressure and density in a turbulent multiphase medium: this naturally favors large fluctuations of the UV radiation field.

        The cosmic ray ionization rate inferred from submillimeter observations of several molecular ions, including OH$^+$, H$_2$O$^+$, ArH$^+$, and H$_3^+$, shows a wide distribution across the Galactic disk \citep{indriolo_herschel_2015}. A recent estimation performed by \citet{neufeld_cosmic-ray_2017} suggests that this rate could vary by a factor of five in the gas located between 5 and 12 kpc from the Galactic center.

        Finally, potential variations of the composition and the size distribution of grains in the local diffuse gas should also be considered. While the extinction curve is found to be surprisingly uniform in the Milky Way \citep{schlafly_optical-infrared_2016}, the local distribution of grains could change, not only along the line of sight but between the atomic and ionized phase and the molecular clouds. This would modify the efficiency of the photoelectric effect and the equilibrium between the two stable states of the neutral ISM, and would also have an impact the H$_2$ formation rate.

		Interestingly, if uncorrelated, the expected variations of $\overline{n_{\rm H}}$ and $G_0$ alone would help to explain the slight discrepancies described in Sect. \ref{sec:impact_dens} but it would also lead to a dispersion of lines of sight far larger than those observed (see Figs. \ref{Fig-compar-sim-mod-Ht} and \ref{Fig-compar-sim-mod-H2}). The fact that the predicted statistics of the HI-to-H$_2$ transition is close to the observed sample therefore suggests that the variations of all the parameters described above cannot be considered independently but must follow strong correlations which apply locally (as discussed, for instance, by \citealt{Bialy_2019a}) but also across the Galactic disk.

	\subsection{Fraction of ionized gas}
		The fraction of volume $\varphi$ occupied by the ionized phases of the ISM, the warm ionized medium (WIM) and the hot ionized medium (HIM), plays an important role in our reconstruction algorithm. As illustrated in Fig. \ref{fig:model}, this parameter controls the filling factor of the neutral medium along a line of sight of length $l_{\rm los}$. Unfortunately, its value in the Milky Way is highly uncertain and still debated.
		
		The consensus is that the volume of the HIM far exceeds that of the WIM and results from an interplay between supernovae explosions, which regularly produce hot gas in the Galactic disk, and buoyancy, which lifts this gas into the halo, releasing the pressure in the midplane. Early analytical studies neglecting buoyancy \citep{mckee_theory_1977} predicted a large	fraction of HIM in the midplane ($\varphi \sim 95$\%). In contrast, early numerical simulations, including the cycle of matter and energy between the disk and the halo, led to considerably smaller predictions with $\varphi \sim 25$\% \citep{de_avillez_volume_2004}. This value is now considered as a lower limit by the most recent numerical simulations which reveal the importance of the driving mode of supernovae explosions \citep{walch_silcc_2015} and of the photoelectric heating \citep{li_supernova_2015, hill_effect_2018} on the volume of the HIM. These latest studies estimate that $20\% \leqslant \varphi \leqslant 90\%$ in the Galactic midplane.
		
		Highly uncertain, the volume filling factor of the HIM can also vary from one line of sight to another (Fig. 1 of \citealt{hill_effect_2018}). In this paper, we adopt a constant and conservative value $\varphi=0.5$ for every line of sight. Changing $\varphi$ would have the effect of either squeeze or stretch the predicted 2D PHs displayed in Fig. \ref{Fig-compar-sim-obs-L2p2} and \ref{Fig-compar-sim-obs-L5p1} along the x axis, and to modify the balance of probabilities of occurrence of lines of sight at high and low molecular fraction.
		Taking into account a realistic distribution of $\varphi$ would require to simulate the Galactic disk and halo over a scale of several kiloparsecs and to properly model and follow the impact of supernovae explosions. This is far beyond the scope of the present paper.

	\subsection{Galactic vertical structure}
		The simulated 2D PH of the HI-to-H$_2$ transition are found to slightly depend on the Galactic gravitational potential. This odd result is nothing but an artifact of the physical scales considered here. Since the sizes of all simulations are below 200 pc, the gas expands, at the most, over 100 pc above the Galactic plane, a distance far smaller than the characteristic scale of variation of the thermal pressure expected for a turbulent gas in hydrostatic equilibrium. It follows that the column densities show no significant variation as a function of the position of the line of sight or its angle with the Galactic plane.
		
		This setup, initially chosen to favor the physical resolution, is a strong shortcoming which prevents us from using and studying the information carried by the Galactic latitude of	each observation. Indeed, the comparison between the FUSE halo survey and the data collected by FUSE and \textit{Copernicus} in the Galactic disk indicates that the HI-to-H$_2$ transition	at high latitude occurs at a total hydrogen column density $\sim$ 2 times smaller than that in the disk \citep{gillmon_fuse_2006}. Similarly to the previous section, studying this effect would requires to compute the local vertical structure of our Galaxy over several	hundreds of parsecs, taking into account the hot and ionized component of the ISM.

	\subsection{Doppler broadening parameter}
		The self-shielding of H$_2$ depends on the dispersion of the Lyman and Werner lines which is usually modeled with a turbulent Doppler broadening parameter $b_D$ (\citealt{draine_structure_1996} and Eq. \ref{eq:f_shield}). In single cloud models, this parameter has the effect of shifting the HI-to-H$_2$ transition toward larger total column density without modifying any of the asymptotic states (e.g., \citealt{bialy_h_2017}). In this paper, we identify this parameter with the turbulent velocity dispersion of the gas at large scales and therefore adopt $b_D = 8$ km s$^{-1}$ for the fiducial simulation. This is done to prevent an overestimation of the H$_2$ self-shielding at large scales, at the cost of underestimating the self-shielding at the scale of a CNM clouds.
		
		To estimate the effect of this parameter, all the grids presented in this paper were also run assuming $b_D = 2$ km s$^{-1}$, which roughly corresponds to the velocity dispersion expected at a scale of 10 pc for the fiducial simulation. While locally important, $b_D$ is found to have a relatively small impact on the 2D PHs of $N_{\rm H}$ and $N({\rm H}_2)$: increasing $b_D$ by a factor of four slightly increases the width of the HI-to-H$_2$ transition and the fraction of lines of sight in region B. We interpret this limited effect as a consequence of the fact that the asymptotic molecular states of any cloud are independent of $b_D$.
		
		Even so, it should be stressed that $b_D$ has a strong impact on the local molecular fraction in transition regions. Therefore, and while inconsequential for the global statistics of H and H$_2$, the value of the Doppler broadening parameter might be paramount for any chemical species preferentially formed at the border of molecular clouds. As proposed by \citet{Bialy_2019a}, the H$_2$ self-shielding could be computed self-consistently using the velocity and density fields of the simulation and a cost effective radiative transfer method. This would prevent the dilemma of favoring large-scale or small-scale self-shielding.

	\subsection{H$_2$ self-shielding at high temperature}
		The prescription of H$_2$ self-shielding used in this paper (Eq. \ref{eq:f_shield}) is taken from \citet{draine_structure_1996}. As discussed by \citet{Wolcott-Green_2011}, such a prescription is reliable for diffuse gas at low temperature but becomes less and less reliable for high temperature environments ($T>500$ K) where efficient collisional excitation of H$_2$ in its rovibrational levels reduces the self-shielding. To estimate the impact of this process, we ran the fiducial simulation with the alternative self-shielding function proposed by \citet{Wolcott-Green_2011} (Eq. 12 in their paper). This prescription leads to a similar probability histogram and therefore does not influence the global analysis of the kingfisher diagram presented in this paper. However, we note that adopting this alternative prescription slightly increases the width of the HI-to-H$_2$ transition, and induces more lines of sight at intermediate molecular fraction (region B, see Fig. \ref{fig:regions}), in better agreement with the observations.

\section{Summary \& conclusions} \label{sec:conclusion}

This paper presents an exhaustive parametric study of the transition from atomic to molecular hydrogen in the local diffuse ISM. Using state-of-the-art MHD simulations, and an ensemble of 305 runs, we quantify separately the impacts of the mean density, the UV radiation field, the integral scale, the resolution, the turbulent forcing, the magnetic field, and the gravity on the molecular content of multiphase environments. The original feature of this work is to not only focus on the production of individual column densities but also on their statistics, meaning the probabilities of occurrence of these column densities along random lines of sight. For the first time, both the chemical and statistical information are used concomitantly, through the so-called kingfisher diagrams, to interpret the distribution of H and H$_2$ observed toward 360 lines of sight across the local interstellar matter.

The results of the simulations are interpreted with a semi-analytical model which attempts to separate the effects of local conditions from those induced by the probabilistic reconstruction of individual lines of sight. To compare the kingfisher diagrams to the observational sample, we propose a new version of the Kolmogorov-Smirnov test which can be generalized and used for the comparison of two probability histograms or distribution functions in any dimension larger than one.

Taking into account the distance of each background source and simulating random lines of sight over the same distribution of distances is paramount to explain the range of observed column densities and their corresponding statistics. Once this aspect is included, the joint actions of thermal instability and large-scale turbulence in the standard simulation are found to produce a wealth of lines of sight which reproduce the observed position and width of the HI-to-H$_2$ transition, and whose probabilities of occurrence match those derived from the observations. The agreement is so remarkable that it is almost unnecessary to invoke variations of physical conditions along lines of sight or from one line of sight to another.

The minimal KS distance obtained over the entire grid is $\sim$ 0.5. Such a value implies that there exists a small group of lines of sight in the observational sample whose probability of occurrence is under- or over-predicted by about a factor of three. However, it also implies that the probability of occurrence of any other  group of observed lines of sight, small or large, is reproduced to a better level.

The distribution of column densities computed from the simulations strongly depends on the Galactic midplane density parametrized by the mean density $\overline{n_{\rm H}}$, the density of OB stars parametrized by the UV scaling factor $G_0$, and the scale of neutral diffuse clouds parametrized by the box size $L$. It is so because these three parameters not only regulate the mean pressure of the gas, hence the fractions of mass and volume occupied by the CNM and WNM, but also control the typical scale of density fluctuations in the WNM and the distribution of sizes of the CNM structures where H$_2$ forms. The tightest concordance between the observed and simulated samples is obtained for a mean density $\overline{n_{\rm H}} = 1-2$ cm$^{-3}$ and a UV radiation field scaling factor $G_0=1$ (in Habing units), in good agreement with the values deduced from HI and CO all sky surveys and from direct observations of the UV radiation field in the Solar Neighborhood. The range of observed column densities of H and H$_2$ requires a box size $L=200$ pc which corresponds to the estimated scale of HI superclouds. 

Within this setup, the column densities of HI are inferred to be built up in large-scale WNM and CNM structures correlated in densities over $\sim 20$ pc and $\sim 10$ pc, respectively. In contrast H$_2$ is inferred to be built up at smaller scales. However, the fact that the kingfisher diagram is independent from the resolution of the simulation suggests that most of the mass and volume of H$_2$ is contained in CNM structures between $\sim 3$ and $\sim 10$ pc. All these values are given for the standard simulation ($L=200$ pc and $L_{\rm drive} \sim 100$ pc) but naturally depend on the size of the box and the mean density of the gas.

In spite of the strong influences of $\overline{n_{\rm H}}$, $G_0$, and $L$, the 
statistical properties of the HI-to-H$_2$ transition are otherwise remarkably stable. Admittedly, the kingfisher diagram depends on the strength of the turbulence if most of the forcing is injected in solenoidal modes; however such a configuration prevents to reproduce the observational sample unless the large-scale velocity dispersion of the gas is unrealistically small. In contrast, if most of the kinetic energy is injected in compressive modes, the kingfisher diagram is found to weakly depend on the strength of the forcing. Similarly, the HI-to-H$_2$ transition is almost not affected by gravity and is found to weakly depend on the Doppler broadening parameter and the strength of the magnetic field, as long as $B_x \leqslant $ 4 $\mu$G. The 2D PH of the column densities of H and H$_2$ is therefore a valuable tool to constrain the nature of the turbulent forcing at large scale; however, it provides few or no information regarding the velocity dispersion of the gas, the amount of gravitationally bound environments and the strength of the magnetic field. Other observational tracers are required.

All these results open new perspectives for the study of the chemical state of the ISM in which any observation must be understood through the combination of local physical conditions and the probabilistic ordering of these conditions along the line of sight. In particular, similar studies should be applied to all atomic and molecular species with observational samples large enough to conduct statistical analysis. It also invites to expend the study of PHs to higher dimensions, taking into account simultaneously the joint information contained in the column densities of several species. In this context, the generalization of the Kolmogorov-Smirnov test proposed in this paper will be very valuable. All these aspects are currently under development and will be the subjects of the following papers of this series.

	\begin{acknowledgements}
		
		We thank the referee for his careful reading and his valuable comments and suggestions.		The research leading to these results has received fundings from the European Research Council, under the European Community's Seventh framework Programme, through the Advanced Grant MIST (FP7/2017-2022, No 742719). We would also like to acknowledge the support from the Programme National "Physique et Chimie du Milieu Interstellaire" (PCMI) of CNRS/INSU with INC/INP co-funded by CEA and CNES. We would finally like to thank E. Falgarone and M. Gerin for the stimulating discussions we had and their precious comments regarding this work.
		
	\end{acknowledgements}	

\bibliographystyle{aa}
\bibliography{paper_H2}

\appendix
\section{Observations of HI and H$_2$ in the local diffuse ISM}\label{app:sources}

		The complete observational dataset used in this work is given in Table \ref{table:obs}, including the sources identifiers, the coordinates, and the column densities of HI and H$_2$. The column densities of H$_2$ are obtained through direct observations of its FUV absorption	lines. The sample presented in Table \ref{table:obs} therefore results from a combination of the \textit{Copernicus} survey of nearby stars (e.g., \citealt{savage_survey_1977,bohlin_survey_1983}) and the FUSE surveys of the Galactic disk and the Galactic halo (e.g., \citep{rachford_far_2002,lehner_far_2003,cartledge_homogeneity_2004,pan_cloud_2004,gillmon_fuse_2006,jensen_new_2007,jensen_variation_2007,rachford_molecular_2009}. In contrast, and as reviewed by \citet{fruscione_distribution_1994}, the HI column densities are derived from both direct methods, which include Ly$\alpha$ absorption studies, EUV observations of stellar spectra, and observations of the 21cm line, and indirect methods, which include curve of growth of neutral and singly ionized atoms and optical interstellar absorption lines of NaI which are both found to correlate with $N({\rm H})$ \citep{de_boer_intersstellar_1986,ferlet_na_1985}. 
		The column densities of HI given in Table \ref{table:obs} therefore result from a combination of FUV and optical studies for nearby stars (e.g., \citealt{bohlin_survey_1978,bohlin_survey_1983,diplas_iue_1994,fitzpatrick_analysis_1990, jensen_variation_2007,jensen_new_2007} ) and radio studies of the 21cm line for extragalactic sources at high Galactic latitude \citep{wakker_far_2003,gillmon_fuse_2006}.

		As reported by all these authors, the indirect methods relying on the observations of metals can be subjects to uncertainties mostly due to the assumptions made regarding the elemental abundances. In addition, the column densities of HI derived from the emission profiles of the 21cm line can also be highly uncertain because the measurements are done over a beam far larger than the pencil-beam sampled by H$_2$ data but also because it requires to identify in the HI profiles the components associated to the molecular gas. It should thus be kept in mind that while the errors on the column densities of H$_2$ are somehow limited, those on HI can be sometimes larger than a factor of five, in particular for the lines of sight at high Galactic latitude \citep{gillmon_fuse_2006}.
		The color excess $E(B-V)$ given in Table \ref{table:obs} finally results from a compilation which includes direct measurements of the star reddening compared to its intrinsic $(B-V)_0$ color (e.g., \citealt{savage_survey_1977,fitzpatrick_analysis_1990,diplas_iue_1994,rachford_far_2002}) and dust	emission maps from the IRAS telescope\footnote{Available from the NED 
			(http://nedwww.ipac.caltech.edu) and on the more recent plateform 
			https://irsa.ipac.caltech.edu.
		} \citep{schlegel_maps_1998}.
		
		With all these data at hand, we adopt the following methodology to derive the total proton column densities $N_{\rm H}$: if the column densities of HI and H$_2$ are available, then	$N_{\rm H}$ is computed as $N({\rm H}) + 2 N({\rm H}_2)$~; if not, $N_{\rm H}$ is derived from the reddening $E(B-V)$ as $N_{\rm H} = 5.8 \times 10^{21} E(B-V)$ cm$^{-2}$ assuming a standard Galactic extinction curve and the average interstellar ratio $R_V = A_V/E(B-V) = 3.1$ \citep{Fitzpatrick_1986,fitzpatrick_correcting_1999}. It should be noted that the $N_{\rm H} / E(B-V)$ ratio observed at low column density is larger than the standard value used in this work \citep{Liszt_2014,Lenz_2017}. Because of this and the uncertainties on the HI column densities discussed above, the values of $N_{\rm H}$ derived here should be considered as estimates. Examples of the underlying uncertainties on $N_{\rm H}$ can be seen in Table \ref{table:obs} where $N({\rm H})$ sometimes exceeds slightly the total column density derived from $E(B-V)$.
\onecolumn
\begin{center}
	\captionsetup{width=16cm}
	\begin{longtable}{| c | c | c | c | c | c @{\hspace{0.01cm}} c | c @{\hspace{0.01cm}} c | c |}
		\caption{Observational dataset used in this work. The distance of each source is computed from the parallax measured by Gaia if the data is given in the DR2 catalog \citep{2018A&A...616A...1G}; otherwise, the distance of the source is taken from \citet{gudennavar_compilation_2012}. Column densities are expressed in cm$^{-2}$. The total proton column densities $N_{\rm H}$ are computed as $N({\rm H}) + N({\rm H}_2)$ if the column densities of HI and H$_2$ are available, or derived from the reddening $E(B-V)$ as $N_{\rm H} = 5.8 \times 10^{21} E(B-V)$ cm$^{-2}$ assuming a standard Galactic extinction curve and the average interstellar ratio $R_V = A_V/E(B-V) = 3.1$ \citep{Fitzpatrick_1986,fitzpatrick_correcting_1999}.} \label{table:obs}
		\\
		\hline	
		Source ID & Longitude [$^{\circ}$] & Latitude [$^{\circ}$]  & Distance [kpc] & E(B-V)    && log$_{10}$(N(H))  && log$_{10}$(N(H$_2$)) & log$_{10}$(N$_{\rm H}$) \\
		\hline
		\endfirsthead
		\caption{continued.} \\
		\hline	
		Source ID & Longitude [$^{\circ}$] & Latitude [$^{\circ}$]  & Distance [kpc] & E(B-V)   && log$_{10}$(N(H))  && log$_{10}$(N(H$_2$)) & log$_{10}$(N$_{\rm H}$) \\
		\hline
		\endhead
		\hline
		\endfoot
		\endlastfoot
    BD +35 4258 &   77.190 &   -4.740 &    2.000 &    0.290 $^{(9)}$ &  & 21.28 $^{(41)}$ &  & 19.56 $^{(41)}$ & 21.30 \\
    BD +48 3437 &   93.560 &   -2.060 &    2.639 &    0.350 $^{(5)}$ &  & 21.36 $^{(18)}$ &  & 20.42 $^{(43)}$ & 21.45 \\
    BD +53 2820 &  101.240 &   -1.690 &    3.521 &    0.330 $^{(45)}$ &  & 21.35 $^{(32)}$ &  & 20.01 $^{(32)}$ & 21.39 \\
   CPD -59 2603 &  287.590 &   -0.690 &    4.098 &    0.460 $^{(8)}$ &  & 21.46 $^{(18)}$ &  & 20.15 $^{(39)}$ & 21.50 \\
   CPD -69 1743 &  303.710 &   -7.350 &    3.817 &    0.300 $^{(18)}$ &  & 21.12 $^{(18)}$ &  & 19.99 $^{(43)}$ & 21.18 \\
    ESO 141-G55 &  338.180 &  -26.710 &     $-$  &    0.111 $^{(21)}$ &  & 20.70 $^{(36)}$ &  & 19.32 $^{(36)}$ & 20.73 \\
      HD 000886 &  109.430 &  -46.680 &    0.255 &    0.010 $^{(3)}$ &  & 20.04 $^{(3)}$ & <& 14.20 $^{(2)}$ & 19.76 \\
      HD 001383 &  119.020 &   -0.890 &    3.344 &    0.510 $^{(18)}$ &  & 21.36 $^{(18)}$ &  & 20.45 $^{(32)}$ & 21.46 \\
      HD 002905 &  120.840 &    0.140 &    0.521 &    0.350 $^{(3)}$ &  & 21.20 $^{(3)}$ &  & 20.27 $^{(2)}$ & 21.29 \\
      HD 005394 &  123.580 &   -2.150 &    0.190 &    0.210 $^{(45)}$ &  & 19.99 $^{(13)}$ &  & 17.51 $^{(32)}$ & 19.99 \\
      HD 010516 &  131.320 &  -11.330 &    0.151 &    0.200 $^{(3)}$ &  & 20.54 $^{(3)}$ &  & 19.08 $^{(2)}$ & 20.57 \\
      HD 012323 &  132.910 &   -5.870 &    2.809 &    0.410 $^{(32)}$ &  & 21.18 $^{(32)}$ &  & 20.32 $^{(43)}$ & 21.29 \\
      HD 013268 &  133.960 &   -4.990 &    1.692 &    0.411 $^{(21)}$ &  & 21.34 $^{(18)}$ &  & 20.51 $^{(43)}$ & 21.45 \\
      HD 013745 &  134.580 &   -4.960 &    2.268 &    0.460 $^{(18)}$ &  & 21.26 $^{(18)}$ &  & 20.67 $^{(43)}$ & 21.44 \\
      HD 014434 &  135.080 &   -3.820 &    2.558 &    0.480 $^{(18)}$ &  & 21.45 $^{(18)}$ &  & 20.43 $^{(43)}$ & 21.53 \\
      HD 014633 &  140.780 &  -18.200 &    5.051 &    0.070 $^{(45)}$ &  & 20.56 $^{(3)}$ & <& 19.11 $^{(2)}$ & 20.61 \\
      HD 015137 &  137.460 &   -7.580 &    3.704 &    0.310 $^{(18)}$ &  & 21.11 $^{(18)}$ &  & 20.32 $^{(43)}$ & 21.23 \\
      HD 015558 &  134.720 &    0.920 &    2.151 &    0.830 $^{(44)}$ &  & 21.52 $^{(18)}$ &  & 20.89 $^{(44)}$ & 21.69 \\
      HD 017040 &  198.380 &  -62.380 &    0.211 &    0.480 $^{(16)}$ &  &   $-$          &  & 20.81 $^{(16)}$ & 21.44 \\
      HD 021278 &  147.520 &   -6.190 &    0.178 &    0.100 $^{(8)}$ &  & 21.28 $^{(8)}$ &  & 19.48 $^{(2)}$ & 21.29 \\
      HD 021483 &  158.870 &  -21.300 &    0.533 &    0.560 $^{(17)}$ &  &   $-$          &  & 20.81 $^{(2)}$ & 21.51 \\
      HD 021856 &  156.320 &  -16.750 &    0.466 &    0.190 $^{(3)}$ &  & 21.04 $^{(3)}$ &  & 20.04 $^{(2)}$ & 21.12 \\
      HD 022928 &  150.280 &   -5.770 &    0.113 &    0.040 $^{(8)}$ & <& 21.11 $^{(8)}$ &  & 19.30 $^{(2)}$ & 20.37 \\
      HD 022951 &  158.920 &  -16.700 &    0.330 &    0.240 $^{(3)}$ &  & 21.04 $^{(3)}$ &  & 20.46 $^{(2)}$ & 21.22 \\
      HD 023180 &  160.360 &  -17.740 &    0.245 &    0.300 $^{(3)}$ &  & 20.90 $^{(3)}$ &  & 20.61 $^{(2)}$ & 21.21 \\
      HD 023408 &  166.170 &  -23.510 &    0.106 &    0.070 $^{(37)}$ &  &   $-$          &  & 19.75 $^{(2)}$ & 20.61 \\
      HD 023478 &  160.760 &  -17.420 &    0.288 &    0.250 $^{(23)}$ &  & 21.01 $^{(42)}$ &  & 20.48 $^{(42)}$ & 21.21 \\
      HD 023480 &  166.570 &  -23.750 &    0.106 &    0.100 $^{(37)}$ &  &   $-$          &  & 20.12 $^{(2)}$ & 20.76 \\
      HD 023630 &  166.670 &  -23.460 &    0.125 &    0.050 $^{(45)}$ &  & 20.08 $^{(14)}$ &  & 19.54 $^{(2)}$ & 20.28 \\
      HD 024190 &  160.390 &  -15.180 &    0.413 &    0.300 $^{(5)}$ &  & 21.18 $^{(42)}$ &  & 20.38 $^{(42)}$ & 21.30 \\
      HD 024398 &  162.290 &  -16.690 &    0.294 &    0.273 $^{(35)}$ &  & 20.81 $^{(10)}$ &  & 20.67 $^{(2)}$ & 21.20 \\
      HD 024534 &  163.080 &  -17.140 &    0.810 &    0.560 $^{(11)}$ &  & 20.73 $^{(18)}$ &  & 20.92 $^{(27)}$ & 21.34 \\
      HD 024760 &  157.350 &  -10.090 &    0.082 &    0.100 $^{(8)}$ &  & 20.45 $^{(18)}$ &  & 19.52 $^{(10)}$ & 20.54 \\
      HD 024912 &  160.370 &  -13.110 &    0.725 &    0.291 $^{(35)}$ &  & 21.11 $^{(10)}$ &  & 20.53 $^{(2)}$ & 21.29 \\
      HD 026571 &  172.420 &  -20.550 &    0.274 &    0.290 $^{(17)}$ &  & 19.65 $^{(16)}$ &  & 20.81 $^{(16)}$ & 21.13 \\
      HD 027778 &  172.760 &  -17.390 &    0.224 &    0.400 $^{(17)}$ &  & 21.10 $^{(27)}$ &  & 20.79 $^{(27)}$ & 21.40 \\
      HD 028497 &  208.780 &  -37.400 &    0.468 &    0.020 $^{(3)}$ &  & 20.30 $^{(8)}$ &  & 14.82 $^{(2)}$ & 20.30 \\
      HD 029248 &  199.310 &  -31.380 &    0.212 &    0.020 $^{(7)}$ &  & 20.45 $^{(10)}$ & <& 17.41 $^{(7)}$ & 20.06 \\
      HD 029647 &  174.050 &  -13.350 &    0.155 &    1.040 $^{(16)}$ &  & 20.16 $^{(16)}$ &  & 21.54 $^{(16)}$ & 21.85 \\
      HD 030122 &  176.620 &  -14.030 &    0.257 &    0.603 $^{(21)}$ &  &   $-$          &  & 20.70 $^{(43)}$ & 21.54 \\
      HD 030614 &  144.070 &   14.040 &    0.730 &    0.320 $^{(3)}$ &  & 20.90 $^{(3)}$ &  & 20.34 $^{(2)}$ & 21.09 \\
      HD 031237 &  196.270 &  -24.560 &    0.263 &    0.060 $^{(7)}$ &  & 20.41 $^{(7)}$ & <& 17.45 $^{(7)}$ & 20.54 \\
      HD 034078 &  172.080 &   -2.260 &    0.406 &    0.760 $^{(8)}$ &  & 21.30 $^{(8)}$ &  & 20.88 $^{(43)}$ & 21.55 \\
      HD 034816 &  214.830 &  -26.240 &    0.270 &    0.030 $^{(7)}$ &  & 20.30 $^{(8)}$ & <& 15.04 $^{(7)}$ & 20.24 \\
      HD 034989 &  194.620 &  -15.610 &    0.534 &    0.130 $^{(3)}$ &  & 21.11 $^{(3)}$ & <& 18.45 $^{(2)}$ & 20.88 \\
      HD 035149 &  199.160 &  -17.860 &    0.368 &    0.110 $^{(3)}$ &  & 20.74 $^{(3)}$ &  & 18.30 $^{(32)}$ & 20.74 \\
      HD 035439 &  201.960 &  -18.290 &    0.257 &    0.050 $^{(7)}$ &  & 20.46 $^{(10)}$ &  & 14.78 $^{(10)}$ & 20.46 \\
      HD 035715 &  200.090 &  -17.220 &    0.259 &    0.060 $^{(10)}$ &  & 20.57 $^{(8)}$ &  & 14.78 $^{(10)}$ & 20.57 \\
      HD 036166 &  201.670 &  -17.190 &    0.371 &    0.030 $^{(7)}$ &  & 20.32 $^{(7)}$ & <& 15.00 $^{(7)}$ & 20.24 \\
      HD 036486 &  203.900 &  -17.700 &    0.420 &    0.070 $^{(3)}$ &  & 20.18 $^{(8)}$ &  & 14.68 $^{(2)}$ & 20.18 \\
      HD 036822 &  195.400 &  -12.290 &    0.348 &    0.110 $^{(3)}$ &  & 20.81 $^{(3)}$ &  & 19.32 $^{(2)}$ & 20.84 \\
      HD 036861 &  195.050 &  -12.000 &    0.271 &    0.100 $^{(45)}$ &  & 20.87 $^{(8)}$ &  & 19.12 $^{(10)}$ & 20.89 \\
      HD 037022 &  209.010 &  -19.380 &    0.369 &    0.320 $^{(18)}$ &  & 20.66 $^{(19)}$ & <& 17.55 $^{(2)}$ & 21.27 \\
      HD 037043 &  209.520 &  -19.580 &    0.501 &    0.060 $^{(8)}$ &  & 20.30 $^{(8)}$ &  & 14.69 $^{(2)}$ & 20.30 \\
      HD 037128 &  205.210 &  -17.240 &    0.463 &    0.080 $^{(3)}$ &  & 20.45 $^{(3)}$ &  & 16.57 $^{(2)}$ & 20.45 \\
      HD 037202 &  185.690 &   -5.640 &    0.145 &    0.050 $^{(3)}$ &  & 20.04 $^{(3)}$ & <& 17.67 $^{(2)}$ & 20.46 \\
      HD 037367 &  179.040 &   -1.030 &    0.989 &    0.400 $^{(32)}$ &  & 21.28 $^{(32)}$ &  & 20.61 $^{(43)}$ & 21.43 \\
      HD 037468 &  206.820 &  -17.340 &    0.358 &    0.060 $^{(7)}$ &  & 20.52 $^{(10)}$ & <& 18.30 $^{(7)}$ & 20.54 \\
      HD 037742 &  206.450 &  -16.590 &    0.352 &    0.080 $^{(3)}$ &  & 20.41 $^{(3)}$ &  & 15.73 $^{(2)}$ & 20.41 \\
      HD 037903 &  206.850 &  -16.540 &    0.401 &    0.350 $^{(18)}$ &  & 21.16 $^{(32)}$ &  & 20.85 $^{(31)}$ & 21.46 \\
      HD 038087 &  207.070 &  -16.260 &    0.339 &    0.717 $^{(21)}$ &  & 20.91 $^{(41)}$ &  & 20.64 $^{(46)}$ & 21.23 \\
      HD 038666 &  237.290 &  -27.100 &    0.466 &    0.046 $^{(35)}$ &  & 19.75 $^{(3)}$ &  & 15.51 $^{(2)}$ & 19.75 \\
      HD 038771 &  214.510 &  -18.500 &    0.520 &    0.070 $^{(3)}$ &  & 20.52 $^{(3)}$ &  & 15.68 $^{(2)}$ & 20.52 \\
      HD 039680 &  194.070 &   -5.880 &    3.378 &    0.300 $^{(45)}$ &  & 21.30 $^{(18)}$ &  & 19.53 $^{(38)}$ & 21.31 \\
      HD 040111 &  183.970 &    0.840 &    0.663 &    0.150 $^{(3)}$ &  & 21.08 $^{(8)}$ &  & 19.73 $^{(2)}$ & 21.12 \\
      HD 040893 &  180.090 &    4.340 &    4.000 &    0.460 $^{(46)}$ &  & 21.50 $^{(41)}$ &  & 20.58 $^{(41)}$ & 21.59 \\
      HD 041117 &  189.650 &   -0.860 &    1.000 &    0.450 $^{(8)}$ &  & 21.40 $^{(8)}$ &  & 20.69 $^{(41)}$ & 21.54 \\
      HD 042087 &  187.750 &    1.770 &    1.400 &    0.290 $^{(45)}$ &  & 21.40 $^{(18)}$ &  & 20.52 $^{(41)}$ & 21.50 \\
      HD 043384 &  187.990 &    3.530 &    2.494 &    0.580 $^{(46)}$ &  & 21.27 $^{(46)}$ &  & 20.87 $^{(46)}$ & 21.52 \\
      HD 044506 &  241.630 &  -20.780 &    0.616 &    0.020 $^{(45)}$ &  & 20.30 $^{(7)}$ & <& 14.85 $^{(7)}$ & 20.06 \\
      HD 044743 &  226.060 &  -14.270 &    0.153 &    0.030 $^{(35)}$ & <& 18.70 $^{(3)}$ & <& 17.30 $^{(2)}$ & 20.24 \\
      HD 045314 &  196.960 &    1.520 &    0.827 &    0.370 $^{(45)}$ &  & 21.04 $^{(18)}$ &  & 20.60 $^{(44)}$ & 21.28 \\
      HD 046056 &  206.340 &   -2.250 &    1.524 &    0.490 $^{(8)}$ &  & 21.38 $^{(18)}$ &  & 20.68 $^{(41)}$ & 21.53 \\
      HD 046202 &  206.310 &   -2.000 &    1.350 &    0.380 $^{(45)}$ &  & 21.58 $^{(18)}$ &  & 20.68 $^{(41)}$ & 21.68 \\
      HD 047129 &  205.880 &   -0.310 &    1.520 &    0.360 $^{(3)}$ &  & 21.08 $^{(3)}$ &  & 20.54 $^{(2)}$ & 21.28 \\
      HD 047839 &  202.940 &    2.200 &    0.950 &    0.070 $^{(3)}$ &  & 20.31 $^{(3)}$ &  & 15.54 $^{(2)}$ & 20.31 \\
      HD 048099 &  206.210 &    0.800 &    1.916 &    0.270 $^{(3)}$ &  & 21.15 $^{(3)}$ &  & 20.29 $^{(2)}$ & 21.26 \\
      HD 050896 &  234.760 &  -10.080 &    2.427 &    0.140 $^{(3)}$ &  & 20.54 $^{(3)}$ &  & 19.30 $^{(2)}$ & 20.59 \\
      HD 052089 &  239.830 &  -11.330 &    0.188 &    0.010 $^{(3)}$ &  & 17.95 $^{(24)}$ & <& 17.66 $^{(2)}$ & 19.76 \\
      HD 052918 &  218.010 &    0.610 &    0.384 &    0.060 $^{(7)}$ &  & 20.35 $^{(19)}$ &  & 14.78 $^{(7)}$ & 20.35 \\
      HD 053367 &  223.710 &   -1.900 &    0.129 &    0.740 $^{(17)}$ &  & 21.32 $^{(41)}$ &  & 21.04 $^{(41)}$ & 21.63 \\
      HD 053975 &  225.680 &   -2.320 &    1.247 &    0.220 $^{(3)}$ &  & 21.15 $^{(3)}$ &  & 19.23 $^{(2)}$ & 21.16 \\
      HD 054662 &  224.170 &   -0.780 &    1.170 &    0.350 $^{(3)}$ &  & 21.38 $^{(3)}$ &  & 20.00 $^{(2)}$ & 21.41 \\
      HD 055879 &  224.730 &    0.350 &    1.011 &    0.120 $^{(8)}$ &  & 20.85 $^{(8)}$ & <& 18.90 $^{(2)}$ & 20.84 \\
      HD 057060 &  237.820 &   -5.370 &    1.477 &    0.180 $^{(3)}$ &  & 20.81 $^{(8)}$ &  & 15.78 $^{(2)}$ & 20.81 \\
      HD 057061 &  238.180 &   -5.540 &    1.514 &    0.130 $^{(35)}$ &  & 20.74 $^{(8)}$ &  & 15.48 $^{(2)}$ & 20.74 \\
      HD 057682 &  224.420 &    2.630 &    1.241 &    0.120 $^{(3)}$ &  & 20.96 $^{(8)}$ & <& 18.95 $^{(2)}$ & 20.84 \\
      HD 058510 &  235.520 &   -2.470 &    3.333 &    0.240 $^{(45)}$ &  & 21.23 $^{(19)}$ &  & 20.23 $^{(43)}$ & 21.31 \\
      HD 062542 &  255.920 &   -9.240 &    0.390 &    0.370 $^{(27)}$ &  & 20.90 $^{(27)}$ &  & 20.81 $^{(27)}$ & 21.32 \\
      HD 063005 &  242.470 &   -0.930 &   13.699 &    0.300 $^{(45)}$ &  & 21.24 $^{(32)}$ &  & 20.23 $^{(32)}$ & 21.32 \\
      HD 064740 &  263.380 &  -11.190 &    0.214 &    0.010 $^{(45)}$ &  & 20.26 $^{(7)}$ &  & 14.95 $^{(7)}$ & 20.26 \\
      HD 064760 &  262.060 &  -10.420 &    0.363 &    0.050 $^{(45)}$ &  & 20.26 $^{(7)}$ & <& 14.60 $^{(7)}$ & 20.46 \\
      HD 065575 &  266.680 &  -12.320 &    0.139 &    0.020 $^{(7)}$ & <& 20.74 $^{(7)}$ & <& 14.78 $^{(7)}$ & 20.06 \\
      HD 065818 &  263.480 &  -10.280 &    1.117 &    0.080 $^{(7)}$ &  & 20.54 $^{(7)}$ &  & 15.08 $^{(7)}$ & 20.54 \\
      HD 066788 &  245.430 &    2.050 &    5.747 &    0.200 $^{(41)}$ &  & 21.23 $^{(41)}$ &  & 19.72 $^{(41)}$ & 21.26 \\
      HD 066811 &  255.980 &   -4.710 &    0.668 &    0.040 $^{(3)}$ &  & 19.95 $^{(18)}$ &  & 14.45 $^{(2)}$ & 19.95 \\
      HD 068273 &  262.800 &   -7.690 &    0.479 &    0.040 $^{(10)}$ &  & 19.78 $^{(10)}$ &  & 14.23 $^{(2)}$ & 19.78 \\
      HD 069106 &  254.520 &   -1.330 &    1.292 &    0.180 $^{(26)}$ &  & 21.06 $^{(42)}$ &  & 19.73 $^{(41)}$ & 21.10 \\
      HD 072754 &  266.830 &   -5.820 &    1.718 &    0.360 $^{(8)}$ &  & 21.18 $^{(32)}$ &  & 20.35 $^{(32)}$ & 21.29 \\
      HD 073182 &  245.090 &   11.060 &    0.131 &    0.710 $^{(48)}$ &  &   $-$          &  & 20.94 $^{(48)}$ & 21.61 \\
      HD 073882 &  260.180 &    0.640 &    0.347 &    0.720 $^{(17)}$ &  & 21.11 $^{(27)}$ &  & 21.11 $^{(27)}$ & 21.59 \\
      HD 074375 &  275.820 &  -10.860 &    0.330 &    0.100 $^{(3)}$ &  & 20.82 $^{(3)}$ & <& 18.34 $^{(2)}$ & 20.76 \\
      HD 074575 &  254.990 &    5.770 &    0.235 &    0.070 $^{(7)}$ &  & 20.60 $^{(8)}$ & <& 15.04 $^{(7)}$ & 20.61 \\
      HD 074711 &  265.740 &   -2.610 &    1.326 &    0.250 $^{(45)}$ &  &   $-$          &  & 20.30 $^{(38)}$ & 21.16 \\
      HD 074920 &  265.290 &   -1.950 &    2.874 &    0.280 $^{(45)}$ &  & 21.15 $^{(18)}$ &  & 20.26 $^{(38)}$ & 21.25 \\
      HD 075309 &  265.860 &   -1.900 &    2.041 &    0.270 $^{(32)}$ &  & 21.08 $^{(32)}$ &  & 20.20 $^{(32)}$ & 21.18 \\
      HD 079186 &  267.360 &    2.250 &    1.299 &    0.300 $^{(32)}$ &  & 21.20 $^{(32)}$ &  & 20.72 $^{(32)}$ & 21.42 \\
      HD 079351 &  277.690 &   -7.370 &    0.151 &    0.040 $^{(7)}$ &  & 20.78 $^{(7)}$ & <& 17.90 $^{(7)}$ & 20.37 \\
      HD 080077 &  271.630 &   -0.670 &    2.551 &    1.520 $^{(16)}$ &  &   $-$          &  & 21.40 $^{(16)}$ & 21.95 \\
      HD 081188 &  275.880 &   -3.540 &    0.102 &    0.050 $^{(10)}$ &  & 20.48 $^{(7)}$ & <& 17.70 $^{(7)}$ & 20.46 \\
      HD 087901 &  226.430 &   48.930 &    0.020 &    0.100 $^{(45)}$ & <& 18.08 $^{(13)}$ & <& 14.98 $^{(2)}$ & 20.76 \\
      HD 088115 &  285.320 &   -5.530 &    4.808 &    0.120 $^{(45)}$ &  & 21.00 $^{(18)}$ &  & 19.30 $^{(29)}$ & 21.02 \\
      HD 090087 &  285.160 &   -2.130 &    3.205 &    0.280 $^{(26)}$ &  & 21.15 $^{(18)}$ &  & 19.92 $^{(41)}$ & 21.20 \\
      HD 091316 &  234.890 &   52.770 &    0.505 &    0.080 $^{(3)}$ &  & 20.26 $^{(3)}$ &  & 15.61 $^{(2)}$ & 20.26 \\
      HD 091597 &  286.860 &   -2.370 &    8.696 &    0.300 $^{(45)}$ &  & 21.34 $^{(8)}$ &  & 19.70 $^{(41)}$ & 21.36 \\
      HD 091651 &  286.550 &   -1.720 &    1.934 &    0.300 $^{(18)}$ &  & 21.15 $^{(18)}$ &  & 19.07 $^{(41)}$ & 21.16 \\
      HD 091824 &  285.700 &    0.070 &    2.331 &    0.270 $^{(32)}$ &  & 21.12 $^{(32)}$ &  & 19.85 $^{(32)}$ & 21.16 \\
      HD 091983 &  285.880 &    0.050 &    4.255 &    0.260 $^{(32)}$ &  & 21.17 $^{(32)}$ &  & 20.14 $^{(32)}$ & 21.24 \\
      HD 092554 &  287.600 &   -2.020 &    4.587 &    0.340 $^{(45)}$ &  & 21.28 $^{(18)}$ &  & 18.93 $^{(41)}$ & 21.28 \\
      HD 092740 &  287.170 &   -0.850 &    2.532 &    0.330 $^{(3)}$ &  & 21.20 $^{(3)}$ &  & 19.97 $^{(2)}$ & 21.25 \\
      HD 092809 &  286.780 &   -0.030 &    2.801 &    0.220 $^{(21)}$ &  &   $-$          &  & 20.23 $^{(38)}$ & 21.11 \\
      HD 093030 &  289.600 &   -4.900 &    0.207 &    0.060 $^{(3)}$ &  & 20.28 $^{(3)}$ & <& 17.65 $^{(2)}$ & 20.54 \\
      HD 093162 &  287.510 &   -0.710 &    2.101 &    0.620 $^{(18)}$ &  & 21.55 $^{(18)}$ &  & 19.83 $^{(38)}$ & 21.57 \\
      HD 093204 &  287.570 &   -0.710 &    2.227 &    0.420 $^{(8)}$ &  & 21.40 $^{(8)}$ &  & 19.77 $^{(44)}$ & 21.42 \\
      HD 093205 &  287.570 &   -0.710 &    2.688 &    0.370 $^{(18)}$ &  & 21.34 $^{(8)}$ &  & 19.83 $^{(43)}$ & 21.37 \\
      HD 093206 &  287.670 &   -0.940 &    1.101 &    0.330 $^{(45)}$ &  & 21.34 $^{(18)}$ &  & 19.52 $^{(44)}$ & 21.35 \\
      HD 093222 &  287.740 &   -1.020 &    2.941 &    0.370 $^{(26)}$ &  & 21.40 $^{(29)}$ &  & 19.84 $^{(43)}$ & 21.42 \\
      HD 093237 &  297.180 &  -18.390 &    0.318 &    0.090 $^{(28)}$ &  &   $-$          &  & 19.80 $^{(43)}$ & 20.72 \\
      HD 093521 &  183.140 &   62.150 &    1.949 &    0.050 $^{(45)}$ &  & 20.15 $^{(8)}$ & <& 18.54 $^{(2)}$ & 20.46 \\
      HD 093840 &  282.140 &   11.100 &    3.521 &    0.160 $^{(5)}$ &  & 21.04 $^{(18)}$ &  & 19.28 $^{(43)}$ & 21.05 \\
      HD 093843 &  288.240 &   -0.900 &    2.625 &    0.340 $^{(8)}$ &  & 21.33 $^{(18)}$ &  & 19.61 $^{(41)}$ & 21.35 \\
      HD 094454 &  295.690 &  -14.730 &    0.267 &    0.180 $^{(28)}$ &  &   $-$          &  & 20.70 $^{(42)}$ & 21.02 \\
      HD 094473 &  272.830 &   29.170 &    0.387 &    0.140 $^{(25)}$ &  & 20.90 $^{(25)}$ &  & 19.06 $^{(25)}$ & 20.91 \\
      HD 094493 &  289.010 &   -1.180 &    1.852 &    0.200 $^{(18)}$ &  & 21.11 $^{(18)}$ &  & 20.12 $^{(38)}$ & 21.19 \\
      HD 096675 &  296.620 &  -14.570 &    0.163 &    0.310 $^{(27)}$ &  & 20.66 $^{(27)}$ &  & 20.82 $^{(27)}$ & 21.25 \\
      HD 099171 &  286.330 &   17.380 &    0.555 &    0.050 $^{(3)}$ &  & 20.65 $^{(3)}$ &  & 15.25 $^{(2)}$ & 20.65 \\
      HD 099857 &  294.780 &   -4.940 &    2.326 &    0.330 $^{(18)}$ &  & 21.31 $^{(18)}$ &  & 20.25 $^{(41)}$ & 21.38 \\
      HD 099872 &  296.690 &  -10.620 &    0.230 &    0.360 $^{(43)}$ &  &   $-$          &  & 20.55 $^{(42)}$ & 21.32 \\
      HD 099890 &  291.750 &    4.430 &    1.957 &    0.150 $^{(45)}$ &  & 20.85 $^{(19)}$ &  & 19.47 $^{(41)}$ & 20.88 \\
      HD 101131 &  294.780 &   -1.620 &    2.632 &    0.280 $^{(45)}$ &  &   $-$          &  & 20.27 $^{(44)}$ & 21.21 \\
      HD 101190 &  294.780 &   -1.490 &    3.367 &    0.300 $^{(45)}$ &  & 21.04 $^{(8)}$ &  & 20.42 $^{(44)}$ & 21.21 \\
      HD 101413 &  295.030 &   -1.710 &    1.887 &    0.320 $^{(45)}$ &  & 21.23 $^{(18)}$ &  & 20.38 $^{(44)}$ & 21.34 \\
      HD 101436 &  295.040 &   -1.710 &    3.067 &    0.310 $^{(45)}$ &  & 21.23 $^{(18)}$ &  & 20.38 $^{(44)}$ & 21.34 \\
      HD 102065 &  300.030 &  -18.000 &    0.194 &    0.170 $^{(46)}$ &  & 20.54 $^{(27)}$ &  & 20.50 $^{(27)}$ & 20.99 \\
      HD 103779 &  296.850 &   -1.020 &    2.381 &    0.210 $^{(18)}$ &  & 21.16 $^{(18)}$ &  & 19.82 $^{(41)}$ & 21.20 \\
      HD 104705 &  297.450 &   -0.340 &    2.315 &    0.220 $^{(18)}$ &  & 21.11 $^{(18)}$ &  & 19.98 $^{(39)}$ & 21.17 \\
      HD 106490 &  298.230 &    3.790 &    0.086 &    0.020 $^{(45)}$ &  & 20.04 $^{(7)}$ & <& 14.08 $^{(7)}$ & 20.06 \\
      HD 106943 &  298.960 &    1.140 &    0.353 &    0.145 $^{(5)}$ &  &   $-$      &  & 19.81 $^{(43)}$ & 20.92 \\
      HD 108002 &  300.160 &   -2.480 &    2.770 &    0.316 $^{(5)}$ &  &   $-$          &  & 20.34 $^{(43)}$ & 21.26 \\
      HD 108248 &  300.130 &   -0.360 &    0.114 &    0.030 $^{(7)}$ &  & 19.60 $^{(19)}$ & <& 14.18 $^{(7)}$ & 20.24 \\
      HD 108610 &  300.280 &    0.880 &    0.503 &    0.155 $^{(5)}$ &  &   $-$          &  & 19.86 $^{(43)}$ & 20.95 \\
      HD 108639 &  300.220 &    1.950 &    1.825 &    0.250 $^{(45)}$ &  & 21.35 $^{(42)}$ &  & 19.95 $^{(42)}$ & 21.38 \\
      HD 108927 &  301.920 &  -15.360 &    0.341 &    0.220 $^{(46)}$ &  & 20.86 $^{(27)}$ &  & 20.49 $^{(27)}$ & 21.13 \\
      HD 109399 &  301.710 &   -9.880 &    2.755 &    0.260 $^{(18)}$ &  & 21.04 $^{(19)}$ &  & 20.04 $^{(41)}$ & 21.12 \\
      HD 110432 &  301.960 &   -0.200 &    0.420 &    0.520 $^{(12)}$ &  & 20.85 $^{(27)}$ &  & 20.64 $^{(27)}$ & 21.20 \\
      HD 110434 &  302.070 &   -3.600 &    0.423 &    0.050 $^{(28)}$ &  &   $-$          &  & 19.90 $^{(43)}$ & 20.46 \\
      HD 112244 &  303.550 &    6.030 &    1.167 &    0.340 $^{(3)}$ &  & 21.11 $^{(8)}$ &  & 20.14 $^{(2)}$ & 21.19 \\
      HD 112999 &  304.170 &    2.180 &    0.747 &    0.161 $^{(5)}$ &  &   $-$          &  & 19.99 $^{(42)}$ & 20.97 \\
      HD 113904 &  304.670 &   -2.490 &    2.786 &    0.290 $^{(3)}$ &  & 21.08 $^{(3)}$ &  & 19.83 $^{(2)}$ & 21.13 \\
      HD 114886 &  305.520 &   -0.830 &    1.045 &    0.400 $^{(9)}$ &  & 21.34 $^{(42)}$ &  & 20.34 $^{(43)}$ & 21.42 \\
      HD 115071 &  305.760 &    0.150 &    2.101 &    0.490 $^{(18)}$ &  & 21.36 $^{(42)}$ &  & 20.63 $^{(42)}$ & 21.50 \\
      HD 115455 &  306.060 &    0.220 &    2.268 &    0.400 $^{(45)}$ &  & 21.41 $^{(18)}$ &  & 20.58 $^{(43)}$ & 21.52 \\
      HD 116538 &  308.230 &   10.680 &    1.675 &    0.130 $^{(45)}$ &  & 21.04 $^{(18)}$ &  & 19.63 $^{(38)}$ & 21.07 \\
      HD 116658 &  316.000 &   51.000 &    0.084 &    0.030 $^{(3)}$ &  & 18.83 $^{(13)}$ &  & 12.95 $^{(2)}$ & 18.83 \\
      HD 116781 &  307.050 &   -0.070 &    2.045 &    0.340 $^{(41)}$ &  & 21.18 $^{(41)}$ &  & 20.08 $^{(41)}$ & 21.24 \\
      HD 116852 &  304.880 &  -16.130 &   22.727 &    0.210 $^{(31)}$ &  & 20.96 $^{(32)}$ &  & 19.83 $^{(43)}$ & 21.02 \\
      HD 118716 &  310.190 &    8.720 &    0.168 &    0.040 $^{(10)}$ &  & 19.90 $^{(10)}$ & <& 14.08 $^{(7)}$ & 20.37 \\
      HD 120315 &  100.700 &   65.320 &    0.030 &    0.080 $^{(45)}$ & <& 20.90 $^{(8)}$ &  & 13.38 $^{(2)}$ & 20.67 \\
      HD 120324 &  314.240 &   19.120 &    0.119 &    0.100 $^{(7)}$ &  & 20.40 $^{(7)}$ & <& 14.78 $^{(7)}$ & 20.76 \\
      HD 121263 &  314.070 &   14.190 &    0.120 &    0.020 $^{(45)}$ &  & 19.28 $^{(19)}$ &  & 12.80 $^{(2)}$ & 19.28 \\
      HD 121968 &  333.970 &   55.840 &    3.425 &    0.090 $^{(26)}$ &  & 20.71 $^{(18)}$ &  & 18.70 $^{(39)}$ & 20.72 \\
      HD 122451 &  311.770 &    1.250 &    0.160 &    0.060 $^{(45)}$ &  & 19.52 $^{(3)}$ &  & 12.80 $^{(2)}$ & 19.52 \\
      HD 122879 &  312.260 &    1.790 &    2.387 &    0.298 $^{(35)}$ &  & 21.26 $^{(32)}$ &  & 20.31 $^{(42)}$ & 21.35 \\
      HD 124314 &  312.670 &   -0.420 &    1.808 &    0.530 $^{(18)}$ &  & 21.39 $^{(42)}$ &  & 20.52 $^{(43)}$ & 21.49 \\
      HD 127972 &  322.770 &   16.670 &    0.095 &    0.050 $^{(7)}$ &  & 20.11 $^{(7)}$ & <& 14.18 $^{(7)}$ & 20.46 \\
      HD 135591 &  320.130 &   -2.640 &    0.835 &    0.220 $^{(3)}$ &  & 21.08 $^{(3)}$ &  & 19.77 $^{(2)}$ & 21.12 \\
      HD 135742 &  352.020 &   39.230 &    0.093 &          $-$      &  & 19.38 $^{(19)}$ &  & 14.34 $^{(6)}$ & 19.38 \\
      HD 136298 &  331.320 &   13.820 &    0.115 &    0.020 $^{(10)}$ &  & 20.18 $^{(7)}$ & <& 14.26 $^{(7)}$ & 20.06 \\
      HD 137595 &  336.720 &   18.860 &    0.822 &    0.250 $^{(5)}$ &  & 21.00 $^{(42)}$ &  & 20.56 $^{(42)}$ & 21.24 \\
      HD 138690 &  333.190 &   11.890 &    0.137 &    0.030 $^{(7)}$ &  & 20.23 $^{(7)}$ & <& 14.26 $^{(7)}$ & 20.24 \\
      HD 140037 &  340.150 &   18.040 &    0.402 &    0.090 $^{(28)}$ &  &   $-$          &  & 19.34 $^{(43)}$ & 20.72 \\
      HD 141637 &  346.100 &   21.710 &    0.145 &    0.200 $^{(2)}$ &  & 21.18 $^{(18)}$ &  & 19.23 $^{(2)}$ & 21.19 \\
      HD 143018 &  347.210 &   20.230 &    0.580 &    0.070 $^{(8)}$ &  & 20.74 $^{(8)}$ &  & 19.32 $^{(2)}$ & 20.77 \\
      HD 143275 &  350.100 &   22.490 &    0.155 &    0.190 $^{(8)}$ &  & 21.15 $^{(10)}$ &  & 19.41 $^{(2)}$ & 21.17 \\
      HD 144217 &  353.190 &   23.600 &    0.161 &    0.210 $^{(35)}$ &  & 21.09 $^{(10)}$ &  & 19.83 $^{(2)}$ & 21.13 \\
      HD 144470 &  352.750 &   22.770 &    0.142 &    0.220 $^{(10)}$ &  & 21.18 $^{(10)}$ &  & 20.04 $^{(2)}$ & 21.24 \\
      HD 144965 &  339.040 &    8.420 &    0.266 &    0.350 $^{(28)}$ &  & 21.07 $^{(42)}$ &  & 20.77 $^{(42)}$ & 21.37 \\
      HD 145502 &  354.610 &   22.700 &    0.135 &    0.270 $^{(3)}$ &  & 21.15 $^{(3)}$ &  & 19.89 $^{(2)}$ & 21.20 \\
      HD 147165 &  351.310 &   17.000 &    0.100 &    0.400 $^{(8)}$ &  & 21.38 $^{(18)}$ &  & 19.79 $^{(2)}$ & 21.40 \\
      HD 147343 &  352.450 &   17.630 &    0.181 &    0.640 $^{(16)}$ &  & 21.43 $^{(16)}$ &  & 20.78 $^{(16)}$ & 21.59 \\
      HD 147683 &  344.860 &   10.090 &    0.295 &    0.390 $^{(5)}$ &  & 21.41 $^{(42)}$ &  & 20.68 $^{(42)}$ & 21.55 \\
      HD 147701 &  352.250 &   16.850 &    0.140 &    0.740 $^{(16)}$ &  & 21.50 $^{(16)}$ &  & 20.90 $^{(16)}$ & 21.68 \\
      HD 147888 &  353.650 &   17.710 &    0.092 &    0.520 $^{(32)}$ &  & 21.71 $^{(32)}$ &  & 20.57 $^{(32)}$ & 21.77 \\
      HD 147889 &  352.860 &   17.040 &    0.139 &    1.090 $^{(16)}$ &  & 21.46 $^{(16)}$ &  & 21.37 $^{(16)}$ & 21.88 \\
      HD 147933 &  353.690 &   17.690 &    0.174 &    0.470 $^{(3)}$ &  & 21.81 $^{(10)}$ &  & 20.57 $^{(2)}$ & 21.86 \\
      HD 148184 &  357.930 &   20.680 &    0.122 &    0.530 $^{(3)}$ &  & 21.15 $^{(3)}$ &  & 20.63 $^{(2)}$ & 21.36 \\
      HD 148379 &  337.250 &    1.580 &    3.012 &    0.720 $^{(16)}$ &  &   $-$          &  & 20.41 $^{(16)}$ & 21.62 \\
      HD 148422 &  329.920 &   -5.600 &    6.944 &    0.230 $^{(45)}$ &  & 21.15 $^{(26)}$ &  & 20.13 $^{(44)}$ & 21.23 \\
      HD 148594 &  350.930 &   13.940 &    0.193 &    0.210 $^{(32)}$ &  & 21.80 $^{(32)}$ &  & 19.88 $^{(32)}$ & 21.81 \\
      HD 148605 &  353.100 &   15.800 &    0.117 &    0.100 $^{(10)}$ &  & 20.95 $^{(10)}$ &  & 18.74 $^{(2)}$ & 20.96 \\
      HD 148937 &  336.370 &   -0.220 &    1.135 &    0.660 $^{(18)}$ &  & 21.60 $^{(18)}$ &  & 20.71 $^{(40)}$ & 21.70 \\
      HD 149038 &  339.380 &    2.510 &    0.842 &    0.370 $^{(3)}$ &  & 21.12 $^{(18)}$ &  & 20.44 $^{(2)}$ & 21.27 \\
      HD 149404 &  340.540 &    3.010 &    1.316 &    0.680 $^{(8)}$ &  & 21.40 $^{(8)}$ &  & 20.79 $^{(46)}$ & 21.57 \\
      HD 149438 &  351.530 &   12.810 &    0.195 &    0.060 $^{(3)}$ &  & 20.43 $^{(8)}$ &  & 15.50 $^{(2)}$ & 20.43 \\
      HD 149757 &    6.280 &   23.590 &    0.172 &    0.320 $^{(3)}$ &  & 20.78 $^{(8)}$ &  & 20.65 $^{(2)}$ & 21.17 \\
      HD 149881 &   31.370 &   36.230 &    2.439 &    0.050 $^{(45)}$ &  & 20.65 $^{(3)}$ & <& 19.00 $^{(2)}$ & 20.46 \\
      HD 150898 &  329.980 &   -8.470 &    0.882 &    0.110 $^{(45)}$ &  & 20.95 $^{(3)}$ &  & 19.81 $^{(2)}$ & 21.01 \\
      HD 151804 &  343.620 &    1.940 &    1.629 &    0.400 $^{(3)}$ &  & 21.08 $^{(3)}$ &  & 20.26 $^{(2)}$ & 21.19 \\
      HD 151805 &  343.200 &    1.590 &    1.672 &    0.190 $^{(45)}$ &  & 21.32 $^{(42)}$ &  & 20.36 $^{(42)}$ & 21.41 \\
      HD 151890 &  346.120 &    3.910 &    0.268 &    0.050 $^{(7)}$ &  & 20.40 $^{(7)}$ & <& 14.26 $^{(7)}$ & 20.46 \\
      HD 152233 &  343.480 &    1.220 &    2.300 &    0.400 $^{(45)}$ &  & 21.35 $^{(8)}$ &  & 20.29 $^{(44)}$ & 21.42 \\
      HD 152236 &  343.030 &    0.870 &    1.403 &    0.680 $^{(8)}$ &  & 21.77 $^{(18)}$ &  & 20.73 $^{(46)}$ & 21.84 \\
      HD 152248 &  343.460 &    1.180 &    1.698 &    0.420 $^{(45)}$ &  &   $-$          &  & 20.29 $^{(44)}$ & 21.39 \\
      HD 152408 &  344.080 &    1.490 &    2.242 &    0.480 $^{(3)}$ &  & 21.26 $^{(3)}$ &  & 20.38 $^{(2)}$ & 21.36 \\
      HD 152590 &  344.840 &    1.830 &    1.637 &    0.380 $^{(32)}$ &  & 21.37 $^{(32)}$ &  & 20.47 $^{(32)}$ & 21.47 \\
      HD 152623 &  344.620 &    1.610 &    1.500 &    0.330 $^{(45)}$ &  & 21.28 $^{(18)}$ &  & 20.21 $^{(38)}$ & 21.35 \\
      HD 152723 &  344.810 &    1.610 &   16.667 &    0.460 $^{(18)}$ &  & 21.43 $^{(18)}$ &  & 20.33 $^{(39)}$ & 21.49 \\
      HD 154368 &  349.970 &    3.220 &    1.217 &    0.820 $^{(16)}$ &  & 21.00 $^{(27)}$ &  & 21.16 $^{(27)}$ & 21.59 \\
      HD 155806 &  352.590 &    2.870 &    0.994 &    0.230 $^{(45)}$ &  & 21.08 $^{(10)}$ &  & 19.92 $^{(2)}$ & 21.14 \\
      HD 157246 &  334.640 &  -11.480 &    0.267 &    0.050 $^{(8)}$ &  & 20.74 $^{(18)}$ &  & 19.24 $^{(2)}$ & 20.77 \\
      HD 157857 &   12.970 &   13.310 &    3.968 &    0.370 $^{(45)}$ &  & 21.30 $^{(18)}$ &  & 20.69 $^{(43)}$ & 21.47 \\
      HD 158408 &  351.270 &   -1.840 &    0.134 &    0.020 $^{(3)}$ & <& 19.26 $^{(3)}$ & <& 14.11 $^{(2)}$ & 20.06 \\
      HD 158926 &  351.740 &   -2.210 &    0.220 &    0.080 $^{(45)}$ &  & 19.23 $^{(20)}$ &  & 12.70 $^{(2)}$ & 19.23 \\
      HD 160578 &  301.040 &   -4.720 &    0.202 &    0.083 $^{(35)}$ &  & 20.19 $^{(18)}$ & <& 14.23 $^{(7)}$ & 20.68 \\
      HD 161807 &  351.780 &   -5.850 &    1.319 &    0.140 $^{(45)}$ &  &   $-$          &  & 19.86 $^{(44)}$ & 20.91 \\
      HD 163758 &  355.360 &   -6.100 &    3.876 &    0.350 $^{(8)}$ &  & 21.23 $^{(18)}$ &  & 19.85 $^{(39)}$ & 21.26 \\
      HD 164284 &   30.990 &   13.370 &    0.143 &    0.190 $^{(7)}$ &  & 20.82 $^{(7)}$ &  & 19.85 $^{(7)}$ & 20.90 \\
      HD 164353 &   29.730 &   12.630 &    0.566 &    0.110 $^{(3)}$ &  & 21.00 $^{(3)}$ &  & 20.26 $^{(2)}$ & 21.13 \\
      HD 164402 &    7.160 &   -0.030 &    1.672 &    0.280 $^{(3)}$ &  & 21.11 $^{(3)}$ &  & 19.49 $^{(2)}$ & 21.13 \\
      HD 164740 &    5.970 &   -1.170 &    1.109 &    0.870 $^{(46)}$ &  & 21.95 $^{(46)}$ &  & 20.19 $^{(41)}$ & 21.96 \\
      HD 164816 &    6.060 &   -1.200 &    1.185 &    0.310 $^{(26)}$ &  & 21.18 $^{(18)}$ &  & 20.00 $^{(38)}$ & 21.23 \\
      HD 164906 &    6.050 &   -1.330 &    1.235 &    0.380 $^{(45)}$ &  & 21.20 $^{(26)}$ &  & 20.22 $^{(38)}$ & 21.28 \\
      HD 165024 &  343.330 &  -13.820 &    0.279 &    0.060 $^{(45)}$ &  & 20.85 $^{(3)}$ &  & 18.95 $^{(2)}$ & 20.86 \\
      HD 165052 &    6.120 &   -1.480 &    1.276 &    0.360 $^{(45)}$ &  & 21.36 $^{(18)}$ &  & 20.20 $^{(38)}$ & 21.42 \\
      HD 165246 &    6.400 &   -1.560 &    1.996 &     $-$           &  & 21.41 $^{(42)}$ &  & 20.15 $^{(42)}$ & 21.46 \\
      HD 165955 &  357.410 &   -7.430 &    1.205 &    0.120 $^{(45)}$ &  & 21.11 $^{(32)}$ &  & 16.53 $^{(32)}$ & 21.11 \\
      HD 167263 &   10.760 &   -1.580 &    2.079 &    0.310 $^{(3)}$ &  & 21.08 $^{(3)}$ &  & 20.18 $^{(2)}$ & 21.18 \\
      HD 167264 &   10.460 &   -1.740 &    0.861 &    0.330 $^{(18)}$ &  & 21.15 $^{(10)}$ &  & 20.28 $^{(2)}$ & 21.25 \\
      HD 167971 &   18.250 &    1.680 &    2.033 &    1.040 $^{(27)}$ &  & 21.60 $^{(27)}$ &  & 20.85 $^{(27)}$ & 21.73 \\
      HD 168076 &   16.940 &    0.840 &    2.100 &    0.790 $^{(27)}$ &  & 21.65 $^{(18)}$ &  & 20.68 $^{(27)}$ & 21.73 \\
      HD 168941 &    5.820 &   -6.310 &    2.488 &    0.240 $^{(45)}$ &  & 21.11 $^{(18)}$ &  & 20.10 $^{(41)}$ & 21.19 \\
      HD 169454 &   17.540 &   -0.670 &    2.128 &    1.120 $^{(16)}$ & >& 19.95 $^{(16)}$ &  & 21.16 $^{(16)}$ & 21.81 \\
      HD 170740 &   21.060 &   -0.530 &    0.231 &    0.480 $^{(46)}$ &  & 21.15 $^{(18)}$ &  & 20.86 $^{(27)}$ & 21.46 \\
      HD 175191 &    9.560 &  -12.440 &    0.070 &    0.050 $^{(45)}$ & <& 19.48 $^{(3)}$ & <& 14.00 $^{(2)}$ & 20.46 \\
      HD 177989 &   17.810 &  -11.880 &    2.538 &    0.250 $^{(18)}$ &  & 20.95 $^{(18)}$ &  & 20.23 $^{(39)}$ & 21.09 \\
      HD 179406 &   28.230 &   -8.310 &    0.283 &    0.499 $^{(21)}$ &  & 21.23 $^{(46)}$ &  & 20.73 $^{(41)}$ & 21.44 \\
      HD 184915 &   31.770 &  -13.290 &    0.466 &    0.270 $^{(18)}$ &  & 20.85 $^{(18)}$ &  & 20.31 $^{(2)}$ & 21.05 \\
      HD 185418 &   53.600 &   -2.170 &    0.755 &    0.380 $^{(45)}$ &  & 21.11 $^{(27)}$ &  & 20.76 $^{(27)}$ & 21.39 \\
      HD 186994 &   78.620 &   10.060 &    1.965 &    0.130 $^{(45)}$ &  & 20.90 $^{(3)}$ &  & 19.59 $^{(41)}$ & 20.94 \\
      HD 188209 &   80.990 &   10.060 &    1.497 &    0.210 $^{(3)}$ &  & 20.90 $^{(3)}$ &  & 20.01 $^{(2)}$ & 21.00 \\
      HD 188439 &   81.770 &   10.320 &    1.147 &    0.140 $^{(3)}$ &  & 20.85 $^{(8)}$ &  & 19.95 $^{(2)}$ & 20.95 \\
      HD 190918 &   72.650 &    2.070 &    1.953 &    0.400 $^{(26)}$ &  & 21.40 $^{(18)}$ &  & 19.95 $^{(43)}$ & 21.43 \\
      HD 191765 &   73.450 &    1.550 &    1.845 &    0.450 $^{(38)}$ &  & 21.56 $^{(18)}$ &  & 20.27 $^{(38)}$ & 21.60 \\
      HD 191877 &   61.570 &   -6.450 &    1.401 &    0.140 $^{(45)}$ &  & 20.90 $^{(18)}$ &  & 20.02 $^{(38)}$ & 21.00 \\
      HD 192035 &   83.330 &    7.760 &    2.252 &    0.350 $^{(18)}$ &  & 21.20 $^{(19)}$ &  & 20.68 $^{(43)}$ & 21.41 \\
      HD 192639 &   74.900 &    1.480 &    2.597 &    0.560 $^{(45)}$ &  & 21.32 $^{(18)}$ &  & 20.69 $^{(27)}$ & 21.49 \\
      HD 193322 &   78.100 &    2.780 &    0.989 &    0.400 $^{(3)}$ &  & 21.15 $^{(8)}$ &  & 20.08 $^{(2)}$ & 21.22 \\
      HD 193924 &  340.900 &  -35.190 &    0.056 &    0.020 $^{(3)}$ & <& 19.30 $^{(3)}$ & <& 14.30 $^{(2)}$ & 20.06 \\
      HD 195965 &   85.710 &    5.000 &    0.861 &    0.190 $^{(45)}$ &  & 20.90 $^{(18)}$ &  & 20.36 $^{(38)}$ & 21.10 \\
      HD 197512 &   87.890 &    4.630 &    1.664 &    0.330 $^{(27)}$ &  & 21.26 $^{(27)}$ &  & 20.66 $^{(27)}$ & 21.44 \\
      HD 198478 &   85.750 &    1.490 &    1.176 &    0.439 $^{(35)}$ &  & 21.32 $^{(32)}$ &  & 20.87 $^{(32)}$ & 21.55 \\
      HD 198781 &   99.940 &   12.610 &    0.935 &    0.350 $^{(32)}$ &  & 20.91 $^{(32)}$ &  & 20.48 $^{(32)}$ & 21.15 \\
      HD 199579 &   85.700 &   -0.300 &    0.941 &    0.310 $^{(45)}$ &  & 21.04 $^{(8)}$ &  & 20.53 $^{(27)}$ & 21.25 \\
      HD 200120 &   88.030 &    0.970 &    0.399 &    0.180 $^{(3)}$ &  & 20.26 $^{(3)}$ &  & 19.30 $^{(10)}$ & 20.34 \\
      HD 200775 &  104.060 &   14.190 &    0.361 &    0.570 $^{(23)}$ &  &   $-$          &  & 21.15 $^{(43)}$ & 21.52 \\
      HD 201345 &   78.440 &   -9.540 &    3.195 &    0.191 $^{(21)}$ &  & 20.87 $^{(18)}$ &  & 19.43 $^{(39)}$ & 20.90 \\
      HD 202347 &   88.220 &   -2.080 &    0.931 &    0.170 $^{(9)}$ &  & 20.99 $^{(41)}$ &  & 20.00 $^{(38)}$ & 21.07 \\
      HD 202904 &   80.980 &  -10.050 &    0.187 &    0.130 $^{(35)}$ &  & 20.68 $^{(7)}$ &  & 19.15 $^{(7)}$ & 20.70 \\
      HD 203064 &   87.610 &   -3.840 &    0.587 &    0.320 $^{(8)}$ &  & 21.00 $^{(3)}$ &  & 20.29 $^{(2)}$ & 21.14 \\
      HD 203374 &  100.510 &    8.620 &    2.611 &    0.600 $^{(18)}$ &  & 21.20 $^{(42)}$ &  & 20.60 $^{(42)}$ & 21.38 \\
      HD 203532 &  309.460 &  -31.740 &    0.292 &    0.280 $^{(47)}$ &  & 21.27 $^{(32)}$ &  & 20.64 $^{(32)}$ & 21.44 \\
      HD 203938 &   90.560 &   -2.330 &    0.223 &    0.720 $^{(27)}$ &  & 21.48 $^{(27)}$ &  & 21.00 $^{(27)}$ & 21.70 \\
      HD 204172 &   83.390 &   -9.960 &    1.927 &    0.170 $^{(3)}$ &  & 21.00 $^{(3)}$ &  & 19.60 $^{(2)}$ & 21.03 \\
      HD 206165 &  102.270 &    7.250 &    0.746 &    0.470 $^{(17)}$ &  &   $-$          &  & 20.78 $^{(34)}$ & 21.44 \\
      HD 206267 &   99.290 &    3.740 &    1.117 &    0.510 $^{(16)}$ &  & 21.30 $^{(27)}$ &  & 20.86 $^{(27)}$ & 21.54 \\
      HD 206773 &   99.800 &    3.620 &    0.958 &    0.440 $^{(32)}$ &  & 21.09 $^{(32)}$ &  & 20.44 $^{(32)}$ & 21.25 \\
      HD 207198 &  103.140 &    6.990 &    1.025 &    0.590 $^{(8)}$ &  & 21.34 $^{(8)}$ &  & 20.83 $^{(27)}$ & 21.55 \\
      HD 207308 &  103.110 &    6.820 &    1.026 &    0.520 $^{(15)}$ &  & 21.20 $^{(41)}$ &  & 20.76 $^{(41)}$ & 21.44 \\
      HD 207538 &  101.600 &    4.670 &    0.838 &    0.640 $^{(46)}$ &  & 21.34 $^{(18)}$ &  & 20.91 $^{(27)}$ & 21.58 \\
      HD 208266 &  102.710 &    4.980 &    0.911 &    0.520 $^{(40)}$ &  &   $-$          &  & 20.87 $^{(34)}$ & 21.48 \\
      HD 208440 &  104.030 &    6.440 &    0.829 &    0.290 $^{(31)}$ &  & 21.23 $^{(18)}$ &  & 20.34 $^{(34)}$ & 21.33 \\
      HD 208905 &  103.530 &    5.170 &    1.030 &    0.370 $^{(5)}$ &  &   $-$          &  & 20.43 $^{(43)}$ & 21.33 \\
      HD 209339 &  104.580 &    5.870 &    0.845 &    0.380 $^{(15)}$ &  & 21.16 $^{(41)}$ &  & 20.21 $^{(38)}$ & 21.25 \\
      HD 209481 &  101.010 &    2.180 &    1.101 &    0.370 $^{(5)}$ &  & 21.11 $^{(18)}$ &  & 20.54 $^{(43)}$ & 21.30 \\
      HD 209833 &   84.490 &  -21.270 &    0.098 &    0.089 $^{(21)}$ &  &   $-$          & <& 13.68 $^{(48)}$ & 20.10 \\
      HD 209952 &  350.000 &  -52.470 &    0.030 &    0.060 $^{(45)}$ &  & 19.04 $^{(19)}$ & <& 13.68 $^{(2)}$ & 20.54 \\
      HD 209975 &  104.870 &    5.390 &    0.858 &    0.360 $^{(18)}$ &  & 21.11 $^{(10)}$ &  & 20.15 $^{(43)}$ & 21.20 \\
      HD 210121 &   56.880 &  -44.460 &    0.342 &    0.380 $^{(27)}$ &  & 20.63 $^{(27)}$ &  & 20.75 $^{(27)}$ & 21.19 \\
      HD 210191 &   37.150 &  -51.760 &    0.662 &    0.070 $^{(3)}$ &  & 20.70 $^{(8)}$ & <& 18.60 $^{(2)}$ & 20.61 \\
      HD 210809 &   99.850 &   -3.130 &    4.329 &    0.330 $^{(32)}$ &  & 21.25 $^{(32)}$ &  & 20.00 $^{(32)}$ & 21.30 \\
      HD 210839 &  103.830 &    2.610 &    0.617 &    0.570 $^{(3)}$ &  & 21.11 $^{(18)}$ &  & 20.84 $^{(27)}$ & 21.43 \\
      HD 212791 &  101.640 &   -4.300 &    0.998 &    0.050 $^{(32)}$ &  & 21.21 $^{(32)}$ &  & 19.42 $^{(32)}$ & 21.22 \\
      HD 214080 &   44.810 &  -56.920 &    1.203 &    0.050 $^{(8)}$ &  & 20.60 $^{(8)}$ & <& 19.00 $^{(2)}$ & 20.46 \\
      HD 214680 &   96.650 &  -16.980 &    0.359 &    0.110 $^{(3)}$ &  & 20.70 $^{(3)}$ &  & 19.22 $^{(2)}$ & 20.73 \\
      HD 214993 &   97.650 &  -16.180 &    0.321 &    0.110 $^{(7)}$ &  & 20.79 $^{(7)}$ &  & 19.63 $^{(7)}$ & 20.85 \\
      HD 216532 &  109.650 &    2.680 &    0.751 &    0.860 $^{(17)}$ &  &   $-$          &  & 21.10 $^{(34)}$ & 21.70 \\
      HD 216898 &  109.930 &    2.390 &    0.840 &    0.850 $^{(17)}$ &  &   $-$          &  & 21.05 $^{(34)}$ & 21.69 \\
      HD 217035 &  110.250 &    2.860 &    0.829 &    0.760 $^{(18)}$ &  & 21.46 $^{(18)}$ &  & 20.95 $^{(34)}$ & 21.67 \\
      HD 217312 &  110.560 &    2.950 &    1.631 &    0.660 $^{(17)}$ &  & 21.48 $^{(18)}$ &  & 20.80 $^{(34)}$ & 21.63 \\
      HD 217615 &  332.370 &  -56.680 &    0.279 &    0.247 $^{(21)}$ &  &   $-$          &  & 19.67 $^{(48)}$ & 20.94 \\
      HD 217675 &  102.210 &  -16.100 &    0.109 &    0.050 $^{(10)}$ &  &   $-$          &  & 19.67 $^{(2)}$ & 20.46 \\
      HD 218376 &  109.950 &   -0.780 &    0.374 &    0.220 $^{(3)}$ &  & 20.95 $^{(3)}$ &  & 20.15 $^{(2)}$ & 21.07 \\
      HD 218915 &  108.060 &   -6.890 &    8.065 &    0.210 $^{(45)}$ &  & 21.20 $^{(8)}$ &  & 20.15 $^{(39)}$ & 21.27 \\
      HD 219188 &   83.030 &  -50.170 &    2.268 &    0.080 $^{(10)}$ &  & 20.85 $^{(3)}$ &  & 19.34 $^{(2)}$ & 20.88 \\
      HD 220057 &  112.130 &    0.210 &    0.392 &    0.270 $^{(32)}$ &  & 21.17 $^{(32)}$ &  & 20.28 $^{(32)}$ & 21.27 \\
      HD 224151 &  115.440 &   -4.640 &    1.898 &    0.420 $^{(44)}$ &  & 21.32 $^{(18)}$ &  & 20.57 $^{(41)}$ & 21.45 \\
      HD 224572 &  115.550 &   -6.360 &    0.292 &    0.200 $^{(4)}$ &  & 20.88 $^{(3)}$ &  & 20.23 $^{(2)}$ & 21.04 \\
      HD 232522 &  130.700 &   -6.710 &   11.905 &    0.180 $^{(32)}$ &  & 21.08 $^{(32)}$ &  & 20.22 $^{(32)}$ & 21.19 \\
      HD 303308 &  287.590 &   -0.610 &    2.457 &    0.430 $^{(44)}$ &  & 21.40 $^{(8)}$ &  & 20.24 $^{(44)}$ & 21.46 \\
      HD 308813 &  294.790 &   -1.610 &    5.291 &    0.260 $^{(45)}$ &  & 21.15 $^{(18)}$ &  & 20.29 $^{(44)}$ & 21.26 \\
      HD 315021 &    6.120 &   -1.340 &    1.292 &    0.310 $^{(18)}$ &  & 21.28 $^{(18)}$ &  & 19.99 $^{(38)}$ & 21.32 \\
   HE 0226-4110 &  253.940 &  -65.780 &     $-$  &    0.016 $^{(21)}$ & >& 19.50 $^{(36)}$ & <& 14.29 $^{(36)}$ & 19.97 \\
   HE 1143-1810 &  281.850 &   41.710 &     $-$  &    0.039 $^{(36)}$ &  & 20.47 $^{(36)}$ &  & 16.54 $^{(36)}$ & 20.47 \\
   HS 0624+6907 &  145.710 &   23.350 &     $-$  &    0.098 $^{(21)}$ &  & 20.80 $^{(36)}$ &  & 19.82 $^{(36)}$ & 20.88 \\
   MRC 2251-178 &   46.200 &  -61.330 &     $-$  &    0.039 $^{(36)}$ &  & 20.39 $^{(36)}$ &  & 14.54 $^{(36)}$ & 20.39 \\
       Mrk 0009 &  158.360 &   28.750 &     $-$  &    0.059 $^{(21)}$ &  & 20.64 $^{(36)}$ &  & 19.36 $^{(36)}$ & 20.68 \\
       Mrk 0106 &  161.140 &   42.880 &     $-$  &    0.028 $^{(21)}$ &  & 20.35 $^{(36)}$ &  & 16.23 $^{(36)}$ & 20.35 \\
       Mrk 0116 &  160.530 &   44.840 &     $-$  &    0.032 $^{(36)}$ &  & 20.41 $^{(36)}$ &  & 19.08 $^{(36)}$ & 20.45 \\
       Mrk 0205 &  125.450 &   41.670 &     $-$  &    0.042 $^{(36)}$ &  & 20.40 $^{(36)}$ &  & 16.53 $^{(36)}$ & 20.40 \\
       Mrk 0209 &  134.150 &   68.080 &     $-$  &    0.015 $^{(36)}$ & >& 19.73 $^{(36)}$ & <& 14.48 $^{(36)}$ & 19.94 \\
       Mrk 0290 &   91.490 &   47.950 &     $-$  &    0.015 $^{(36)}$ &  & 20.11 $^{(36)}$ &  & 16.18 $^{(36)}$ & 20.11 \\
       Mrk 0335 &  108.760 &  -41.420 &     $-$  &    0.035 $^{(21)}$ &  & 20.43 $^{(36)}$ &  & 18.83 $^{(36)}$ & 20.45 \\
       Mrk 0421 &  179.830 &   65.030 &     $-$  &    0.013 $^{(21)}$ &  & 19.94 $^{(36)}$ &  & 14.63 $^{(36)}$ & 19.94 \\
       Mrk 0478 &   59.240 &   65.030 &     $-$  &    0.013 $^{(21)}$ & >& 19.21 $^{(36)}$ & <& 14.56 $^{(36)}$ & 19.88 \\
       Mrk 0501 &   63.600 &   38.860 &     $-$  &    0.019 $^{(21)}$ &  & 20.24 $^{(36)}$ &  & 14.78 $^{(36)}$ & 20.24 \\
       Mrk 0509 &   35.970 &  -29.860 &     $-$  &    0.057 $^{(21)}$ &  & 20.58 $^{(36)}$ &  & 17.87 $^{(36)}$ & 20.58 \\
       Mrk 0817 &  100.300 &   53.480 &     $-$  &    0.007 $^{(36)}$ & >& 19.83 $^{(36)}$ & <& 14.03 $^{(36)}$ & 19.61 \\
       Mrk 0876 &   98.270 &   40.380 &     $-$  &    0.027 $^{(21)}$ &  & 20.36 $^{(36)}$ &  & 16.64 $^{(36)}$ & 20.36 \\
       Mrk 1095 &  201.690 &  -21.130 &     $-$  &    0.128 $^{(21)}$ &  & 20.95 $^{(36)}$ &  & 18.76 $^{(36)}$ & 20.96 \\
       Mrk 1383 &  349.220 &   55.130 &     $-$  &    0.032 $^{(21)}$ &  & 20.40 $^{(36)}$ &  & 14.35 $^{(36)}$ & 20.40 \\
       Mrk 1513 &   63.670 &  -29.070 &     $-$  &    0.044 $^{(21)}$ &  & 20.52 $^{(36)}$ &  & 16.42 $^{(36)}$ & 20.52 \\
 MS 0700.7+6338 &  152.470 &   25.630 &     $-$  &    0.051 $^{(21)}$ &  & 20.43 $^{(36)}$ &  & 18.75 $^{(36)}$ & 20.45 \\
       NGC 0985 &  180.840 &  -59.490 &     $-$  &    0.033 $^{(21)}$ &  & 20.52 $^{(36)}$ &  & 16.05 $^{(36)}$ & 20.52 \\
       NGC 1068 &  172.100 &  -51.900 &     $-$  &    0.034 $^{(21)}$ &  & 19.61 $^{(36)}$ &  & 18.13 $^{(36)}$ & 19.64 \\
       NGC 1705 &  261.080 &  -38.740 &     $-$  &    0.008 $^{(21)}$ & >& 19.66 $^{(36)}$ & <& 14.17 $^{(36)}$ & 19.67 \\
       NGC 4151 &  155.080 &   75.060 &     $-$  &    0.028 $^{(21)}$ &  & 20.20 $^{(36)}$ &  & 16.70 $^{(36)}$ & 20.20 \\
       NGC 4670 &  212.690 &   88.630 &     $-$  &    0.015 $^{(21)}$ &  & 19.95 $^{(36)}$ &  & 14.72 $^{(36)}$ & 19.95 \\
       NGC 7469 &   83.100 &  -45.470 &     $-$  &    0.069 $^{(21)}$ &  & 20.59 $^{(36)}$ &  & 19.67 $^{(36)}$ & 20.68 \\
    PG 0804+761 &  138.280 &   31.030 &     $-$  &    0.035 $^{(21)}$ &  & 20.54 $^{(36)}$ &  & 18.66 $^{(36)}$ & 20.55 \\
    PG 0844+349 &  188.560 &   37.970 &     $-$  &    0.037 $^{(21)}$ &  & 20.34 $^{(36)}$ &  & 18.22 $^{(36)}$ & 20.35 \\
    PG 0953+414 &  179.790 &   51.710 &     $-$  &    0.013 $^{(21)}$ &  & 20.00 $^{(36)}$ &  & 15.03 $^{(36)}$ & 20.00 \\
    PG 1116+215 &  223.360 &   68.210 &     $-$  &    0.023 $^{(21)}$ & >& 19.70 $^{(36)}$ & <& 14.16 $^{(36)}$ & 20.13 \\
    PG 1211+143 &  267.550 &   74.320 &     $-$  &    0.035 $^{(21)}$ &  & 20.25 $^{(36)}$ &  & 18.38 $^{(36)}$ & 20.26 \\
    PG 1259+593 &  120.560 &   58.050 &     $-$  &    0.008 $^{(21)}$ &  & 19.67 $^{(36)}$ &  & 14.75 $^{(36)}$ & 19.67 \\
    PG 1302-102 &  308.590 &   52.160 &     $-$  &    0.043 $^{(21)}$ &  & 20.42 $^{(36)}$ &  & 15.62 $^{(36)}$ & 20.42 \\
    PKS 0405-12 &  204.930 &  -41.760 &     $-$  &    0.058 $^{(21)}$ &  & 20.41 $^{(36)}$ &  & 15.79 $^{(36)}$ & 20.41 \\
   PKS 0558-504 &  257.960 &  -28.570 &     $-$  &    0.044 $^{(21)}$ &  & 20.53 $^{(36)}$ &  & 15.44 $^{(36)}$ & 20.53 \\
   PKS 2005-489 &  350.370 &  -32.600 &     $-$  &    0.056 $^{(21)}$ &  & 20.60 $^{(36)}$ &  & 15.07 $^{(36)}$ & 20.60 \\
   PKS 2155-304 &   17.730 &  -52.250 &     $-$  &    0.022 $^{(21)}$ &  & 20.06 $^{(36)}$ &  & 14.17 $^{(36)}$ & 20.06 \\
 QSO B1226+0219 &  289.950 &   64.360 &     $-$  &    0.021 $^{(21)}$ &  & 20.22 $^{(36)}$ &  & 15.74 $^{(36)}$ & 20.22 \\
 QSO J1104+7658 &  130.390 &   38.550 &     $-$  &    0.034 $^{(21)}$ &  & 20.25 $^{(36)}$ &  & 18.98 $^{(36)}$ & 20.29 \\
 QSO J1821+6420 &   94.000 &   27.420 &     $-$  &    0.043 $^{(21)}$ &  & 20.43 $^{(36)}$ &  & 17.91 $^{(36)}$ & 20.43 \\
       Ton S180 &  138.990 &  -85.070 &     $-$  &    0.014 $^{(21)}$ & >& 20.08 $^{(36)}$ & <& 14.37 $^{(36)}$ & 19.91 \\
       Ton S210 &  224.970 &  -83.160 &     $-$  &    0.017 $^{(21)}$ &  & 20.19 $^{(36)}$ &  & 16.57 $^{(36)}$ & 20.19 \\
         TY CrA &  359.990 &  -17.780 &    0.136 &    0.480 $^{(16)}$ &  &    $-$          &  & 21.10 $^{(16)}$ & 21.44 \\
     VII Zw 118 &  151.360 &   25.990 &     $-$  &    0.038 $^{(21)}$ &  & 20.56 $^{(36)}$ &  & 18.84 $^{(36)}$ & 20.58 \\
    WD 0004+330 &  112.480 &  -28.690 &    0.097 &    0.049 $^{(21)}$ &  & 19.68 $^{(19)}$ &  & 14.46 $^{(30)}$ & 19.68 \\
    WD 1636+351 &   56.980 &   41.400 &    0.145 &    0.026 $^{(21)}$ &  & 19.57 $^{(19)}$ &  & 15.05 $^{(30)}$ & 19.57 \\
    WD 1800+685 &   98.730 &   29.780 &    0.159 &    0.054 $^{(21)}$ &  & 18.86 $^{(33)}$ &  & 14.75 $^{(30)}$ & 18.86 \\
    WD 2247+583 &  107.640 &   -0.640 &    0.107 &    1.310 $^{(21)}$ &  & 19.89 $^{(22)}$ &  & 15.11 $^{(30)}$ & 19.89 \\
\hline
\end{longtable}
\begin{list}{}{}
\item 
(1) \cite{1976ApJ...204..750Y}; 
(2) \cite{savage_survey_1977}; 
(3) \cite{bohlin_survey_1978}; 
(4) \cite{1978ApJS...38..129H}; 
(5) \cite{1980A&AS...42..251N}; 
(6) \cite{1982ApJ...257..125F}; 
(7) \cite{bohlin_survey_1983}; 
(8) \cite{1985ApJ...294..599S}; 
(9) \cite{1985ApJS...59..397S}; 
(10) \cite{1986ApJ...301..355J}; 
(11) \cite{1988ApJS...67..225V}; 
(12) \cite{1989ApJ...340..273V}; 
(13) \cite{1990ApJ...358..473W}; 
(14) \cite{1991ApJ...381..462W}; 
(15) \cite{1992A&AS...94..211G}; 
(16) \cite{1992ApJ...398...53P}; 
(17) \cite{1994ApJ...424..772F}; 
(18) \cite{diplas_iue_1994}; 
(19) \cite{fruscione_distribution_1994}; 
(20) \cite{1995ApJ...445L..95R}; 
(21) \cite{schlegel_maps_1998}; 
(22) \cite{1999A&A...346..969W}; 
(23) \cite{1999A&A...347..669L}; 
(24) \cite{2000ApJ...528..756L}; 
(25) \cite{2000ApJ...529..251R}; 
(26) \cite{2001ApJS..136..631S}; 
(27) \cite{rachford_far_2002}; 
(28) \cite{2002MNRAS.334..327A}; 
(29) \cite{2003ApJ...591.1000A}; 
(30) \cite{lehner_far_2003}; 
(31) \cite{2003ApJ...597..408C}; 
(32) \cite{cartledge_homogeneity_2004}; 
(33) \cite{2005ApJ...622..377O}; 
(34) \cite{2005ApJ...633..986P}; 
(35) \cite{2006AcA....56..373G}; 
(36) \cite{gillmon_fuse_2006}; 
(37) \cite{2006ApJ...649..788R}; 
(38) \cite{2007ApJ...655..285S}; 
(39) \cite{2007ApJ...658..446B}; 
(40) \cite{2007ApJ...667.1002S}; 
(41) \cite{jensen_new_2007}; 
(42) \cite{2008ApJ...687.1043C}; 
(43) \cite{2008ApJ...687.1075S}; 
(44) \cite{2008ApJ...688.1124S}; 
(45) \cite{2008ApJS..176...59B}; 
(46) \cite{rachford_molecular_2009}; 
(47) \cite{2010ApJ...708..334B}; 
(48) \cite{gudennavar_compilation_2012}
\end{list}
\end{center}


\twocolumn		
\section{Heating and cooling equations}\label{app:cooling} 

		\begin{figure}[h!]
			\includegraphics[width=9cm,trim = 0.0cm 0.2cm 0.0cm 0.0cm, clip,angle=0]{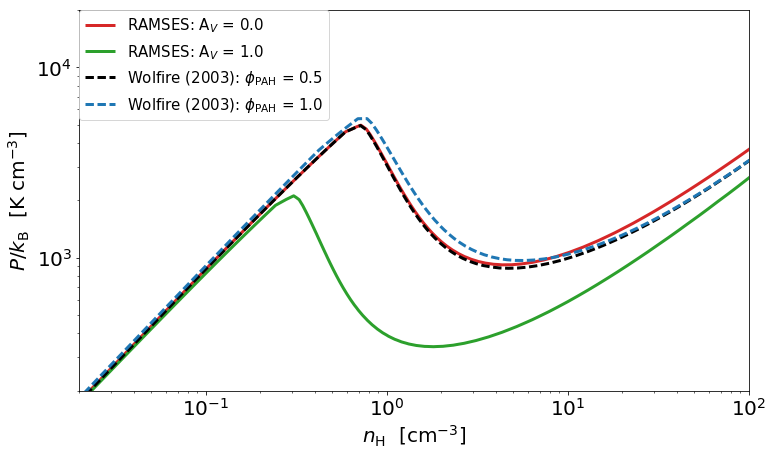}
			\caption{Thermal equilibrium curves ($\mathcal{L} = 0$) computed with RAMSES for $A_V=0$ (red solid curve) and $A_V=1$ (green solid curve) compared to those predicted by \citet{wolfire_neutral_2003} for $\phi_{\rm PAH}$ = 0.5 (black dashed curve) and 1.0 (blue dashed curve).
			}
			\label{fig:Wolfire}
		\end{figure}
		
		The analytical equations of the heating and cooling rates used in this work are taken from \cite{wolfire_neutral_1995} and \cite{wolfire_neutral_2003}.
		
		The electronic density is calculated from the ionization equilibrium
			\begin{equation}
			n_{\rm e} = 2.4\times 10^{-3}  \zeta_{\rm CR}^{1/2} \left(\frac{T}{100\textrm{ K}}\right)^{0.25}  \frac{G_{\rm eff}^{1/2}}{\phi} + n_{\rm H}x_{\rm C^+} \text{ cm}^{-3},
			\end{equation}
			where $\zeta_{\rm CR}$ is the total ionization rate (including primary and secondary ionizations) of H by soft X-rays and cosmic ray particles (expressed in unit of $10^{-16}$ s$^{-1}$) and
			$G_{\rm eff}$ is the local effective radiation field in Habing units (see Sect. \ref{sec:cooling}).
			$\phi_{\rm PAH}$ is a recombination parameter of electrons on PAHs, discussed in \cite{wolfire_neutral_2003}, and set to 0.5. 
			$x_{\rm C^+}$ is the abundance of C$^+$ relative to $n_{\rm H}$, $x_{\rm C^+} = n({\rm C}^+)/n_{\rm H}$. Throughout this work, we adopt $x_{\rm C^+} = 1.4 \times 10^{-4}$, which corresponds to the value derived in the Solar Neighborhood assuming 40\% depletion of carbon onto grains and that the remaining carbon is singly ionized. 
			This equation for the density of electrons differs from that of \cite{wolfire_neutral_2003}
			by the addition of C$^+$ which is the most abundant ion in the diffuse and transluscent CNM \citep{snow_diffuse_2006}.
			
		Following \cite{wolfire_neutral_2003}, we include the heating induced by the photoelectric effect on grains and PAHs and by cosmic ray ionizations. The former is modeled with a rate
			\begin{align}
			{\color{BrickRed}\blacktriangle}\,\,\,\, \Gamma_{\rm ph} &= 1.3\times 10^{-24} \, \epsilon \,  G_{\rm eff} \text{ erg  s}^{-1}, 
			\end{align}
			where the heating efficiency
			\begin{equation}
			\epsilon= \frac{4.9\times 10^{-2}}{1.0 + [\kappa/1925]^{0.73}} 
			+ \frac{3.7\times 10^{-2}(T/10^4 \textrm{ K})^{0.7}}{1.0 + [\kappa/5000]}
			\end{equation}
			and
			\begin{equation}
			\kappa = \frac{G_{\rm eff} T^{1/2}}{n_{\rm e}\phi_{\rm PAH}}.
			\end{equation}
		The latter is modeled with a rate
			\begin{equation}
			{\color{BrickRed}\blacktriangle}\,\,\,\, \Gamma_\text{CR} \sim 10^{-27} \left(\frac{\zeta_{\rm CR}}{10^{-16} {\rm s}^{-1}}\right)\text{ erg  s$^{-1}$}.
			\end{equation}
		
		Regarding the cooling, we include the fine-structure emission of CII and OI, the emission of Lyman $\alpha$ photons by HI, and the recombination of electrons onto charged grains and PAHs.
			The cooling rate due to collisional excitation of the fine-structure levels of C$^+$ by atomic hydrogen and electrons is given by 
			\begin{equation}
			{\color{blue}\blacktriangledown}\,\,\,\, \Lambda_{\rm CII} = \left[ 2.25 \times 10^{-23} + 10^{-20} \left(\frac{T}{100\textrm{ K}}\right)^{-0.5} \frac{n_{\rm e}}{n_{\rm H}} \right]	\textrm{ e}^{-92/ T}  x_{\rm C^+}
			\text{ erg  cm$^{3}$ s$^{-1}$}.
			\end{equation}
			The cooling rate by collisional excitation of the fine-structure level of OI by atomic hydrogen is computed as 
			\begin{equation}
			{\color{blue}\blacktriangledown}\,\,\,\, \Lambda_{\rm OI} = 7.81\times 10^{-24}\left(\frac{T}{100\textrm{ K}}\right)^{0.4}  \textrm{ e}^{-228/ T} x_{\rm O} \text{ erg  cm$^{3}$ s$^{-1}$},
			\end{equation}
			where $x_{\rm O} = n({\rm O})/n_{\rm H}$ is the relative abundance of atomic oxygen. Throughout this paper, we adopt $x_{\rm O} =3.2 \times 10^{-4}$, which corresponds to the value derived in the Solar Neighborhood assuming 37\% depletion of oxygen onto grains and that the remaining oxygen is in its atomic form.
			Those two lines are the dominant cooling terms of the CNM phase.
			The cooling induced by the excitation of the Lyman $\alpha$ line, which is the dominant cooling at T $\gtrsim$ 8000 K (WNM), is taken from \citet{spitzer_physical_1978}
			\begin{equation}
			{\color{blue}\blacktriangledown}\,\,\,\, \Lambda_{\rm HI} = 7.3\times 10^{-19} x_{\rm e} \textrm{ e}^{-118400/T} \text{ erg  cm$^{3}$ s$^{-1}$}.
			\end{equation}
			Finally, the cooling rate due to electron recombination onto charged grains and PAHs is set to
			\begin{equation}
			{\color{blue}\blacktriangledown}\,\,\,\, \Lambda_{\rm rec} = 4.65\times 10^{-30}  T^{0.94}  \kappa^{\beta} x_{\rm e} \phi_{\rm PAH} \text{ erg  cm$^{3}$ s$^{-1}$},
			\end{equation}
			with $\beta = 0.74/T^{0.068}$.
		To validate the calculations of the heating and cooling terms, we compare in Fig. \ref{fig:Wolfire} the thermal equilibrium curve obtained with RAMSES to the predictions of \citet{wolfire_neutral_2003}.

\section{Analytical description of 1D and 2D probability histograms}\label{app:analytical} 

\begin{figure}[!h]
\begin{center}
\includegraphics[width=9cm,trim = 1.5cm 2.5cm 1.4cm 1.5cm, clip,angle=0]{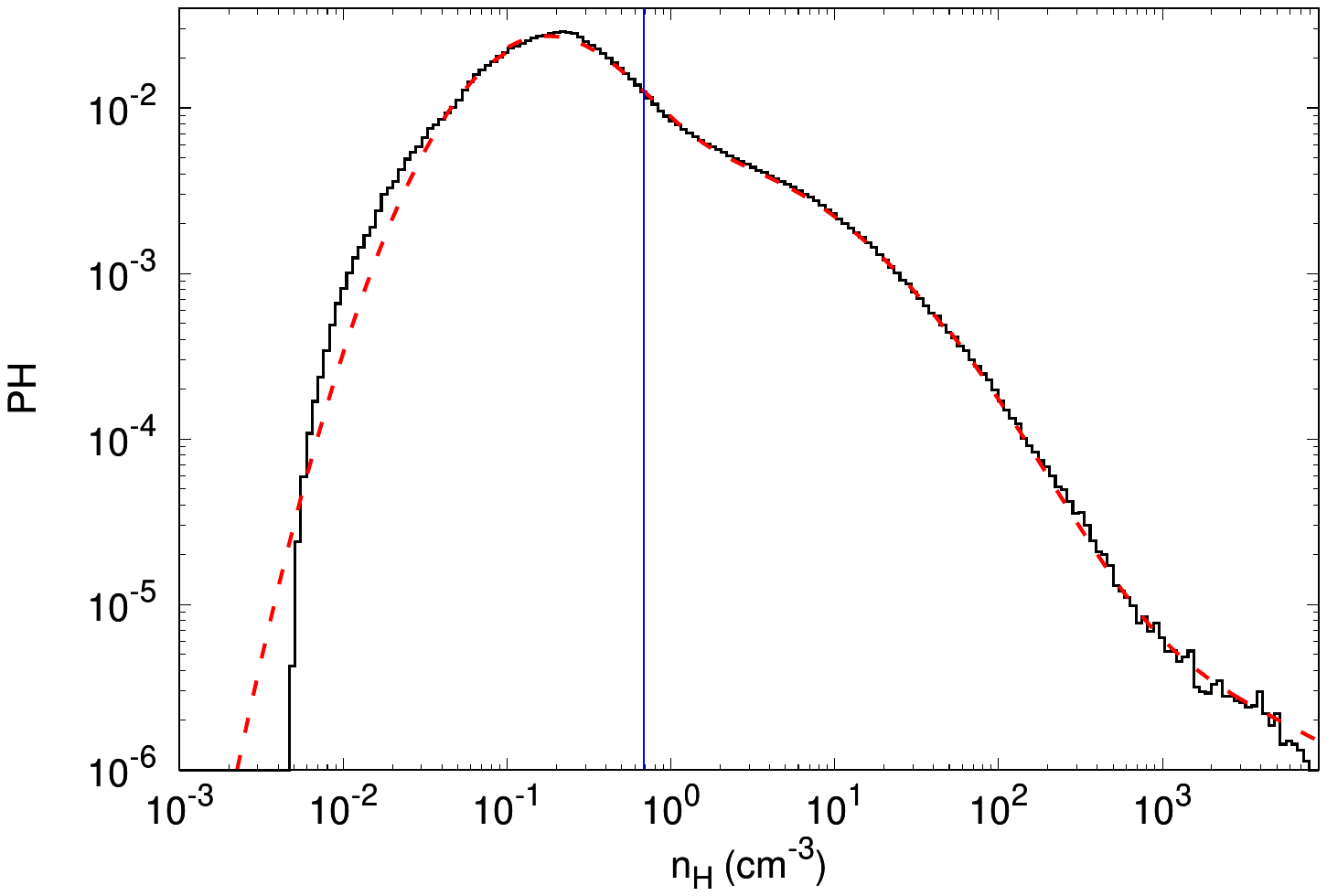}
\caption{Probability histogram of the proton density $n_{\rm H}$ extracted from the fiducial simulation (see Table \ref{table:grids}). The black histogram correspond to the extracted data. The red dashed curve shows an example of the sum of two log-normal components and a power-law tail at high density, for comparison. The blue line indicates the inflection point of the PH between the diffuse and the dense components.}
\label{Fig-PDF-dens}
\end{center}
\end{figure}

In order to interpret the results found in Sect. \ref{sec:results}, we propose a semi-analytical prescription to predict the 1D and 2D PHs of the total column density and the column density of H$_2$ obtained with numerical simulations. This prescription is based on the work of \citet{vazquez-semadeni_probability_2001}, \citet{bialy_h_2017}, and \citet{Bialy_2019a} who showed that lines of sight across isothermal simulations of turbulence can be accurately modeled as a series of random density fluctuations. 

\begin{figure*}[!ht]
	\begin{center}
		\includegraphics[width=9cm,trim = 1.8cm 0.8cm 0.0cm 0.2cm, clip,angle=0]{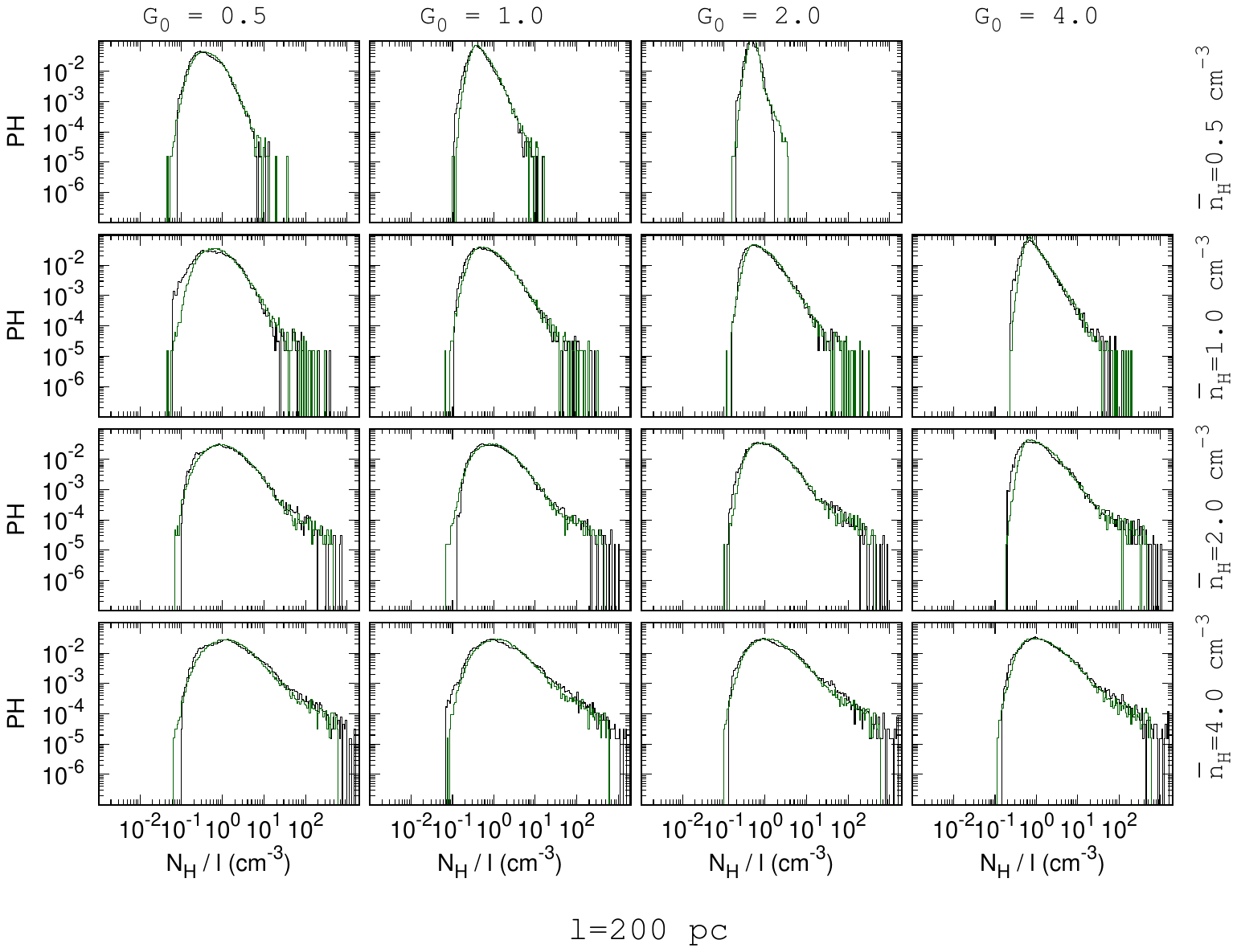}
		\includegraphics[width=9cm,trim = 1.8cm 0.8cm 0.0cm 0.2cm, clip,angle=0]{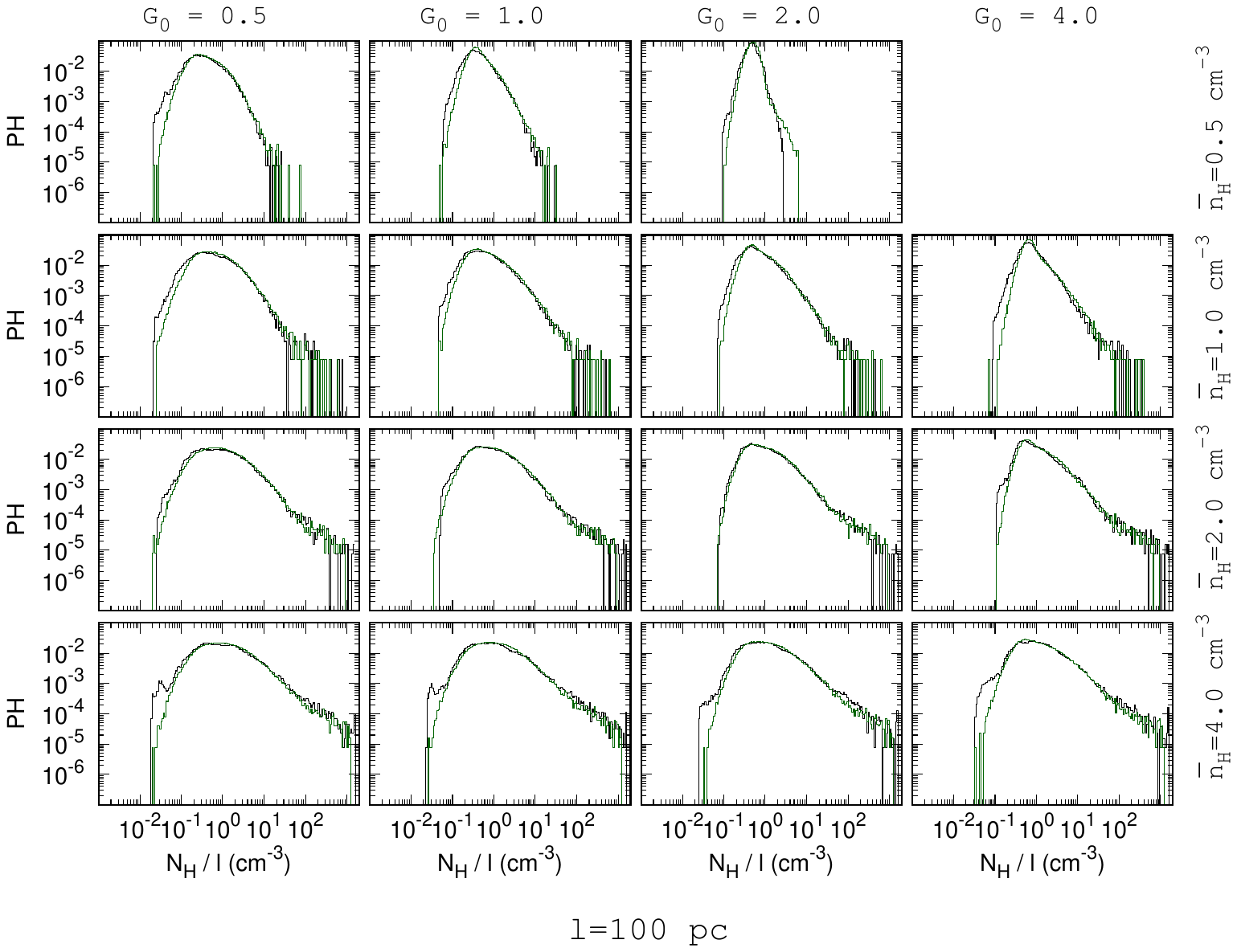}
		\includegraphics[width=9cm,trim = 1.5cm 0.8cm 0.0cm 0.2cm, clip,angle=0]{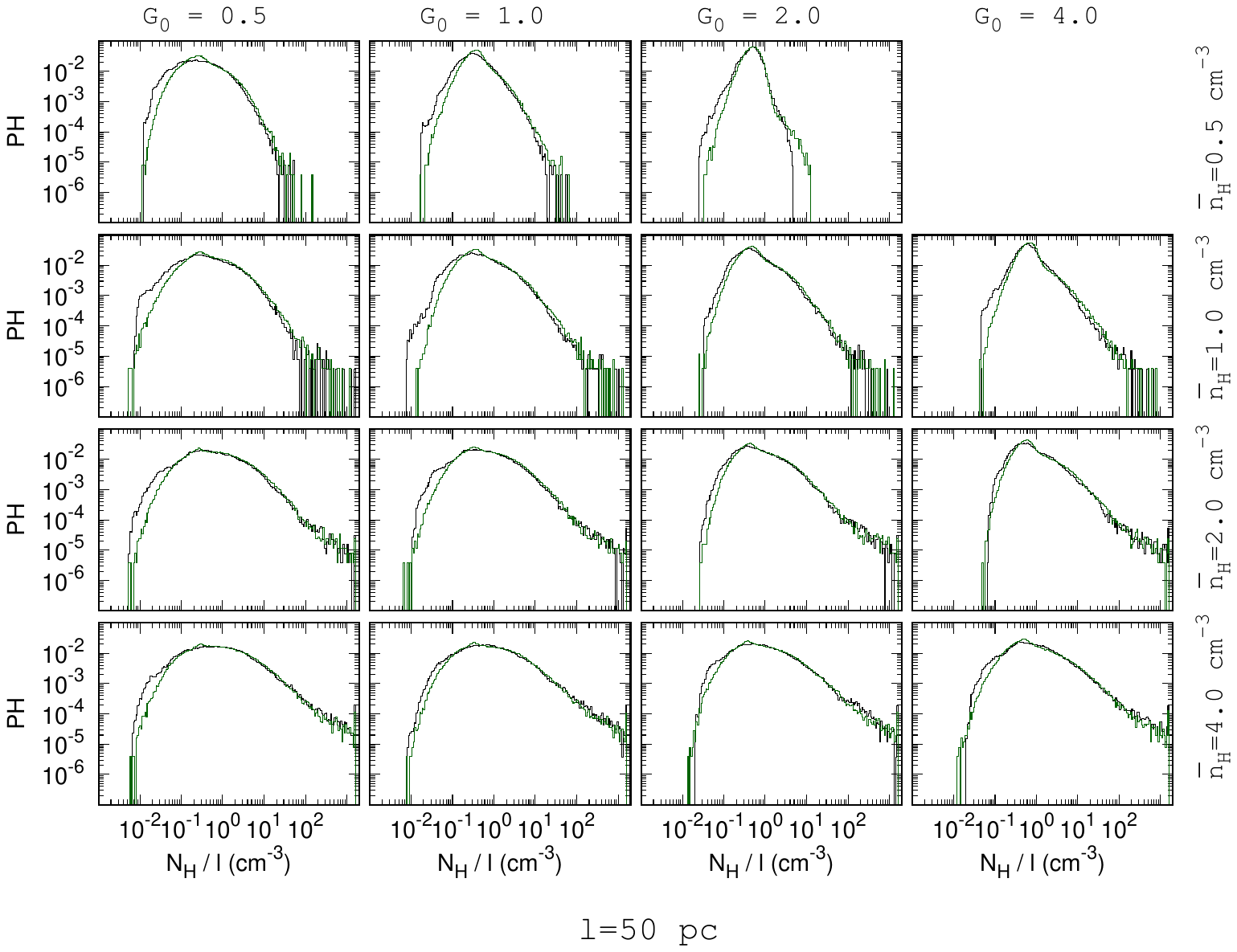}
		\includegraphics[width=9cm,trim = 1.5cm 0.8cm 0.0cm 0.2cm, clip,angle=0]{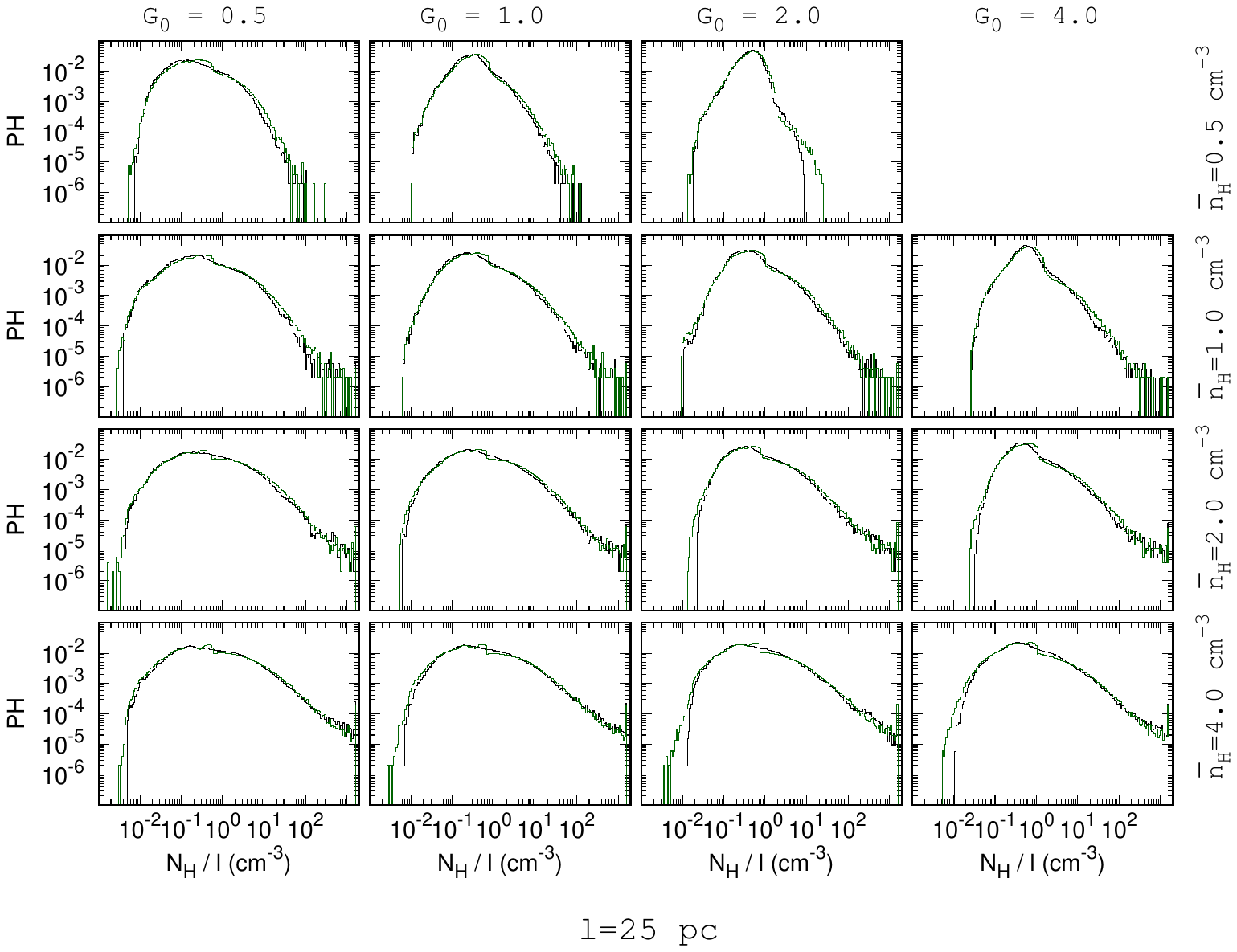}
		\caption{Comparisons of the 1D probability histograms  of the total column 
			density normalized to the integration scale, $N_{\rm H} / l$, extracted from the simulations 
			(black histograms) and constructed with the semi-analytical model described in the main 
			text (green histograms) for $l = 200$ (top left), 100 (top right), 50 (bottom left), and
			25 (bottom right) pc. Each of these four main panels displays the comparisons performed
			for 15 simulations with different $\overline{n_{\rm H}}$ and $G_0$ around the fiducial 
			setup defined by $\overline{n_{\rm H}} = 2$ cm$^{-3}$ and $G_0=1$. All probability histograms inferred
			from the semi-analytical model are obtained assuming a fixed correlation length of
			the diffuse component $y_{\rm dec}^{\rm diff} = 0.2 \times L_{\rm drive} = 20$ pc 
			and a correlation length of the dense component $y_{\rm dec}^{\rm dens} = 10 \times 
			(\overline{n_{\rm H}} / 2 {\rm cm}^{-3} )^{1/3}$ pc (see main text).}
		\label{Fig-compar-sim-mod-Ht}
	\end{center}
\end{figure*}

\begin{figure*}[!ht]
	\begin{center}
		\includegraphics[width=9cm,trim = 0.5cm 0.8cm 0.0cm 0.5cm, clip,angle=0]{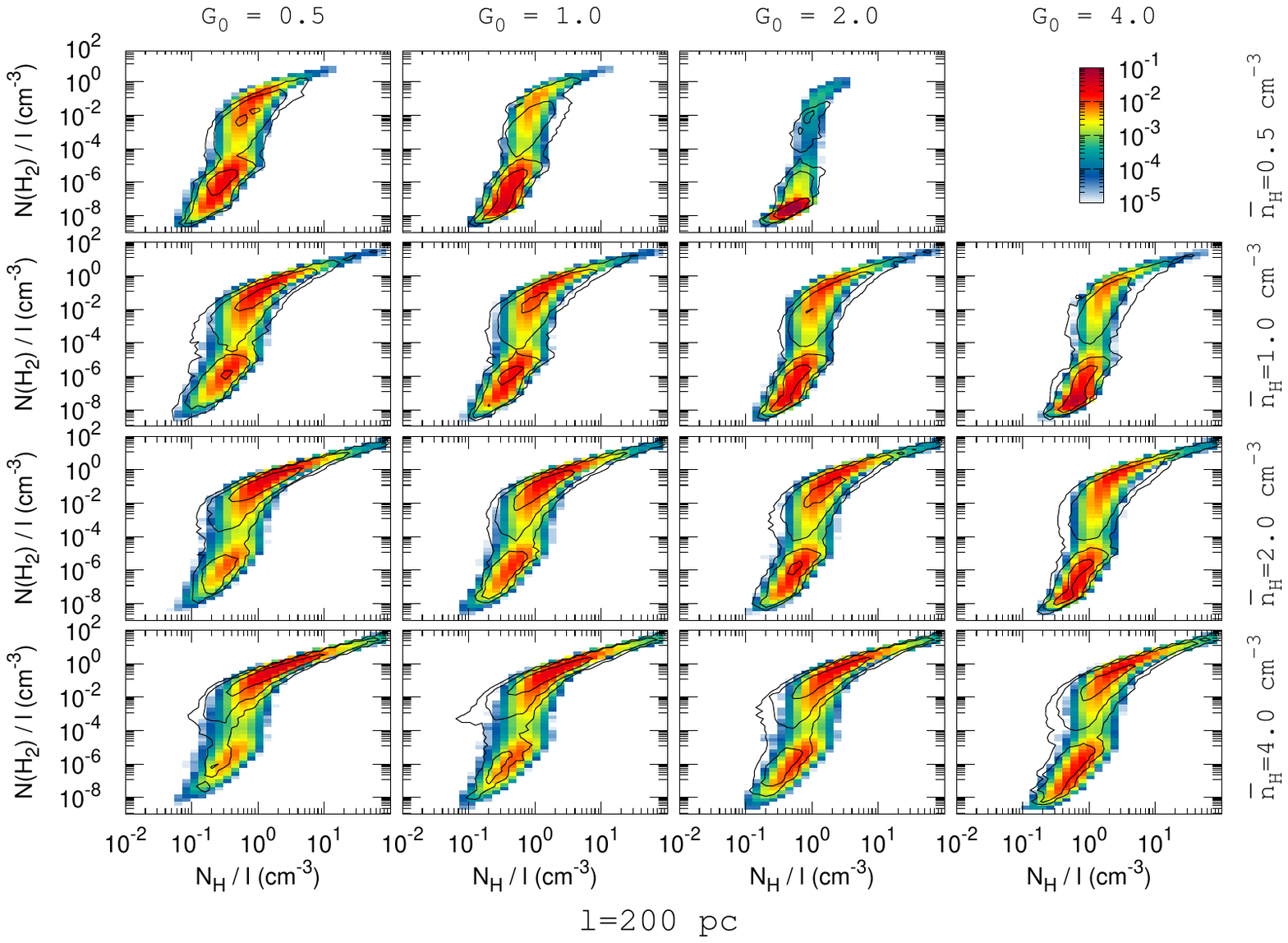}
		\includegraphics[width=9cm,trim = 0.5cm 0.8cm 0.0cm 0.5cm, clip,angle=0]{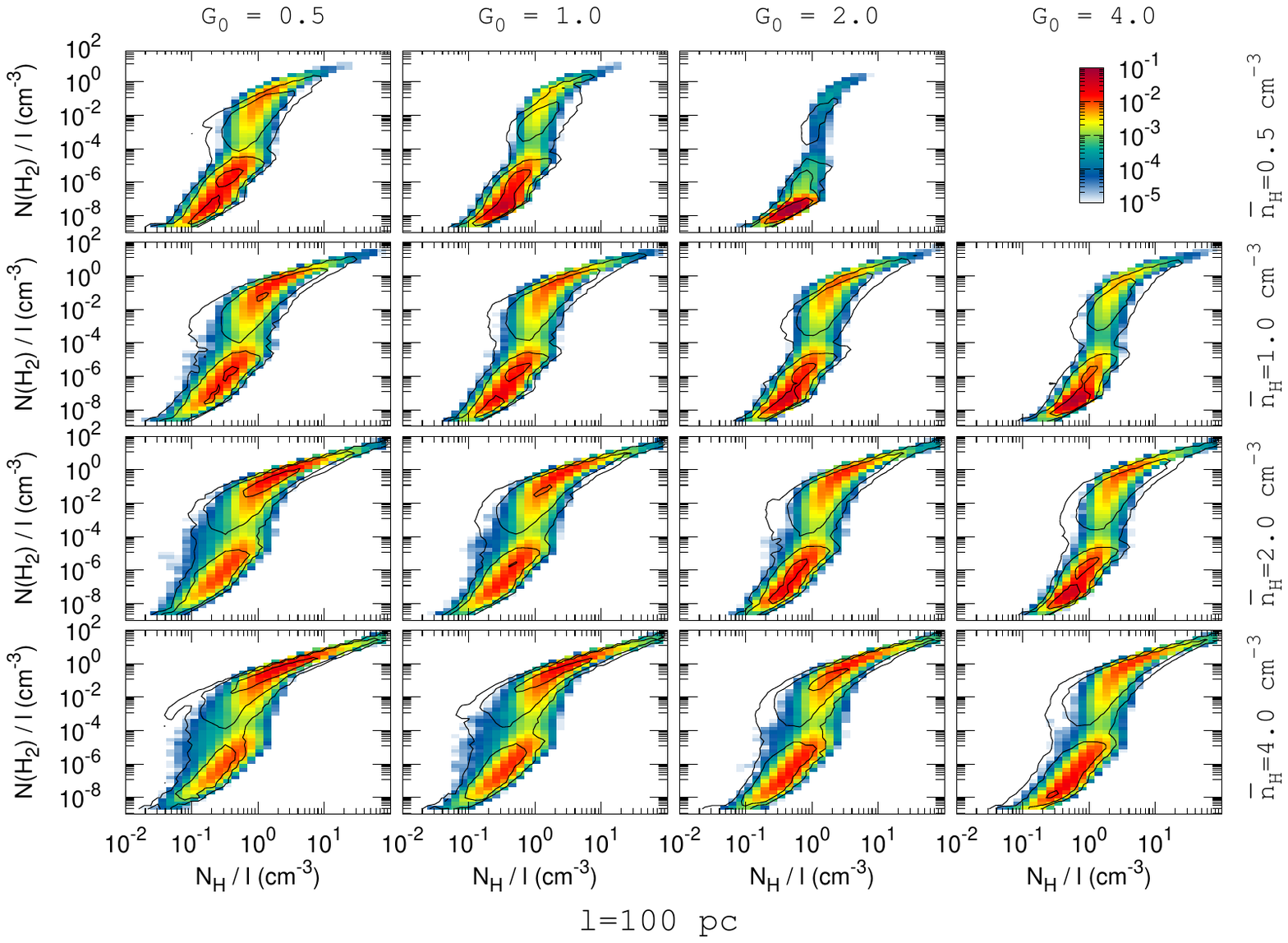}
		\includegraphics[width=9cm,trim = 0.5cm 0.8cm 0.0cm 0.5cm, clip,angle=0]{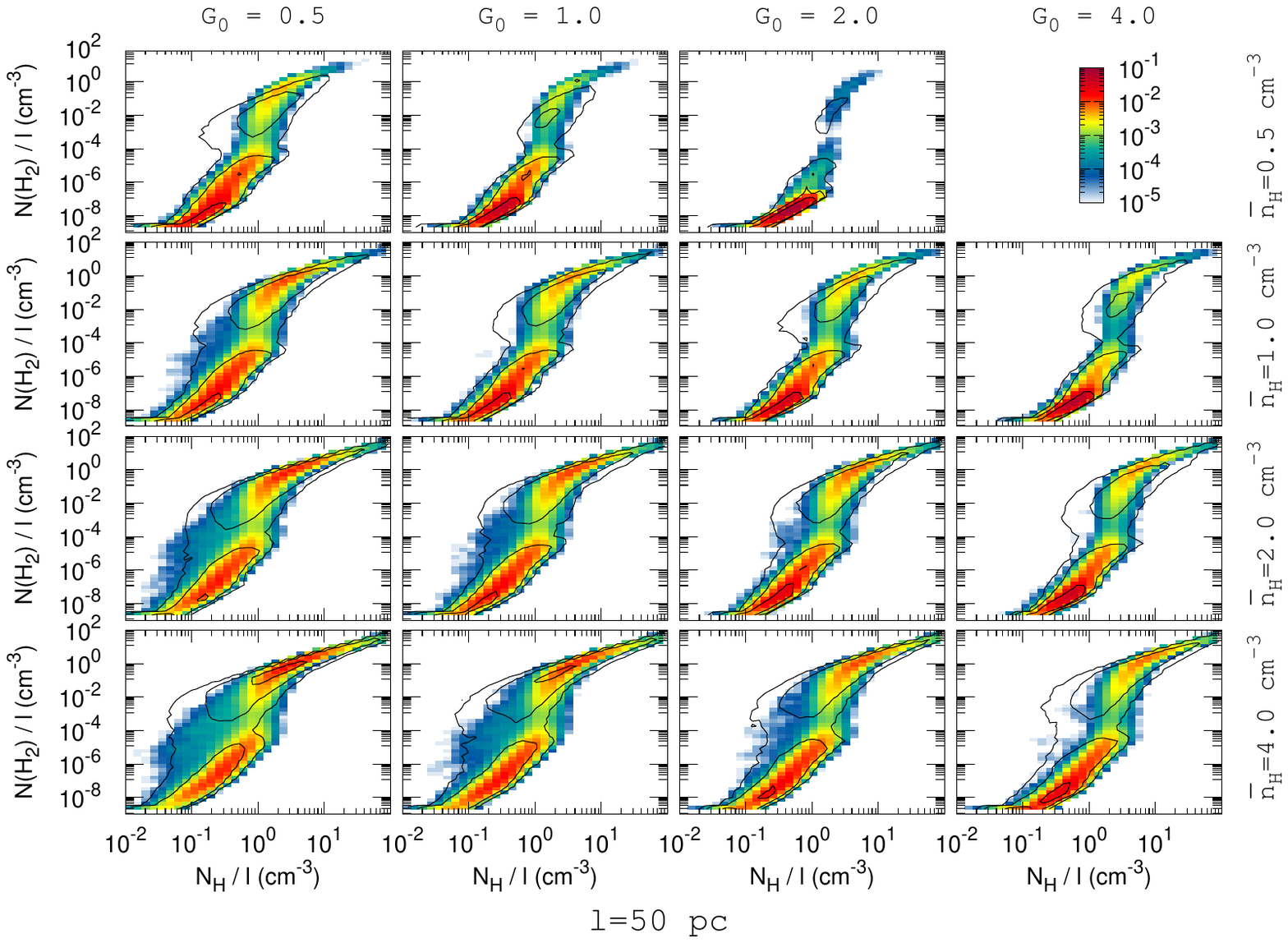}
		\includegraphics[width=9cm,trim = 0.5cm 0.8cm 0.0cm 0.5cm, clip,angle=0]{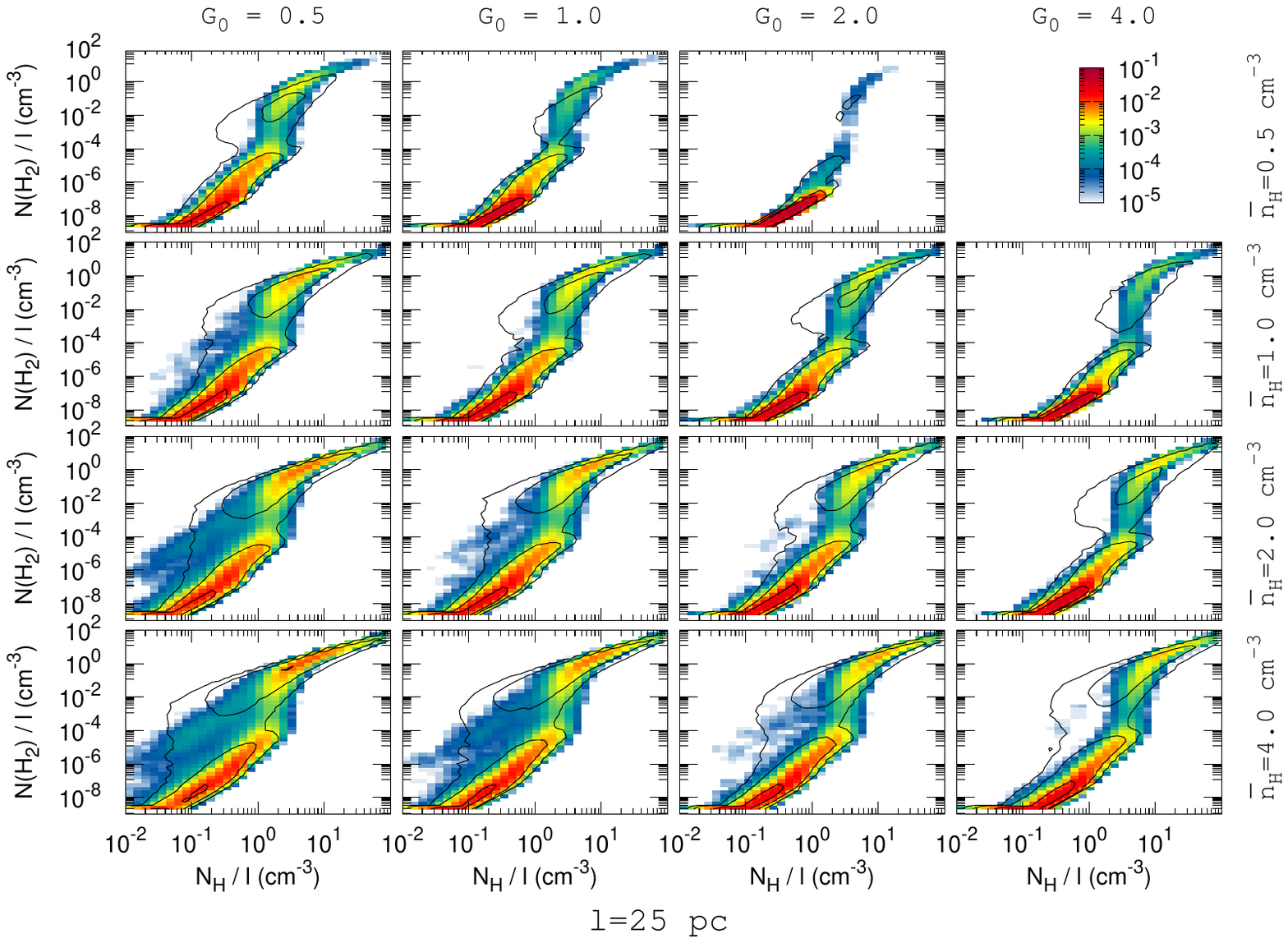}
		\caption{Comparisons of the 2D probability histograms  of the total column 
			density normalized to the integration scale, $N_{\rm H} / l$, and the column density of
			H$_2$ normalized to integration scale, $N({\rm H}_2) / l$ extracted from the simulations 
			and constructed with the semi-analytical model described in the main text for $l = 200$ 
			(top left), 100 (top right), 50 (bottom left), and 25 (bottom right) pc. The results
			of the simulations are indicated with contour plots with isoprobabilities of $10^{-4}$,
			$10^{-3}$, and $10^{-2}$. The colored histograms correspond to the results obtained
			with the semi-analytical model. As in Fig. \ref{Fig-compar-sim-mod-Ht}, each of the 
			four main panels displays the comparisons performed for 15 simulations with different 
			$\overline{n_{\rm H}}$ and $G_0$ around the fiducial setup defined by $\overline{n_{\rm H}} 
			= 2$ cm$^{-3}$ and $G_0=1$. All probability histograms inferred from the semi-analytical model are obtained 
			assuming a fixed correlation length of the diffuse component $y_{\rm dec}^{\rm diff} = 0.2 
			\times L_{\rm drive} = 20$ pc and a correlation length of the dense component 
			$y_{\rm dec}^{\rm dens} = 10 \times (\overline{n_{\rm H}} / 2 {\rm cm}^{-3} )^{1/3}$ pc 
			(see main text).}
		\label{Fig-compar-sim-mod-H2}
	\end{center}
\end{figure*}

\begin{figure*}[!ht]
	\begin{center}
		\includegraphics[width=9cm,trim = 0.5cm 0.8cm 0.0cm 0.5cm, clip,angle=0]{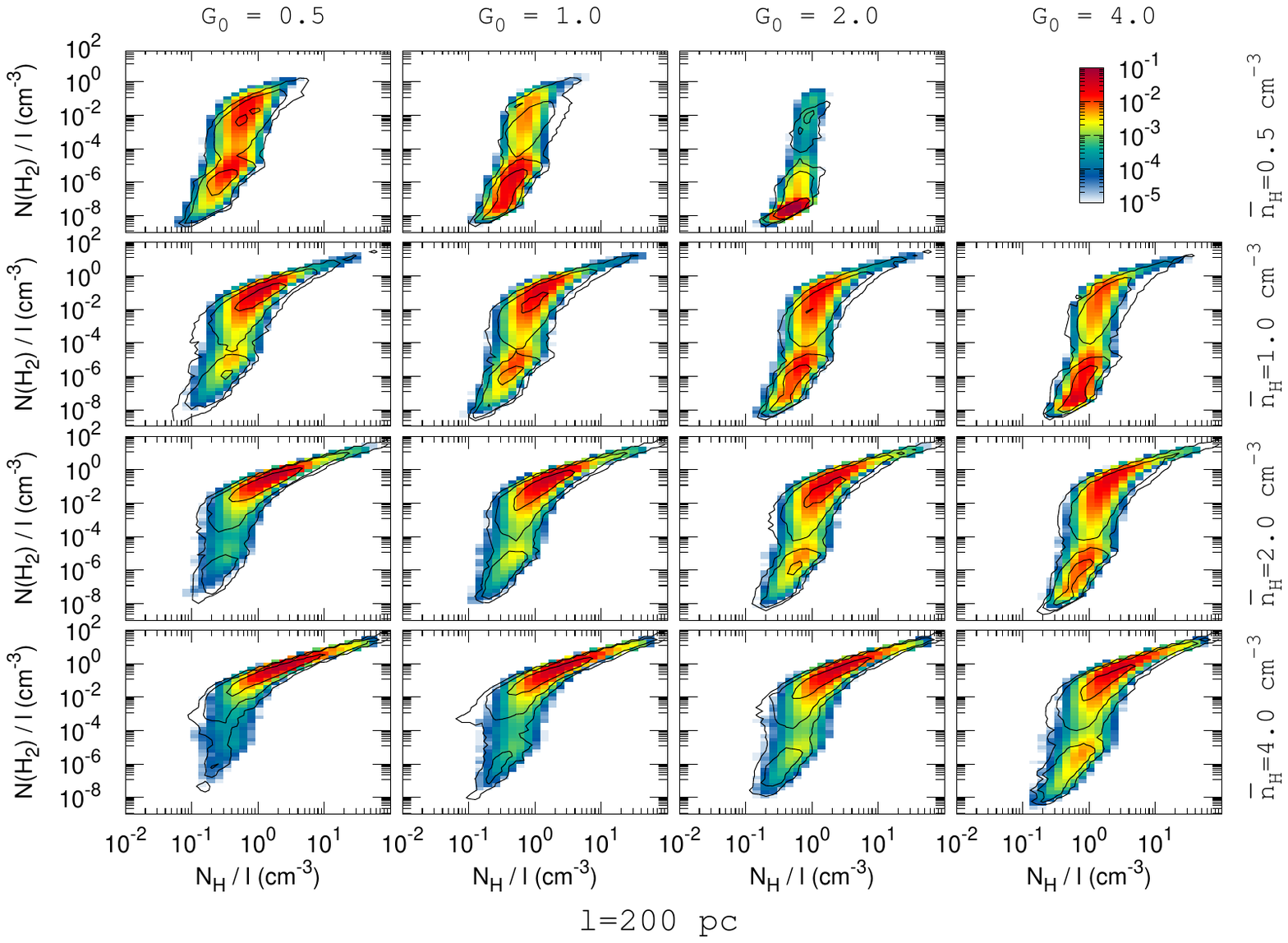}
		\includegraphics[width=9cm,trim = 0.5cm 0.8cm 0.0cm 0.5cm, clip,angle=0]{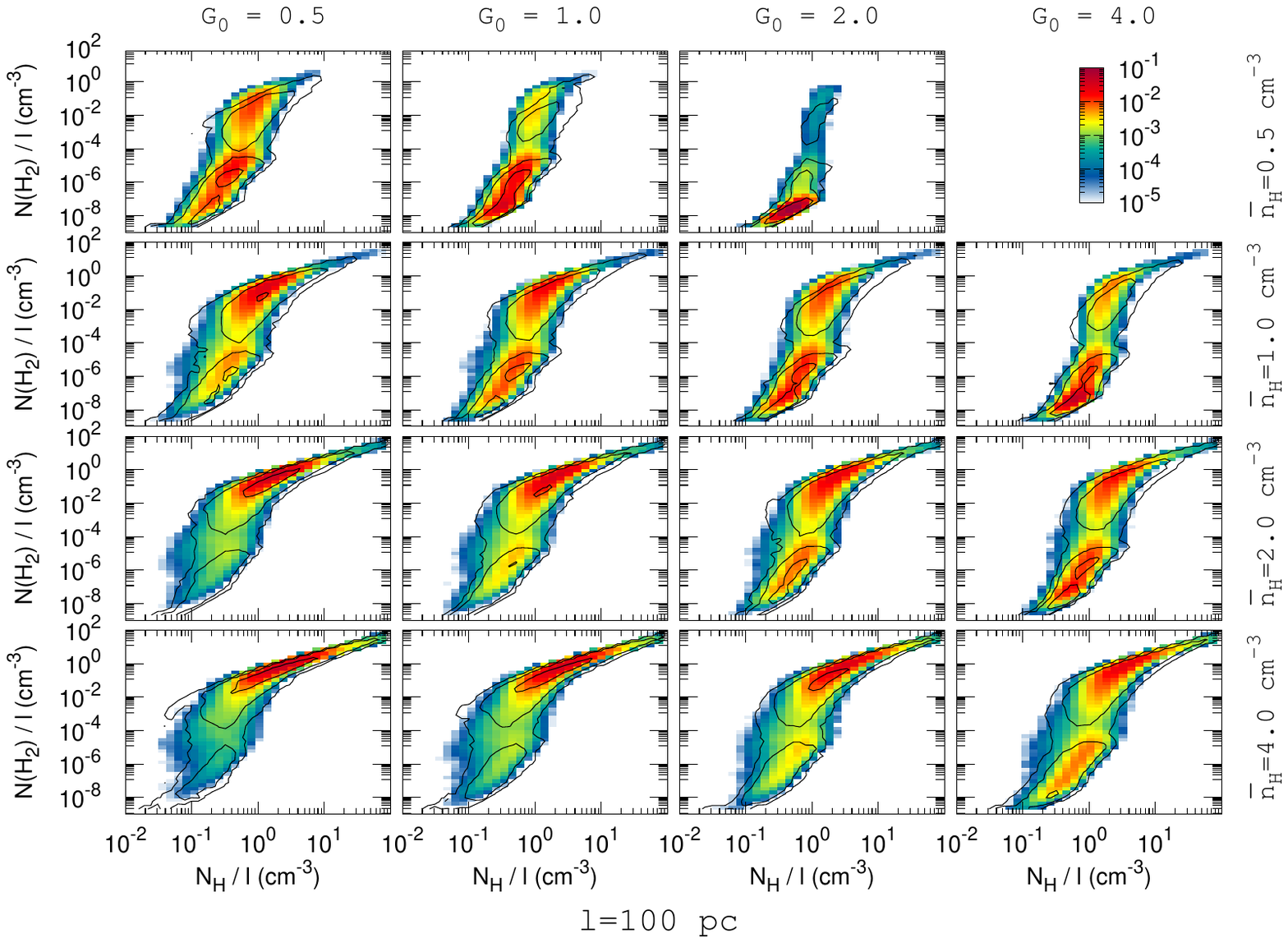}
		\includegraphics[width=9cm,trim = 0.5cm 0.8cm 0.0cm 0.5cm, clip,angle=0]{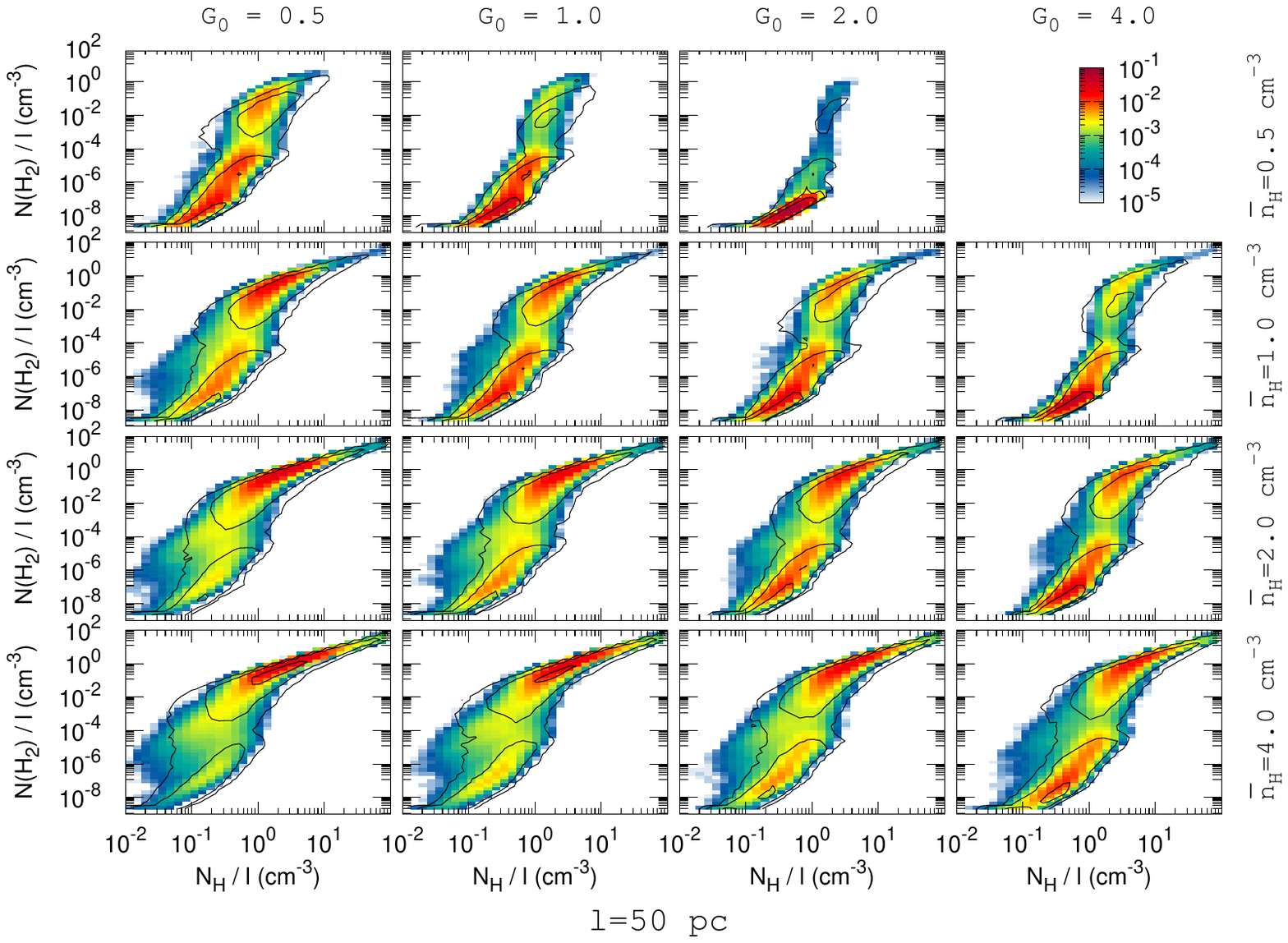}
		\includegraphics[width=9cm,trim = 0.5cm 0.8cm 0.0cm 0.5cm, clip,angle=0]{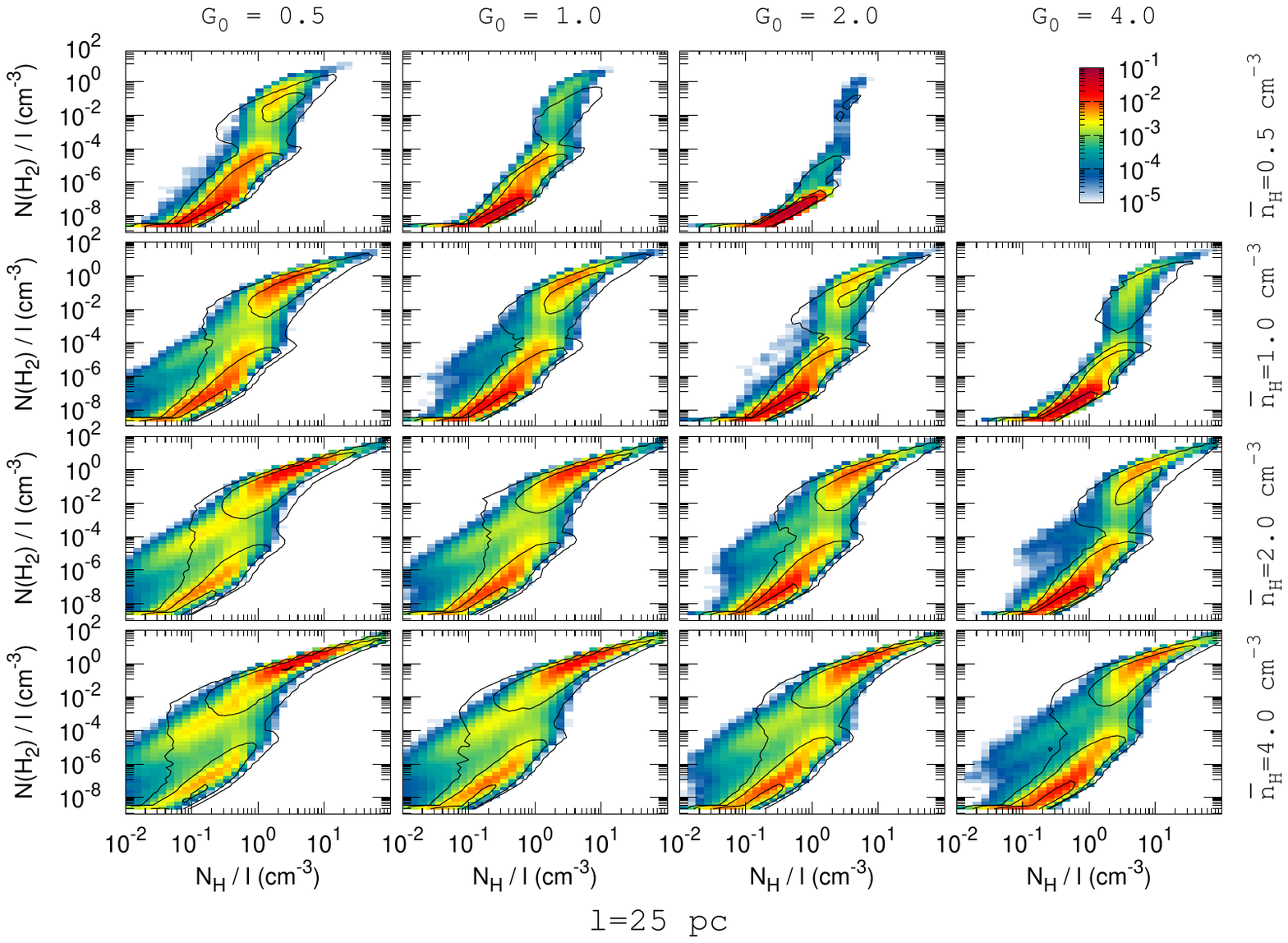}
		\caption{Same as Fig. \ref{Fig-compar-sim-mod-H2}, assuming that $y_{\rm dec}^{\rm dens} = 
			10 \times (\overline{n_{\rm H}} / 2 {\rm cm}^{-3} )^{1/3} (n_{\rm H} / n_{\rm H}^{\rm lim})^{-0.35}$ 
			pc. The exponent used here corresponds to a simple test to show the effect of a dense 
			decorrelation scale $y_{\rm dec}^{\rm dens}$ varying with density.}
		\label{Fig-compar-sim-mod-H2_2}
	\end{center}
\end{figure*}

\subsection{Decorrelation scales}

In Fig. \ref{Fig-PDF-dens}, we display an example of the volume-weighted distribution of the proton density computed for the fiducial simulation (see Table \ref{table:grids}). The total column density integrated along the x direction over a random line of sight of size $l \leqslant L$ is
\begin{equation}
N_{\rm H}(l) = \int_O^l n_{\rm H} dx = \sum_{i=i_0}^{i_l=i_0 + \frac{l}{L}R^{1/3}} n_{\rm H}(i) dx,
\end{equation}
where $R$ is the resolution of the box of size $L$ (see Table \ref{table:grids}), $i_0$ is a random starting index for the integration of the column densities, and $i_l$ is the final index deduced from $l$. Because of spatial correlations of the density $n_{\rm H}$, this computation is not equivalent to the sum of random realizations of $n_{\rm H}$ drawn out of the 1D probability distribution (Fig. \ref{Fig-PDF-dens}). It depends, instead, on how and over which distance the values of $n_{\rm H}$ are correlated. As proposed by \citet{bialy_h_2017}, and \citet{Bialy_2019a}, and since the correlations of density in a turbulent medium decrease over long distances, we assume that these correlations can be modeled with a parameter $y_{\rm dec}$, called the decorrelation scale. The density is supposed to be constant over distances smaller than $y_{\rm dec}$, and uncorrelated over larger distances. 

In this framework, if $y_{\rm dec}$ is the same for all densities, the total column density integrated over a distance $l$ becomes equivalent to the sum of $1 + l/y_{\rm dec}$ random realizations of $n_{\rm H}$. In isothermal simulations, the 1D probability distribution of the gas density is found to follow a lognormal distribution with a dispersion proportional to the Mach number, dependent on the nature of the turbulent driving, and independent on the mean density of the gas (e.g.
\citealt{padoan_universality_1997,federrath_density_2008}). Because of this property, \citet{bialy_h_2017} were able to establish a relation between the dispersion $\sigma_{N_{\rm H}/l}$ of the distribution of the averaged densities and the dispersion $\sigma_{n_{\rm H}}$ of the distribution of the proton density:
\begin{equation}
\frac{\sigma_{N_{\rm H}/l}}{\sigma_{n_{\rm H}}} = \left( 1 + \frac{L_{\rm drive}}{y_{\rm dec}} \frac{l}{L_{\rm drive}} \right)^{-1/2},
\end{equation}
where $L_{\rm drive}$ is the turbulence driving scale. Fitting the $\sigma_{N_{\rm H}/l}/\sigma_{n_{\rm H}}$ ratio as a function  of $l/L_{\rm drive}$, \citet{bialy_h_2017} estimated $y_{\rm dec} = 0.2 \times L_{\rm drive}$.

Unfortunately, this method cannot be applied here. Indeed, oppositely to isothermal 
simulations and as illustrated in Fig. \ref{Fig-PDF-dens}, the  PH of $n_{\rm H}$ derived from simulations of the multiphase ISM is usually described by the sum two log-normal distributions plus a power-law tail at high density that could be a signature of the CNM  which is known to behave like a polytropic gas with an exponent $\gamma < 1$ \citep{passot_density_1998} or a signature of gravity \citep{federrath_universality_2013,girichidis_evolution_2014}. To overcome this issue, and for the sake of simplicity, we assume that a two phase medium is described by two decorrelation scales. All densities below a limit $n_{\rm H}^{\rm lim}$ are supposed to belong to the diffuse log-normal component and to be correlated over a scale $y_{\rm dec}^{\rm diff}$. Similarly, all densities above $n_{\rm H}^{\rm lim}$ are supposed to belong to the dense log-normal component and to be correlated over a smaller scale $y_{\rm dec}^{\rm dens}$. We identified here the limit $n_{\rm H}^{\rm lim}$ between the two log-normal distributions with the inflection point of the 1D probability histogram of the gas density (blue line in Fig. \ref{Fig-PDF-dens}). Since the diffuse component behave like an isothermal gas at the temperature of the WNM, we assume that $y_{\rm dec}^{\rm diff} / L_{\rm drive}$ is constant for all simulations and adopt the value given by \citet{bialy_h_2017}, $y_{\rm dec}^{\rm diff} = 0.2 \times L_{\rm drive}$, where $L_{\rm drive} = L / 2$ is the main driving scale used for the turbulent forcing (see Sect. \ref{sec:turb_forcing}). In contrast, and because the diffuse component occupies most of the volume, we state that $y_{\rm dec}^{\rm dens}$ depends on the total mass of the gas or equivalently its mean density $\overline{n_{\rm H}}$. To simplify, we propose that 
\begin{equation}
y_{\rm dec}^{\rm dens} \propto \overline{n_{\rm H}}^{1/3},
\end{equation}
which means that the impact of changing the total mass of the gas is to change the typical volume of the dense structures by the same factor. Within this framework, the semi-analytical model proposed here therefore depends on a single parameter: the decorrelation scale of the dense component for the fiducial simulation. In the following, we assume $y_{\rm dec}^{\rm dens} = 10$ pc for $\overline{n_{\rm H}}=2$ cm$^{-3}$ which implies a decorrelation length of 6.3, 7.9, and 12.6 pc for $\overline{n_{\rm H}}=0.5$, 1, and 4 cm$^{-3}$, respectively.

It is quite optimistic to believe that the dense and cold component of the ISM can be modeled by a single decorrelation scale $y_{\rm dec}^{\rm dens}$. This component is indeed likely to follow a distribution of sizes which decrease with the local pressure,
hence the local density $n_{\rm H}$. However, and as we show below, such an assumption allows us to highlight important features of the simulations. The essential impact of the distribution of sizes is proven and discussed in Appendix \ref{Sect-distrib-sizes}.

\subsection{Comparison with simulations}

The goal of the model is to offer an explanation on how the PHs of local densities translate into PHs of column densities in a simulation of the multiphase ISM. To test its validity, we generate a series of $\mathcal{N}$ fictitious lines of sight of size $l$ and compare the PHs to those obtained with an equivalent sample of lines of sight extracted from numerical simulations. For each line of sight, we draw a sequence of random realizations of $n_{\rm H}$ out of its known 1D PH using the rejection method. For each draw, the density is supposed to be constant over a distance $y_{\rm dec}^{\rm diff}$ if $n_{\rm H} < n_{\rm H}^{\rm lim}$, and over a distance $y_{\rm dec}^{\rm dens}$ otherwise\footnote{To ensure that the resulting sample matches the original PH, the probability associated to each density is weighted by the inverse of the component size $1/y_{\rm dec}^{\rm diff}$ or $1/y_{\rm dec}^{\rm dens}$.\label{footprob}}. The contribution of this piece of gas to the total column density is therefore computed as $n_{\rm H} y_{\rm dec}^{\rm diff}$ or $n_{\rm H} y_{\rm dec}^{\rm dens}$. In contrast, the contribution of this piece to the column density of H$_2$ is inferred from the expected density profile of H$_2$ over a 1D slab of size $y_{\rm dec}^{\rm diff}$ or $y_{\rm dec}^{\rm dens}$. Throughout the slab, the density of H$_2$ is calculated at equilibrium taking into account local and large-scale extinction and self-shielding as
\begin{equation}
\langle e^{-\sigma_{\rm d} (N_{\rm H}^{\rm loc} + N_{\rm H}^{\rm ext})} \rangle 
\langle f_{\rm shield}\left( \frac{N^{\rm loc}({\rm H}_2) + N^{\rm ext}({\rm H}_2)}{5\times 10^{14} {\rm cm}^{-2}} \right) \rangle, 
\end{equation}
where $N_{\rm H}^{\rm loc}$ and $N^{\rm loc}({\rm H}_2)$ are the local column densities of protons and H$_2$ computed from the border of the slab. The mean values in this expression are calculated from six random realizations of the large-scale column densities $ N_{\rm H}^{\rm ext}$ and $N^{\rm ext}({\rm H}_2)$ drawn out of the sample a lines of sight under construction. The construction of each line of sight ends when its size reaches the integration scale $l$. The entire sample of lines of sight is finally reconstructed until convergence of the large-scale shielding processes described above.

The comparisons between the 1D and 2D PHs reconstructed from the semi-analytical model and extracted from the simulations are shown in Figs. \ref{Fig-compar-sim-mod-Ht} and \ref{Fig-compar-sim-mod-H2} for samples of $\mathcal{N}=64000$ lines of sight, 15 different simulations, and different values of the integration scale $l$, from $L$ down to $y_{\rm dec}^{\rm diff}$. Unexpectedly, setting the decorrelation length of the dense component to $\sim 10$ pc for the fiducial simulation ($\overline{n_{\rm H}} = 2$ cm$^{-3}$) leads to a remarkable agreement between all the fictitious and actuals PHs. Once this parameter is set, the model not only reproduces surprisingly well the shapes of the 1D PHs and of the HI-to-H$_2$ transition, but also their global trends depending on $\overline{n_{\rm H}}$ and $G_0$ and their deformations depending on the chosen integration scale $l$. This result is not straightforward and highly depends on the decorrelation lengths $y_{\rm dec}^{\rm diff}$ and $y_{\rm dec}^{\rm dens}$. The agreement observed in Figs.
\ref{Fig-compar-sim-mod-Ht} and \ref{Fig-compar-sim-mod-H2} therefore suggests that the model somehow captures an essential property of the simulations, namely some characteristic lengths of the diffuse and dense components of a multiphase gas with a turbulence driven at a scale $L_{\rm drive}$.

\subsection{Interpretation of the model}

The statistics of the HI-to-H$_2$ transition derived from the model result from a combination of effects. Locally, the fraction of H$_2$ of a given slab depends on the density and the size of the slab $y_{\rm dec}^{\rm diff}$ and $y_{\rm dec}^{\rm dens}$, which set the local self-shielding, and on the surrounding environment, which sets the large-scale self-shielding. How these local properties contribute to the integrated quantities $N_{\rm H}$ and $N({\rm H}_2)$ depends, in turn, on the sizes of the slabs and on the PH of the density which both control the reconstruction of the line of sight.
\begin{enumerate}
\item Since $y_{\rm dec}^{\rm dens}$ is fixed, the HI-to-H$_2$ transition induced by the local self-shielding alone occurs in any slab with a density larger than 
\begin{equation}
n_{\rm H}^{\rm tr} \propto G_0^{1/2} (y_{\rm dec}^{\rm dens})^{-1/2} \propto G_0^{1/2} (\overline{n_{\rm H}})^{-1/6}.
\end{equation}
This equation can be obtained from the expression of the column densities of HI envelopes (\citealt{sternberg_h_2014}, Eq. 40) in the weak field limit. The large-scale self-shielding not only increases the fraction of H$_2$ in atomic slabs (i.e., for $n_{\rm H} < n_{\rm H}^{\rm tr}$) but also shifts $n_{\rm H}^{\rm tr}$ toward lower values, two effects which depend on the size of the box $L$. For $L=200$ pc, including the large-scale self-shielding is found to reduce $n_{\rm H}^{\rm tr}$ by a factor of two.
\item 
Lines of sight with very low molecular fraction necessarily result from the combination of slabs with $n_{\rm H} < n_{\rm H}^{\rm tr}$. The occurrence of such events depends on the volume filling factor of the diffuse gas, hence on the 1D PH of low density material, and on the integration length $l$: as $l$ increases, their likelihood decreases. Such a scenario occurs for a maximum normalized column density of
\begin{equation}
N_{\rm H}/l = n_{\rm H}^{\rm tr}.
\end{equation}
For $l\sim y_{\rm dec}^{\rm diff}$, such a high normalized column density of atomic gas is a likely event. As $l$ increases, it becomes, however, unlikely to throw a line of sight composed of components with identical densities $n_{\rm H}^{\rm tr}$. Therefore, while the above limit is still valid, the maximum normalized column density of weakly molecular gas appears to decrease.
\item Lines of sight with large molecular fraction necessarily contain at least one slab with $n_{\rm H} > n_{\rm H}^{\rm tr}$. Oppositely to the previous case, the occurrence of such events depends on the volume filling factor of the dense gas, hence on the 1D PH of large density material, and on the integration length $l$: as $l$ increases, their likelihood increases. Such a scenario occurs for a minimum normalized column density 
\begin{equation}
N_{\rm H}/l = n_{\rm H}^{\rm tr} y_{\rm dec}^{\rm dens} / l.
\end{equation}
\item These two limits for lines of sight with low and large $f_{{\rm H}_2}$ (items 2. and 3.), set the width of the HI-to-H$_2$ transition seen in column densities. As shown in Fig. \ref{Fig-compar-sim-mod-H2}, this width is somehow smaller than that obtained from the numerical simulations. We will discuss this point in the next section.
\item At last, lines of sight with intermediate H$_2$ fraction ($10^{-4} \leqslant f_{{\rm H}_2} \leqslant 10^{-2}$) mostly result from a combination of slabs of low and moderate densities ($n_{\rm H} \lesssim n_{\rm H}^{\rm tr}$). Because such events are unlikely, the model predicts a small fraction of lines of sight at intermediate $f_{{\rm H}_2}$, in contradiction with results extracted from the simulations. This point will also be discussed further in the next section.
\end{enumerate}
All these properties fully explain the behaviors of the analytical model observed in Fig. \ref{Fig-compar-sim-mod-H2}. The transition density for the fiducial simulation is $n_{\rm H}^{\rm tr}=4$ cm$^{-3}$. As expected, the corresponding lower and upper limits of $N_{\rm H}/l$ required to activate the HI-to-H$_2$ transition are in agreement with the limits found for the lowest integration scale $l=25$ pc (bottom right panels of Fig. \ref{Fig-compar-sim-mod-H2}). Increasing the integration length has three effects: (a) to squeeze the 2D PH along the x and y axis as both $N_{\rm H}$ and $N({\rm H}_2)$ progressively tend toward Gaussian distributions centered on 
the means, (b) to shift the HI-to-H$_2$ transition to lower $N_{\rm H}/l$ according to the limits derived above, and (c) to increase the occurrence of lines of sight with large $f_{{\rm H}_2}$ to the detriment of lines of sight with low $f_{{\rm H}_2}$ as the probability  of intercepting a slab with $n_{\rm H} > n_{\rm H}^{\rm tr}$ rises. The dependence of the distributions on $G_0$ and $\overline{n_{\rm H}}$ are also straightforward. As $G_0$ increases or $\overline{n_{\rm H}}$ decreases, the HI-to-H$_2$ transition is shifted according to the dependence of $n_{\rm H}^{\rm tr}$ on these parameters. The occurrences of high or low molecular fraction lines of sight simply depend on the volume filling factor of the dense gas, and are therefore a direct consequence of the 1D  PH of the gas density. As $G_0$ increases or $\overline{n_{\rm H}}$ decreases, the mean thermal pressure rises and the mass fraction of the CNM diminishes. This favors the occurrence of lines of sight with low molecular fraction.

\subsection{Discrepancies and conclusions}
\label{Sect-distrib-sizes}

Despite the surprising agreement between the simulated and the modeled PHs, in particular regarding the 1D PH of the total column density $N_{\rm H}$, Figs. \ref{Fig-compar-sim-mod-Ht} and \ref{Fig-compar-sim-mod-H2} also reveal important discrepancies. 
Most notably, and as mentioned above, the modeled 2D PHs systematically underestimate the widths of the HI-to-H$_2$ transition and underestimate the proportion of lines of sight at intermediate integrated molecular fraction. These strong discrepancies are entirely due to the hypothesis of a constant decorrelation scale of the dense component.

Obviously, the dense and cold ISM are not characterized by a unique scale but a distribution of sizes which likely decrease with the gas pressure and local density. Such a distribution would increase the probability of occurrence of small components of high density along any line of sight (see footnote \ref{footprob}) and therefore solve the discrepancy between the analytical model and the simulations. Indeed, as schematized in Fig. \ref{Fig-scheme-los}, small and dense components surrounded by diffuse material favor the occurrence of lines of sight at intermediate molecular fractions. This configuration not only reduces the mean molecular fraction predicted by the model but also necessarily widens the HI-to-H$_2$ transition, in closer agreement with the simulations. To illustrate this point, we display in Fig. \ref{Fig-compar-sim-mod-H2_2} the 2D  PHs reconstructed from the semi-analytical model assuming that $y_{\rm dec}^{\rm dens}$ is a power law function of the gas density. 

In other words, the excellent agreement observed in Fig. \ref{Fig-compar-sim-mod-Ht} indicates that setting constant decorrelation scales $y_{\rm dec}^{\rm diff}$ and $y_{\rm dec}^{\rm dens}$ is sufficient to reproduce the distribution of the total quantity of matter $N_{\rm H}$. It is so because the chosen $y_{\rm dec}^{\rm dens}$ probably describes the largest sizes of the dense
components which capture most of its mass and volume. However, the model also proves that choosing a single value of $y_{\rm dec}^{\rm dens}$ is inappropriate to accurately describe the distribution of H$_2$. It is so because the mass and volume of H$_2$ in the simulation is likely built in smaller components. $y_{\rm dec}^{\rm dens}$ should therefore be interpreted as a maximum length scale of the dense gas.

All these considerations show that the simplistic model developed here is very useful to interpret the results of the simulations. It successfully separates local properties and probabilistic effects in the integration of column densities. Moreover it provides estimations of the decorrelation scale of the diffuse gas and the maximum decorrelation scale of the dense gas. Finally, its flaws clearly highlight the importance of a distribution of sizes of the dense and cold ISM and the necessary existence of small dense clouds which produce lines of sight with intermediate integrated molecular fraction. The findings of this section are synthesized in Sect. \ref{sec:interpretmodel} and Fig. \ref{Fig-scheme-los} and used in the rest of the paper as a major tool for interpreting the results of the simulations.

\section{Kolmogorov-Smirnov test}\label{app:KStest}

\begin{figure}
\centering
\includegraphics[width=1.\linewidth]{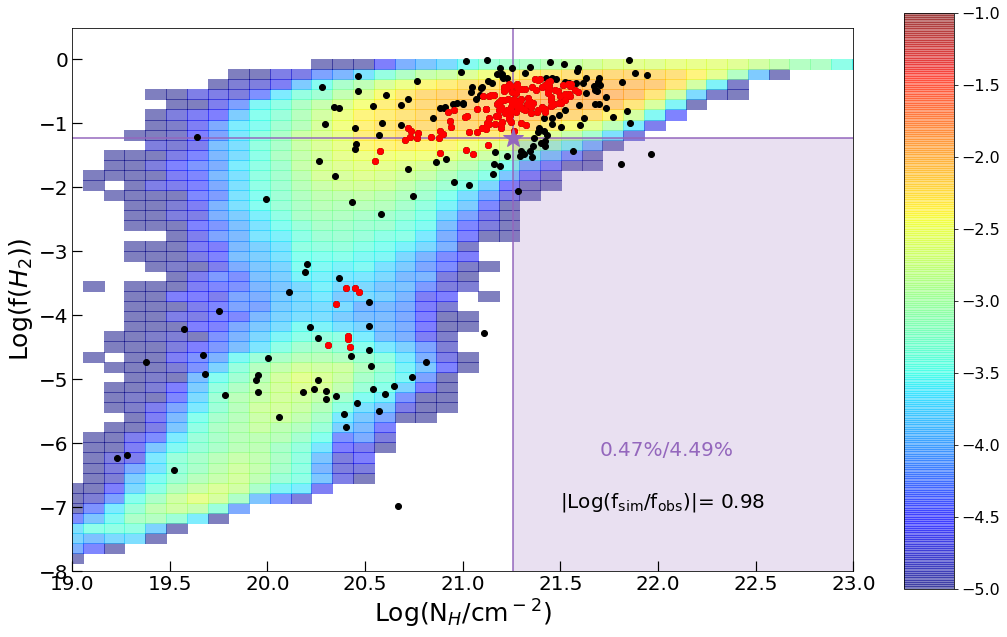}
\caption{Results of the modified KS test applied to the fiducial simulation. The black points are the observational data, the red dots indicate the dataset used for the estimations of the merit function $M$, and the 2D histogram the simulated data. The violet star and rectangle indicate the observational point and the quadrant that maximize $M$ (see main text). The fiducial simulation has a KS distance of 0.98. The corresponding quadrant contains 0.47\% and 4.49\% of the entire simulated and observed datasets.}
\label{fig:KSratio_fiducialmodel}
\end{figure}
		
\begin{figure}
\centering
\includegraphics[width=1\linewidth]{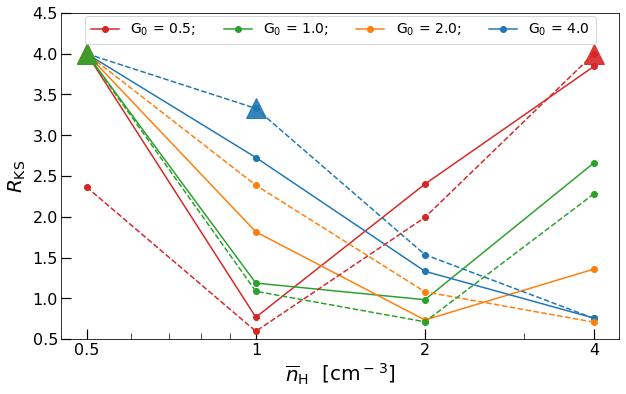}
\caption{KS distance between the simulations and the observational sample as a function  of the mean density $\overline{n_{\rm H}}$, the UV scaling factor $G_0 = 0.5$ (red), 1 (green), 2 (orange), and 4 (blue), and for a resolution of $256^3$ (solid lines) and $128^3$ (dashed lines). All other parameters are set to their standard values (see
Table \ref{table:grids}). Points correspond to reliable measurements of the KS distances. Triangles indicate lower limits corresponding to simulations where the upper error bar on $R_{\rm KS}$ tends toward infinity (see main text).}
\label{fig:KSstat_Resolutions}
\end{figure}

		The results of this paper rely on the comparison of 2D probability histograms of observed and simulated data. To facilitate this comparison and the underlying parametric study, we apply here a modified version of the Kolmogorov-Smirnov (KS) test. This test, originally developed for the study of 1D samples, searches for the maximum cumulative difference between two distributions. \cite{fasano_multidimensional_1987} generalized the KS test to 2D samples following \cite{peacock_two-dimensional_1983} idea of replacing the cumulative probability distribution, which is not well defined in a dimension larger than one, with the integrated probability in each of the four quadrants surrounding a datapoint. Such a consideration allows us to define a KS distance which measures how two 2D distribution functions differ from one another.
		
		As illustrated in Fig. \ref{fig:KSratio_fiducialmodel}, each observational datapoint is identified by a pair of variables ($N_{\rm H}^{\rm obs}$, $f^{\rm obs}({\rm H}_2)$) which divide the space into four quadrants: 
		(1) $N_{\rm H} \leq N_{\rm H}^{\rm obs}$ \& $f_{{\rm H}_2} \leq f^{\rm obs}({\rm H}_2)$, (2) $N_{\rm H} \leq N_{\rm H}^{\rm obs}$ \& $f_{{\rm H}_2} > f^{\rm obs}({\rm H}_2)$, (3) $N_{\rm H} > N_{\rm H}^{\rm obs}$ \& $f_{{\rm H}_2} > f^{\rm obs}({\rm H}_2)$, and (4) $N_{\rm H} > N_{\rm H}^{\rm obs}$ \& $f_{{\rm H}_2} \leq f^{\rm obs}({\rm H}_2)$.
		Each quadrant thus contains two fractions $f_{\rm obs}$ and $f_{\rm sim}$ of the entire observed and simulated datasets. To compare these values, we define a merit function 
		\begin{equation}
		M = \left|\log_{10} \left(\frac{f_{\rm sim}}{f_{\rm obs}}\right)\right|,
		\end{equation}
		and the KS distance between the two distributions as the maximum value of $M$ computed over all quadrants of an observational dataset $\mathcal{O}$,
		\begin{equation}
		R_{\rm KS} = \max_\mathcal{O}(M).
		\end{equation}
		This procedure, not only provides a measurement of the difference between the two distributions, but also the datapoint and the quadrant that maximize the merit function (see Fig. \ref{fig:KSratio_fiducialmodel}). The interpretation is also straightforward. For instance, a KS distance of 1 implies that one of all the quadrants scanned contains 10 times fewer or 10 times more observations than simulated lines of sight, and that all the other quadrants have smaller distances.
		
		The stability of the procedure depends on the errors made on the merit function and therefore on the absolute numbers of observed and simulated lines of sight contained in each quadrant. The observational dataset $\mathcal{O}$ used to compute the KS distance (red points in Fig. \ref{fig:KSratio_fiducialmodel}) is chosen as the subsample such that all quadrants scanned contain at least 10 observed  lines of sight. 
		With this assumption, the error on the merit function is calculated by taking into account only the statistical errors on the number of simulated lines of sight. For each quadrant, we assume that the "true" merit function ranges between 
		\begin{align}
		M_{\rm min} &= \left|\log_{10} \left(\frac{f_{\rm sim}-3 \sqrt{(f_{\rm sim}/S)}}{f_{\rm obs}}\right)\right| \,\, {\rm and}\\
		M_{\rm max} &= \left|\log_{10} \left(\frac{f_{\rm sim}+3 \sqrt{(f_{\rm sim}/S)}}{f_{\rm obs}}\right)\right|,
		\end{align}
		where $S$ is the total number of simulated lines of sight.
		Because the errors are asymmetric, $M_{\rm min}$ can tend toward infinity. If so, the KS distance is a lower limit, even if the infinite error bar is obtained for another quadrant than the one that maximizes $M$.
		In short, for each observational datapoint, we compute $M_{\rm min}$, $M$, $M_{\rm max}$,
		and $R_{\rm KS}$. If one of the $M_{\rm min}$ goes to infinity, $R_{\rm KS}$ is considered
		as a lower limit.
		
		Because the division in quadrants is performed on a cartesian grid, the procedure also depends on the choice of the axes. Mathematically, the best option would be to identify the principal components of the observational sample using proper orthogonal decomposition or singular value decomposition algorithms. Such a method could even be applied to subsamples in order to adaptively rotate the system of axes and follow more precisely the distribution of observations. In any case, the resulting system would be a linear combination of $N_{\rm H}$ and $N({\rm H}_2)$ which could be difficult to relate to the underlying physical properties. Because the molecular fraction is bimodal as a function  of $N_{\rm H}$, we choose here $N_{\rm H}$ and $f_{{\rm H}_2}$ as primary variables. This choice facilitates the physical interpretation of the KS test while ensuring some homogeneity of the spread of the observational data in all quadrants (see Fig. \ref{fig:KSratio_fiducialmodel}).
		
		As a proof of concept, we display in Fig. \ref{fig:KSstat_Resolutions} the results of the KS test applied to a grid of simulations obtained for different values of $\overline{n_{\rm H}}$ and $G_0$ and two different resolutions. Despite its simplicity, this test appears to capture the 
		main behaviors described in Sects. \ref{sec:impact_res} and \ref{sec:impact_dens}. In particular, the KS distance	between the simulations and the observations is found to strongly depend on both	$\overline{n_{\rm H}}$ and $G_0$, and more loosely on the resolution. Moreover, the simulations that minimize $R_{\rm KS}$ are found to be identical to those identified in Sect. \ref{sec:impact_dens} to be in closest agreement with the observations. Interestingly, the	value of $R_{\rm KS}$ obtained for $\overline{n_{\rm H}}=4$ cm$^{-3}$ and $G_0=4$ is relatively
		small, in apparent contradiction with the conclusions of Sect. \ref{sec:impact_dens}. This is due to the limit imposed on the minimum number of observations contained in each quadrant. Because of this limit, several observations at large $N_{\rm H}$ are not
		included in the analysis which reduces the merit function at large column density (region E, see Table \ref{Tab-regionHH2}). Keeping in mind these border effects, the KS test turns out to be a valuable tool for estimating the distance between two distributions 	without performing a detailed comparison of the samples. In this paper, we apply this method to display our results in a synthetic manner (see Sects. \ref{sec:impact_grav}, \ref{sec:impact_turb}, and \ref{sec:impact_mag}) and only
		give additional details when necessary.


\end{document}